\pdfoutput=1
\documentclass[bibyear]{aa}

\usepackage{natbib}
\bibpunct{(}{)}{;}{a}{}{,} 
\usepackage{xcolor}
\usepackage{graphicx}
\usepackage{txfonts}
\begin{document}

   \title{Stellar populations in hosts of giant radio galaxies and their neighbouring galaxies}

   \subtitle{}

   \author{A. Ku\'zmicz
          \inst{1,2,3},
          B. Czerny\inst{1},
          \and
          C. Wildy\inst{1}
          }

\institute{Center for Theoretical Physics, Polish Academy of Sciences, Al. Lotnik\'ow 32/46, 02-668 Warsaw, Poland\\
\email{cygnus@oa.uj.edu.pl}
\and
Astronomical Observatory, Jagiellonian University, ul. Orla 171, 30-244 Krakow, Poland
\and Queen Jadwiga Astronomical Observatory in Rzepiennik Biskupi, 33-163 Rzepiennik Strzy\.zewski, Poland
}

   \date{Received September XX, XXX; accepted XXX, 2018}

 
  \abstract
   {Giant radio galaxies (with projected linear size of radio structure larger than 0.7 Mpc) are very rare and unusual objects. Only $\sim$5\% of extended radio sources reach such sizes. Understanding of the processes responsible for their large sizes is crucial to further our knowledge about the radio source's evolution.}
   {We check the hypothesis that giants become extremely large due to the specific history of their host galaxy formation, as well as in the context of the cluster or group of galaxies where they evolve. Therefore we study the star formation histories in their host galaxies and in galaxies located in their neighbourhood.  }
   {We studied 41 giant-size radio galaxies as well as galaxies located within a radius of 5 Mpc around giants to verify whether the external conditions of the intergalactic medium somehow influence the internal evolution of galaxies in the group/cluster. We compared the results with a control sample of smaller-sized Fanaroff--Riley type II radio galaxies and their neighbouring galaxies. We fit stellar continua in all galaxy spectra using the spectral synthesis code STARLIGHT and provide statistical analysis of the results. }
   {We find that hosts of giant radio galaxies have a larger amount of intermediate age stellar populations compared with smaller-sized FRII radio sources. The same result is also visible when we compare neighbouring galaxies located up to 1.5 Mpc around giants and FRIIs. This may be evidence that star formation in groups with giants was triggered due to global processes occurring in the ambient intergalactic medium. These processes may also contribute to mechanisms responsible for the extremely large sizes of giants.}
   {}


   \maketitle


\section{Introduction}

Amongst many types of extragalactic radio sources, which cover a wide range of radio structures, morphologies and sizes, the giant radio sources (GRS) are very peculiar ones. The linear sizes of their radio structures are defined to be larger than 0.7 Mpc (assuming $H_{0}$=71 km/s/Mpc, $\rm \Omega_{M}$ = 0.27, $\rm \Omega_{vac}$ = 0.73; \cite{spargel2003}), what is comparable with sizes of galaxy clusters. The class of GRSs is not very large. To date we know just 348 confirmed GRSs \citep{kuzmicz2018} but that number is still growing, thanks to low-frequency telescopes, high resolution radio surveys and large spectroscopic surveys.\\
Previous studies focused on the properties of individual objects (e.g. \cite{jamrozy2005, subrahmanyan2006, konar2009, orru2010, machalski2011}), but a few studies also consider larger samples of giants. They concentrate on the role of some factors which could be responsible for the gigantic size of radio structures. In these studies the authors consider the properties of the ambient intergalactic medium (IGM; \cite{machalski2006, subrahmanyan2008, kuligowska2009}), the advanced age of the radio structures (e.g. \cite{mack1998, machalski2009}), recurrent radio activity (e.g. \cite{subrahmanyan1996, schoenmakers2000, machalski2011}), as well as the radio core and the central active galactic nuclei's (AGN) specific properties \citep{ishwara1999, kuzmicz2012}. \\
Studies carried out in recent years show that the GRSs can be used as barometers of the intergalactic medium. When the radio lobes expand, they first interact with the interstellar medium, then the intergalactic, and finally with the intracluster medium. These interactions can be a reason for asymmetries in radio structures. \citet{subrahmanyan2008} and \citet{safouris2009} showed that there is a clear connection between properties of radio lobes and the distribution of neighbouring galaxies. They showed that the asymmetries in radio morphology of some giants can be a result of inhomogeneities in the distribution of ambient intergalactic gas, which follows the large scale structure of the universe. \\
The series of investigations by \citet{chen2011a, chen2011b, chen2012a, chen2012b} focus on the environmental properties around a few giant radio sources. In their studies they analyse the distribution and properties of companion galaxies around giants and find that they tend to lie near the radio lobes. They also show that in some cases (e.g. NGC6251, NGC315) the velocity dispersion of group members is not consistent with that expected from correlation curves of X-ray luminosity versus velocity dispersion \citep{mulchaey1998}. They conclude that the density of X-ray emitting gas is unusually low around studied giants and it can be the explanation of their extremely large sizes of radio structures.\\
GRSs are also very valuable tools for investigating the large-scale structure of the Universe. The authors \citet{malarecki2013, malarecki2015}, \citet{pirya2012}, \citet{peng2015} used their large sizes to probe the distribution of the warm-hot intergalactic medium (WHIM) in filaments of the large-scale structure of the Universe focusing on evidence of radio lobe  interactions with the ambient medium.\\
On the other hand, the role of the environment for the galaxy properties and evolution is not irrelevant. The past mergers influence, for example, the galaxy morphology, star formation, and accretion processes. Also the cluster density is closely related to the morphological types of its galaxies. It has been shown that early-type galaxies dominate high density environments in contrast to late-type galaxies that dominate low-density ones \citep{dressler1980}. There are also a few studies aimed at the connection between cluster environment and star formation history. \citet{moran2005} find that the central galaxies in clusters are older than those at larger distances from cluster centre. Furthermore, \citet{demarco2010} showed that the dense cluster environment stops star formation in low mass galaxies when they enter the cluster. They also find that less massive galaxies formed stars more recently than more massive ones.\\
There are also numerous studies investigating the stellar populations of radio sources (e.g. \cite{holt2007, wills2008}). They concentrated on the identification of the young stellar populations in radio galaxies to establish the timescales of radio activity relative to the merger event. The young stellar populations are observed in $\sim$15--25\% of all powerful extragalactic radio sources \citep{tadhunter2011}. They found that in most of those galaxies the radio activity occurs simultaneously with the starburst and it is explained as a result of a merger event with a gas-rich galaxy. There are also a group of radio galaxies where the radio activity is triggered a long time after the starburst.  \citet{raimannl2005} find that radio galaxies are dominated by intermediate age ($\sim$1 Gyr) stars, suggesting a connection between the radio activity and a starburst which occurred 1 Gyr ago. They also propose that more massive starbursts have led to more powerful radio emission.

In our analysis we have attempted to answer the question of whether or not the history of giant radio galaxy (GRG) host formation may be responsible for the growth of its radio structures. Galaxy formation is related to internal processes such as star formation, but it also depends on the global properties of the ambient medium, intergalactic gas and galaxies which comprise the galaxy cluster. In this paper we present the results of a stellar population analysis for the sample of GRGs, but also extend our studies to galaxies located in the same group/cluster as giants. We used a control sample of smaller-sized radio galaxies to look for systematic differences between stellar populations of giants and non-giants, and to find such properties of GRGs which distinguish them from smaller radio galaxies, which may be responsible for GRGs origin.\\
The paper is organized as follows. In Section 2 we present the sample of galaxies used in our analysis, in  Section 3 we describe the data reduction procedures and methods of spectral synthesis, in Section 4 and 5 we discuss our results, and in Section 6 we present the summary and the conclusions. 

\section{Data and sample selection}

\subsection{Selection of GRGs and comparison sample}
\begin{figure}
\centering
    \includegraphics[width=0.98\columnwidth]{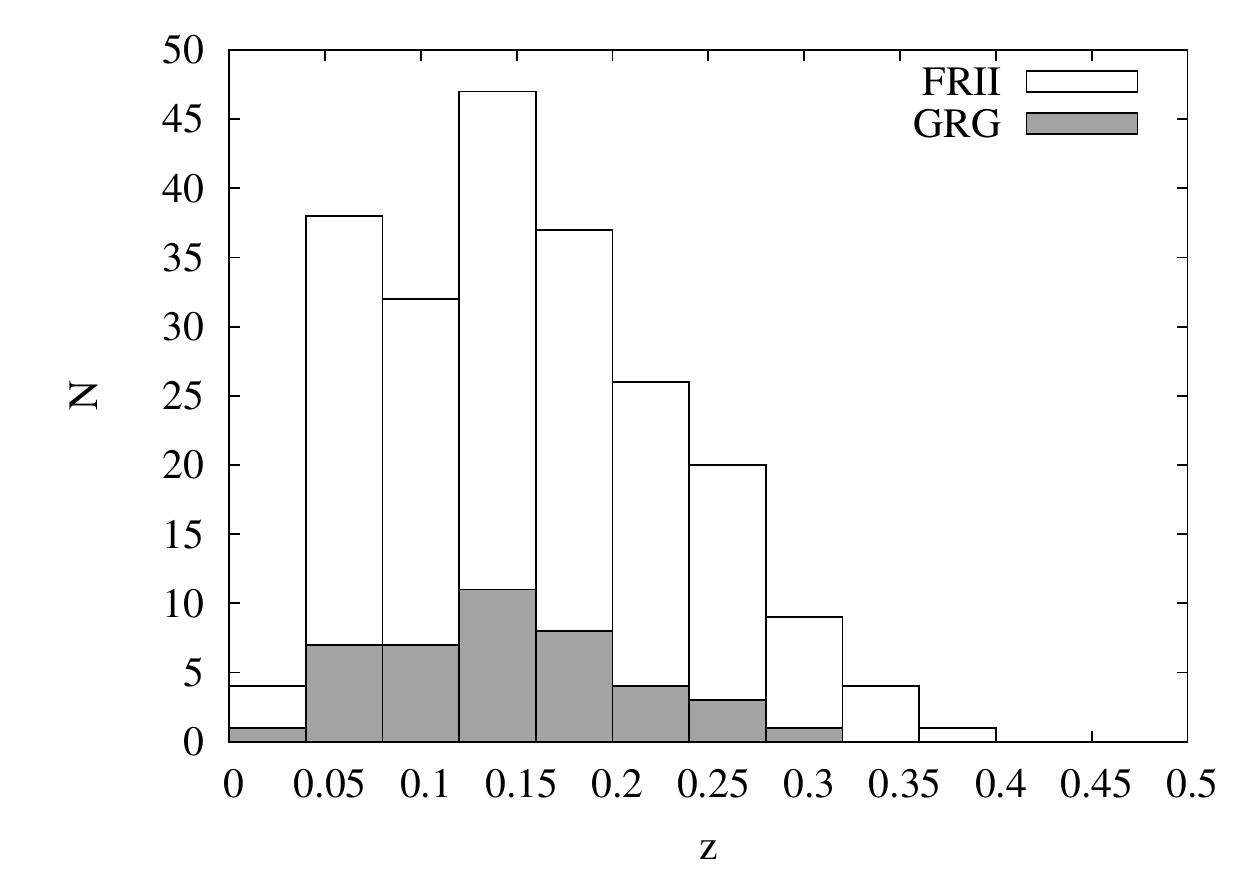}\\
    \includegraphics[width=0.98\columnwidth]{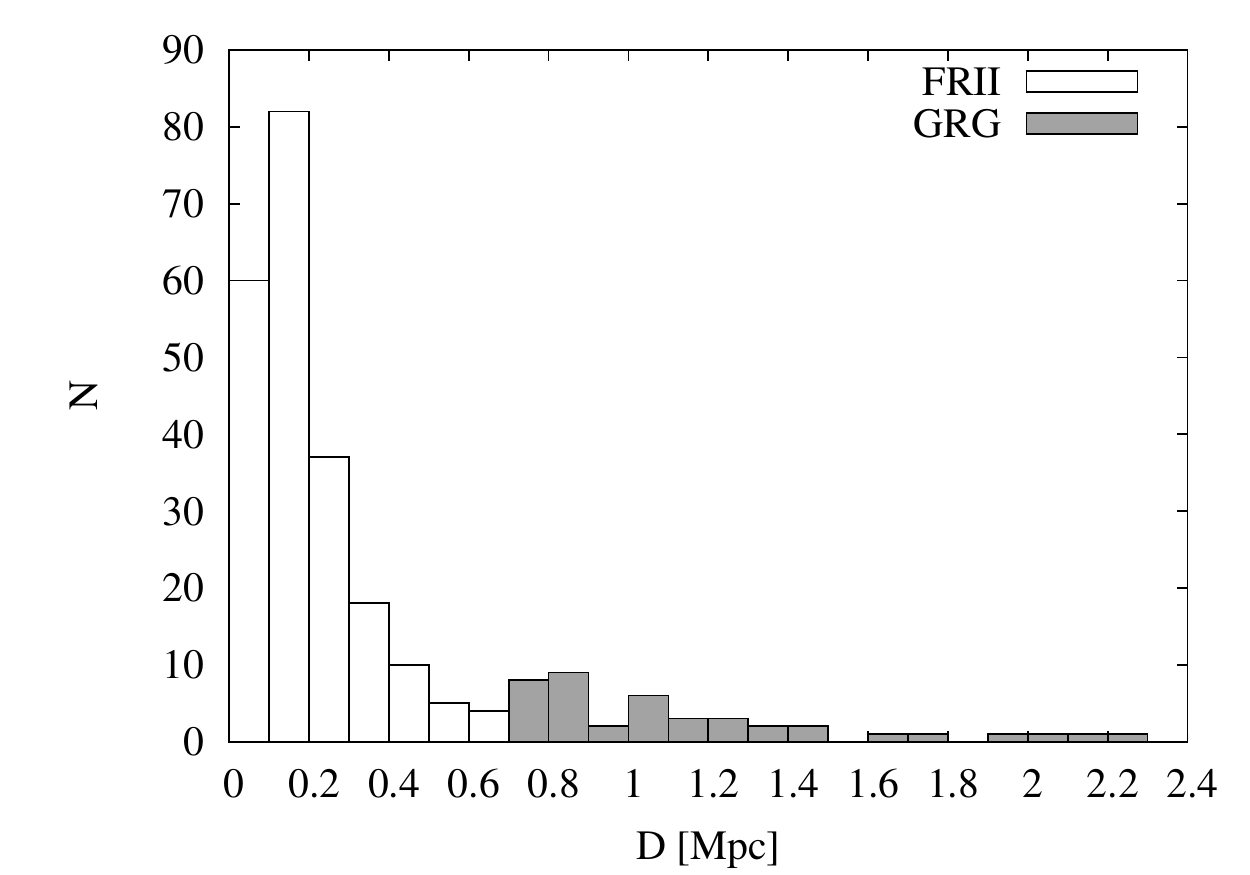}\\
    \includegraphics[width=0.98\columnwidth]{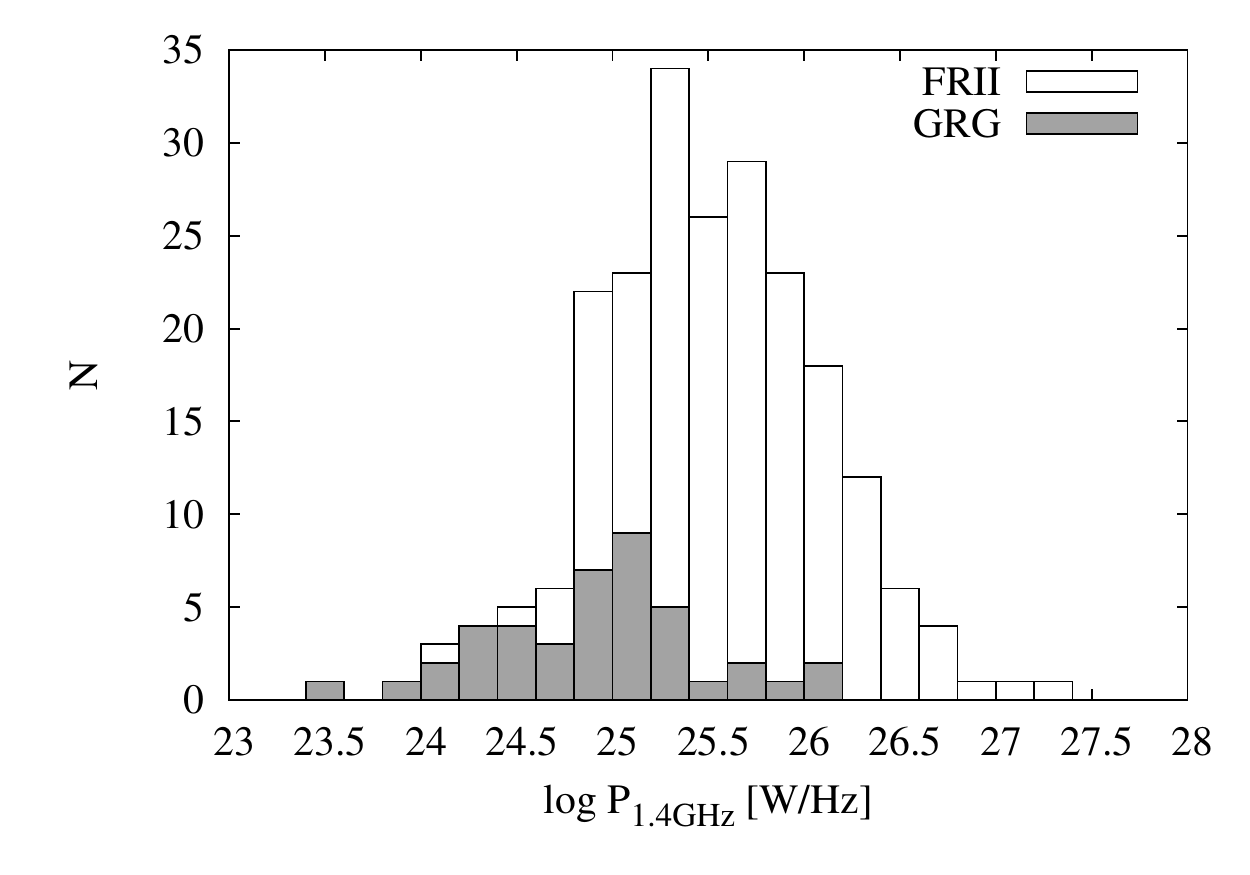}
\caption{Distributions of redshift (top panel), projected linear size (middle panel) and 1.4 GHz total luminosity (bottom panel) for GRG and FRII samples studied in this paper.}
\label{zDP}
\end{figure}

The sample of GRGs is extracted from the catalogue of GRSs by \citet{kuzmicz2018}. From their sample we selected 72 galaxies for which optical spectra were available in Sloan Digital Sky Survey Data Release 13 (SDSS DR13; \cite{albareti2016}). We restrict this number to 41 by including only those galaxies around which we found neighbouring galaxies (at least one galaxy) for which SDSS optical spectra are also available. The details of the selection process for neighbouring galaxies are presented in Section \ref{neighbours}. In our analysis we are required to use only spectra of good quality. The selection criteria restricted the sample to nearby GRGs with redshifts in the range of 0.03$<$z$<$0.31, with mean 1.4 GHz total radio luminosity logP$_{tot}$=24.92 and mean projected linear size D=1.2 Mpc. \\
As a comparison sample, we used the FRII-type radio galaxies from \citet{koziel2011}, in which the authors study properties of 401 FRII radio galaxies with a wide range of radio powers and radio structure sizes. Among them there are also 18 GRGs, and therefore we excluded them from the comparison sample. Similarly to GRGs selected for further analysis, we used only those radio sources for which we found at least one companion galaxy with an available optical spectrum (see Section \ref{neighbours}). As a result, the final comparison sample consists of 217 FRII radio galaxies in a redshift range of 0.008$<$z$<$0.4 and with a mean projected linear size D=0.2 Mpc.\\
In Figure \ref{zDP} we present the characteristics of radio galaxies considered in this paper. We plot the distribution of redshifts, projected linear sizes and 1.4 GHz total radio luminosities for the sample of GRGs and FRIIs. All of the considered radio galaxies are nearby objects (up to z=0.4) with a wide range of radio powers.

\subsection{Selection of neighbouring galaxies}
\label{neighbours} 

We looked for neighbouring galaxies around each radio galaxy (GRG, and FRII from comparison sample) using SDSS DR13. We selected all galaxies within a radius of about $\sim$5 Mpc from radio galaxy host with measured spectroscopic redshifts corresponding to the redshift of radio galaxy host. We adopted the redshift difference between a neighbouring galaxy and the radio galaxy host equal to $\Delta z\leqslant$ 0.003 that corresponds to $\sim$800km/s. The total number of neighbouring galaxies found around all of GRGs was 789 and around FRII radio galaxies was 3692. All of selected galaxies was used in further analysis. The completeness of spectroscopically selected group/cluster members depends on the completeness of SDSS. The SDSS main spectroscopic galaxy sample is complete within the magnitude range 14$<$r$<$18. The hosts of radio galaxies are usually associated with the brightest galaxy in the group and the neighbouring galaxies are up to few magnitude fainter. For bright radio galaxy hosts ($\sim$14 SDSS r-band magnitude) the completeness of fainter spectroscopic group members is $\sim$90\% (in the range of 14$<$r$<$19 mag ), but for weak radio source hosts (with $\sim$18 SDSS r magnitude) the completeness of spectroscopic data below r$>$18 mag is much lower. Therefore, we counted neighbouring galaxies based on the SDSS photometric data to see how spectroscopic selection can be incomplete. We counted all galaxies which are up to five magnitudes fainter than hosts of radio sources requiring their photometric redshift estimations to correspond to the spectroscopic redshift of the radio galaxy host with the $\Delta$z$_{phot}\leqslant$0.02, equal to the error of SDSS photometric redshift estimations \citep{beck2016}. 
 
In Table \ref{tab1} we list the principal parameters of analysed GRGs, arranging it as follows: column 1 -- galaxy name; columns 2 and 3 -- J2000.0 galaxy coordinates; column 4 -- redshift; column 5 -- linear size; column 6 -- number of galaxies in group with available spectroscopic data; column 7 -- number of galaxies within the radius of 0.5 Mpc from GRG; column 8 -- number of galaxies between radius of 0.5 Mpc and 1 Mpc around GRG; column 9 -- number of galaxies between radius of 1 Mpc and 1.5 Mpc around GRG; column 10 -- number of galaxies between radius of 1.5 Mpc and 3 Mpc around GRG; column 11 -- number of galaxies between radius of 3 Mpc and 5 Mpc around GRG; column 12 -- number of galaxies in group which are five magnitudes fainter than host of GRG with $\Delta$z$_{phot}\leqslant$0.02.\\

We can see that the number of galaxies selected from photometric data (column 12) is much larger than number of galaxies selected from spectroscopic data (column 6). However, the number of photometrically selected galaxies should be treated with a caution because of large error of SDSS photometric redshift estimation which causes that some of selected galaxies could not belong to the same galaxy group/cluster. Our studies base only on the spectroscopically selected galaxies despite of the fact that in some groups the completeness can be low. However, we have not studied particular group of galaxies, but groups in general, therefore the low completeness in some groups does not affect the final results in a significant way. All the radio maps of GRGs and the positions of neighbouring galaxies with available spectroscopic and photometric radshifts, are presented in Appendix A.

\begin{table*}
\caption{GRGs studied in this paper.}
\label{tab1}
\centering
\begin{tabular}{c c c c c c c c c c c c }
\hline
IAU             &$\alpha$(2000)        &$\delta$(2000)        & z             &D              & n             & n$_{0.5}$  & n$_{0.5-1}$  & n$_{1-1.5}$  & n$_{1.5-3}$ & n$_{3-5}$ &  n$_{m,z}$          \\
name            &  (h m s)             &($\rm^{o}$ ' ")       &               & (Mpc)         &               &   &   &   &   &  &                                 \\
(1)             & (2)                  & (3)                  &(4)            & (5)           & (6)           & (7)  & (8)  & (9)  & (10)  & (11) & (12)                          \\
\hline
J0003$+$0351    &       00 03 31.50     &       $+$03 51 11.3   &       0.095   &       2.03    &       12              & 2  & 1  & 1  & 4  & 3 & 142     \\
J0010$-$1108    &       00 10 49.69     &       $-$11 08 12.9   &       0.077   &       0.80    &       20              & 3  & 2  & 1  & 6  & 7 & 99      \\
J0042$-$0613    &       00 42 46.85     &       $-$06 13 52.6   &       0.124   &       0.85    &       6               & 0  & 2  & 1  & 1  & 1 & 125     \\
J0115$+$2507    &       01 15 57.24     &       $+$25 07 20.3   &       0.184   &       1.06    &       3               & 0  & 2  & 0  & 0  & 0 & 147     \\
J0120$-$0038    &       01 20 12.51     &       $-$00 38 37.8   &       0.235   &       0.71    &       6               & 0  & 0  & 0  & 3  & 2 & 108     \\
J0134$-$0107    &       01 34 12.80     &       $-$01 07 28.2   &       0.079   &       1.21    &       54              & 9  & 2  & 4  & 12 & 26 &        168     \\
J0135$-$0044    &       01 35 25.66     &       $-$00 44 47.3   &       0.156   &       1.06    &       9               & 0  & 1  & 0  & 2  & 5 & 128     \\
J0259$-$0018    &       02 59 42.88     &       $+$00 18 40.9   &       0.183   &       0.73    &       5               & 1  & 0  & 1  & 0  & 2 & 159     \\
J0751$+$4231    &       07 51 08.79     &       $+$42 31 23.6   &       0.203   &       1.19    &       2               & 0  & 0  & 0  & 0  & 1 & 85      \\
J0857$+$0131&   08 57 01.76     &       $+$01 31 30.9   &       0.273   &       1.30    &       2               & 0  & 0  & 0  & 0  & 1&  95      \\
J0858$+$5620    &       08 58 32.78     &       $+$56 20 14.7   &       0.240   &       0.87    &       4               & 0  & 0  & 0  & 2  & 1 & 92      \\
J0902$+$1737    &       09 02 38.42     &       $+$17 37 51.4   &       0.164   &       1.19    &       3               & 0  & 0  & 1  & 1  & 0 & 90      \\
J0914$+$1006    &       09 14 19.53     &       $+$10 06 40.5   &       0.308   &       1.71    &       5               & 1  & 0  & 0  & 2  & 1 & 135     \\
J0918$+$3151    &       09 18 59.42     &       $+$31 51 40.6   &       0.062   &       0.78    &       31              & 5  & 3  & 4  & 9 & 9 &  93      \\
J0926$+$6519    &       09 26 00.90     &       $+$65 19 23.0   &       0.140   &       0.78    &       12              & 2  & 2  & 0  & 4  & 3 & 79      \\
J0932$+$1611    &       09 32 38.32     &       $+$16 11 57.8   &       0.191   &       0.76    &       2               & 0  & 0  & 0  & 0  & 1&  86      \\
J1004$+$5434    &       10 04 51.83     &       $+$54 34 04.4   &       0.047   &       0.81    &       125             & 9 & 9  & 12 & 36 & 58  &        131     \\
J1006$+$3454    &       10 06 01.77     &       $+$34 54 10.2   &       0.099   &       4.23    &       12              & 0  & 5  & 0  & 4  & 2&  69      \\
J1021$+$1217    &       10 21 24.22     &       $+$12 17 05.3   &       0.129   &       1.97    &       16              & 1  & 2  & 3  & 4  & 5 & 142     \\
J1021$+$0519    &       10 21 31.47     &       $+$05 19 01.0   &       0.156   &       2.23    &       9               & 1  & 1  & 1  & 2  & 3 & 111     \\
J1032$+$2756    &       10 32 14.09     &       $+$27 56 00.2   &       0.085   &       1.04    &       19              & 3  & 2  & 3  & 1  & 9 & 113     \\
J1032$+$5644    &       10 32 59.02     &       $+$56 44 53.8   &       0.045   &       0.97    &       84              & 9  & 9  & 8  & 18 & 39 &        74      \\
J1111$+$2657    &       11 11 24.97     &       $+$26 57 46.6   &       0.034   &       1.12    &       149             & 10 & 14 & 12 & 31 & 81 &        245     \\
J1147$+$3501    &       11 47 22.12     &       $+$35 01 08.0   &       0.063   &       0.85    &       31              & 7  & 5  & 2  & 7  & 9 & 88      \\
J1247$+$6723&   12 47 33.33     &       $+$67 23 16.5   &       0.107   &       1.35    &       12              & 0  & 2  & 1  & 5  & 3 & 121     \\
J1253$+$4041    &       12 53 12.28     &       $+$40 41 23.7   &       0.229   &       1.01    &       2               & 0  & 0  & 0  & 1  & 0 & 90      \\
J1308$+$6154&   13 08 44.75     &       $+$61 54 15.3   &       0.162   &       1.48    &       2               & 0  & 0  & 0  & 0  & 1 & 105     \\
J1311$+$4059    &       13 11 43.06     &       $+$40 59 00.0   &       0.110   &       0.74    &       15              & 3  & 3  & 2  & 1  & 5 & 102     \\
J1327$+$5749    &       13 27 41.32     &       $+$57 49 43.4   &       0.120   &       1.61    &       15              & 3  & 0  & 3  & 3  & 5 & 196     \\
J1328$-$0307    &       13 28 34.33     &       $-$03 07 45.0   &       0.085   &       1.28    &       33              & 1  & 4  & 5  & 7  & 15 &        142     \\
J1345$+$5403    &       13 45 57.50     &       $+$54 03 17.0   &       0.163   &       0.80    &       3               & 0  & 0  & 1  & 0  & 1 & 170     \\
J1400$+$3019    &       14 00 43.44     &       $+$30 19 18.2   &       0.206   &       2.19    &       7               & 2  & 0  & 0  & 1  & 3 & 140     \\
J1409$-$0302    &       14 09 48.85     &       $-$03 02 32.5   &       0.138   &       1.37    &       10              & 2  & 1  & 0  & 4  & 2 & 138     \\
J1418$+$3746    &       14 18 37.65     &       $+$37 46 24.5   &       0.135   &       1.09    &       14              & 0  & 3  & 1  & 4  & 5 & 153     \\
J1428$+$2918    &       14 28 19.24     &       $+$29 18 44.2   &       0.087   &       1.42    &       17              & 0  & 1  & 1  & 7  & 7 & 128     \\
J1429$+$0715    &       14 29 55.38     &       $+$07 15 12.9   &       0.055   &       0.71    &       47              & 3  & 5  & 4  & 10  & 24 &       116     \\
J1507$+$0234    &       15 07 03.78     &       $+$02 34 07.2   &       0.124   &       0.83    &       7               & 0  & 2  & 1  & 2  & 1 & 132     \\
J1540$-$0127    &       15 40 56.82     &       $-$01 27 10.2   &       0.149   &       0.76    &       7               & 1  & 1  & 1  & 1  & 2 & 112     \\
J1555$+$3653    &       15 55 00.42     &       $+$36 53 37.4   &       0.247   &       1.34    &       3               & 0  & 0  & 1  & 1  & 0&  106     \\ 
J1615$+$3826    &       16 15 52.25     &       $+$38 26 31.8   &       0.185   &       0.81    &       5               & 1  & 0  & 1  & 1  & 1&  173     \\
J1635$+$3608    &       16 35 22.54     &       $+$36 08 04.7   &       0.165   &       0.90    &       10              & 1  & 3  & 2  & 2  & 1 & 148     \\
\hline
\end{tabular} 
\end{table*} 

\section{Optical analysis}

\begin{figure}
\centering
    \includegraphics[width=0.92\columnwidth]{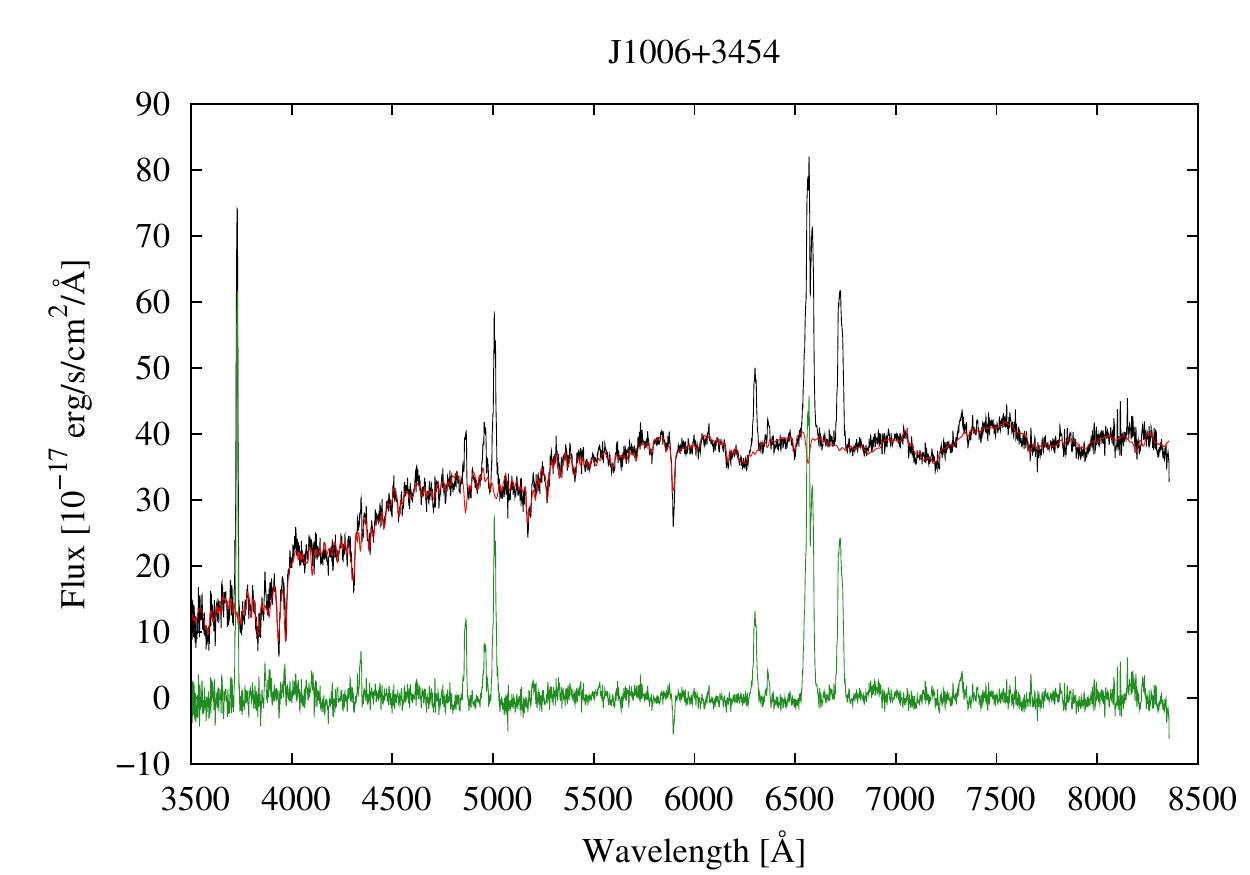}\\
    \includegraphics[width=0.92\columnwidth]{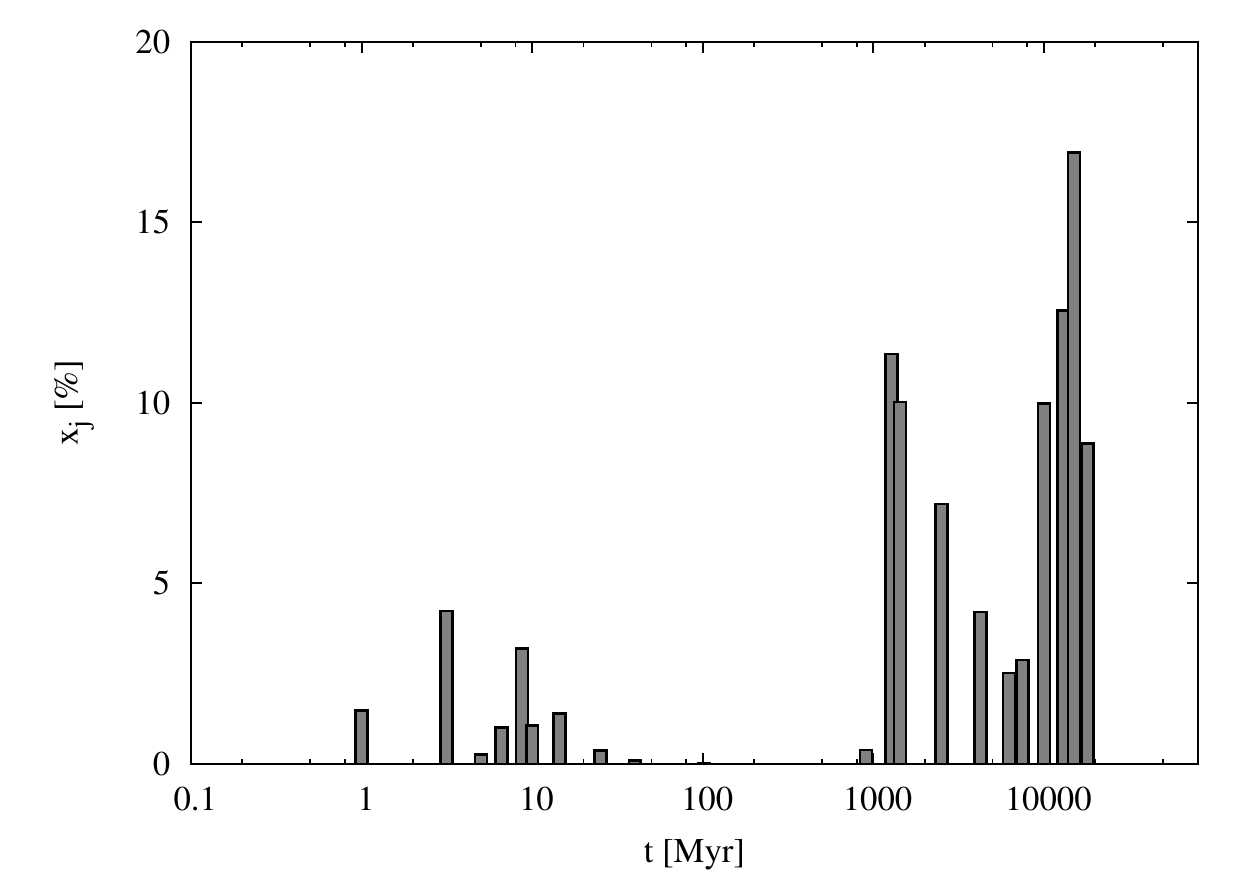}\\
\caption{Spectral modelling using STARLIGHT synthesis code for J1006+3454 GRG. {\bf Top: }Observed spectrum is ploted by black colour, modelled spectrum by red and residual spectrum by the green colour. {\bf Bottom:} Age distribution in the light fraction population vector.}
\label{1006}
\end{figure}

The spectra of giant radio galaxies, as well as galaxies from the comparison sample and all neighbouring galaxies, were processed through the standard procedures of the Image Reduction and Analysis Facility{\footnote{http://iraf.noao.edu}} (IRAF). Each spectrum was corrected for Galactic extinction $A_V$ taken from the NASA/IPAC extragalactic database. The extinction-corrected spectra were then transformed to the rest frame in each case using the redshift values given in the SDSS. For all analysed spectra we applied the simple stellar population (SSP) synthesis code STARLIGHT \citep{cidfernandes2005} to model the observed spectra through fitting a galaxy spectral continuum. 
STARLIGHT code combines N spectra from a base of individual stellar populations in search of linear combinations matching an observed spectrum. The base consists of stellar spectra with different ages and metallicities extracted from the evolutionary synthesis models of \citet{bruzual2003}. The modelled spectrum is fitted using a Metropolis and Markov chain Monte Carlo techniques which explore the parameter space and searches for the minimum of $\chi^2$ between observed and modelled spectrum. For more details see \cite{cidfernandes2005}.
In our modelling we used a base of 150 SSPs with 25 values of stellar ages (between 1 Myr and 18 Gyr) and six metallicities (from 0.005 to 2.5 Z$\odot$). Each SSP with a given age and metallicity contributes to the model flux, and it can be expressed as a light fraction population vector x$_j$, and mass fraction population vector $\mu_j$. As a result of modelling we obtain mean stellar ages, present-day stellar mass, mean metallicities, velocity dispersion, star formation and chemical evolution histories.
In Figure \ref{1006} we present an example of modelled spectrum and the light fraction population vector as a function of stellar age which represents the stellar composition of galaxy.
  
\section{Analysis and results}

\subsection{Stellar populations}
\label{Stellar populations}
In our analysis we compared the parameters obtained for GRGs with those of smaller FRII radio galaxies, as well as their hosts with neighbouring galaxies and neighbours around those two groups of radio galaxies between each other. To examine whether there are any differences in the stellar populations, we firstly model the SSPs for each galaxy and then we average the resultant SSPs for each class of galaxies: for GRGs, FRII radio galaxies, neighbours of giants, and neighbours of FRIIs. 

\begin{figure*}
\centering
    \includegraphics[width=0.92\columnwidth]{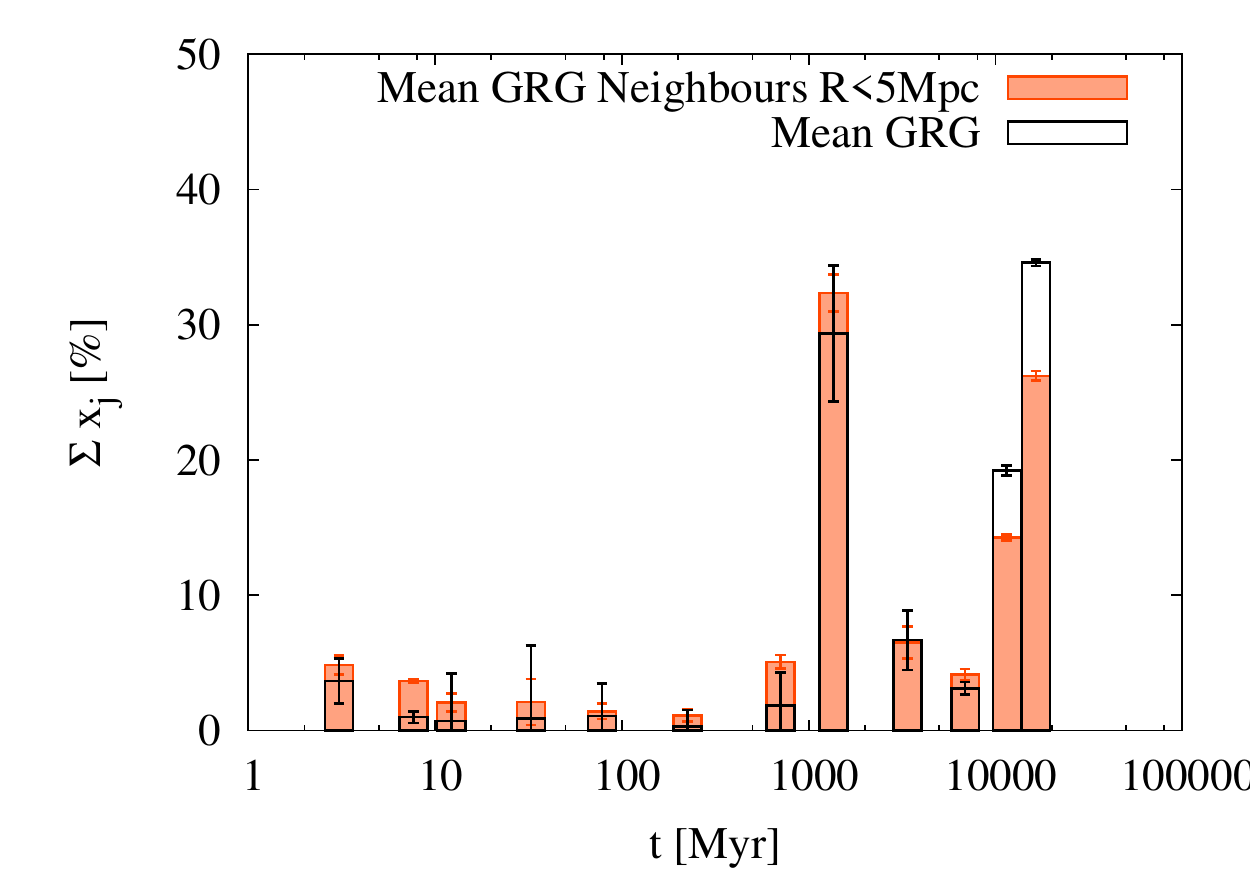}
    \includegraphics[width=0.92\columnwidth]{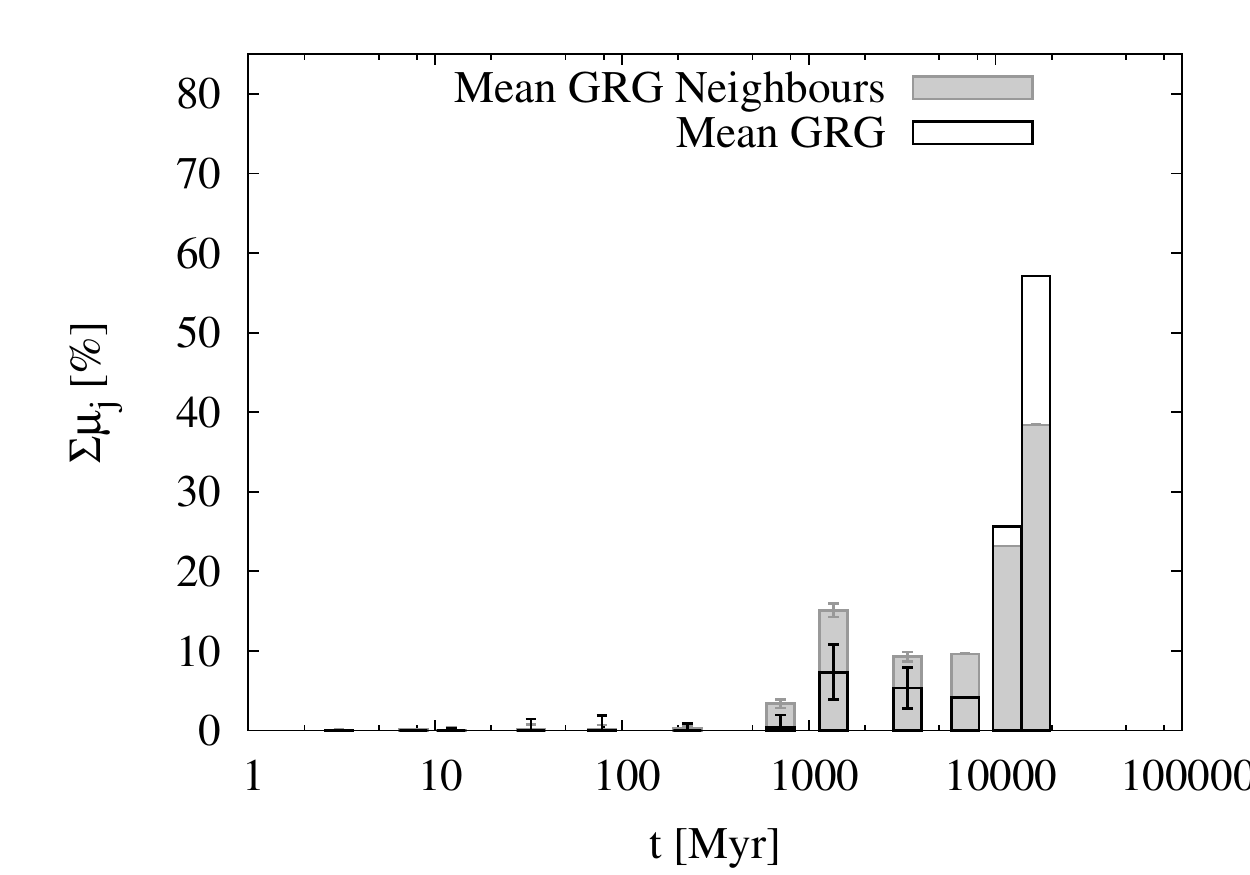}\\
    \includegraphics[width=0.92\columnwidth]{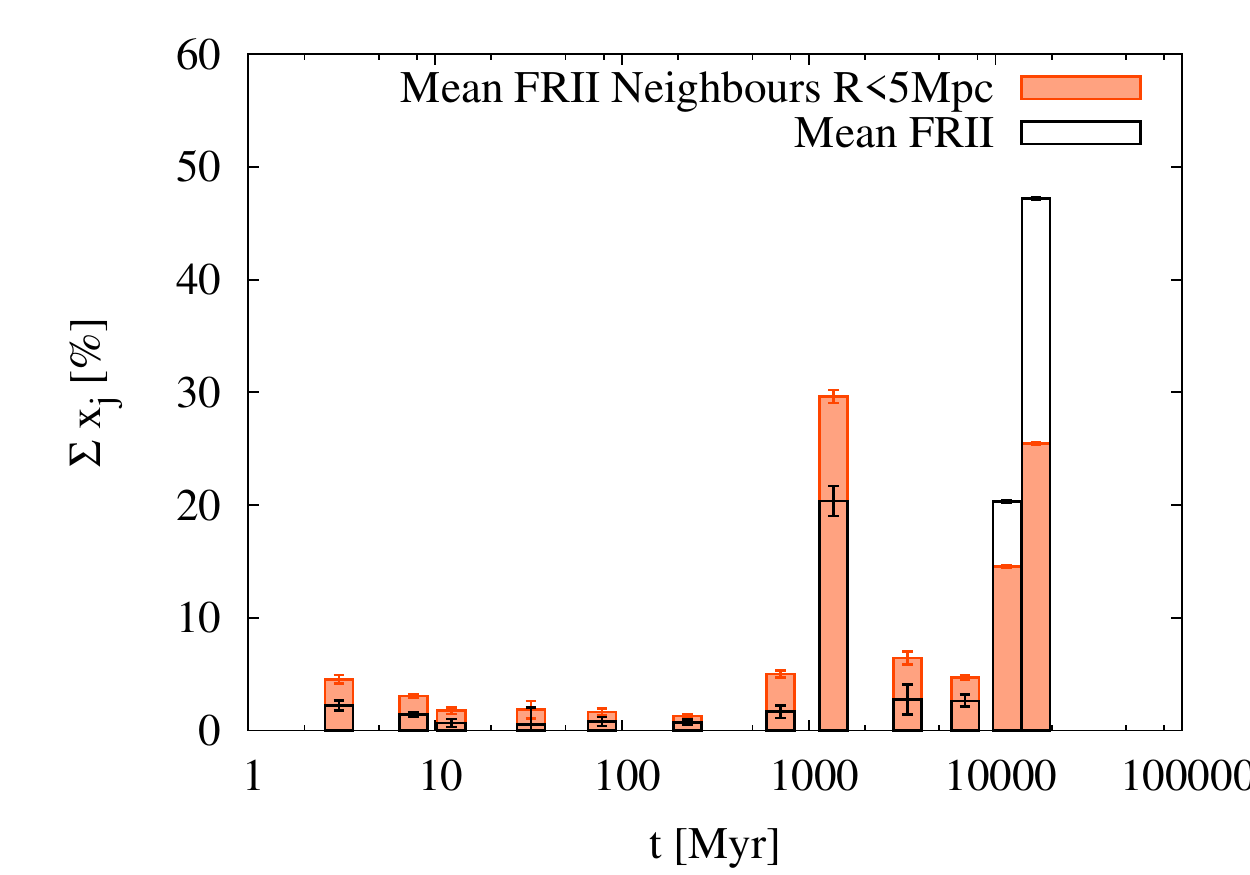}
    \includegraphics[width=0.92\columnwidth]{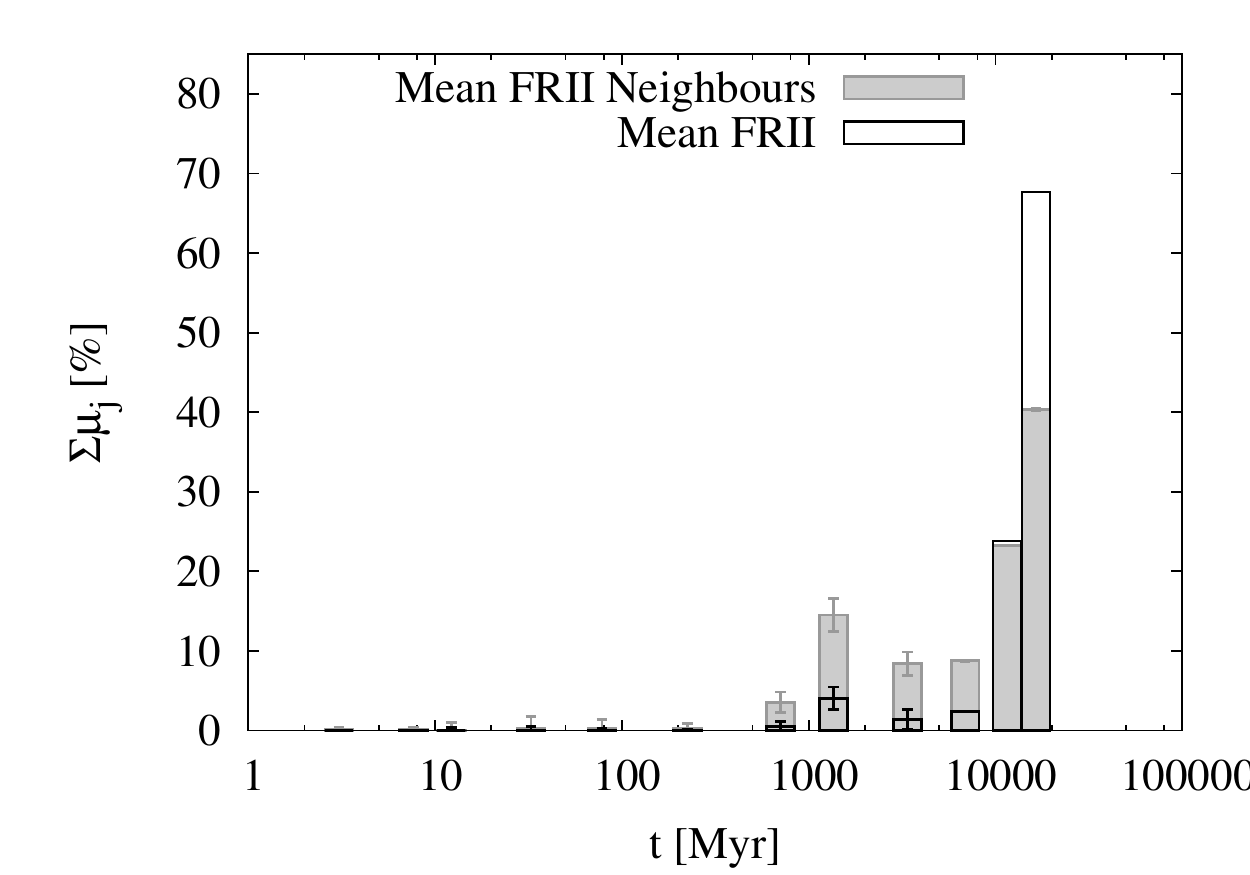}\\
    \includegraphics[width=0.92\columnwidth]{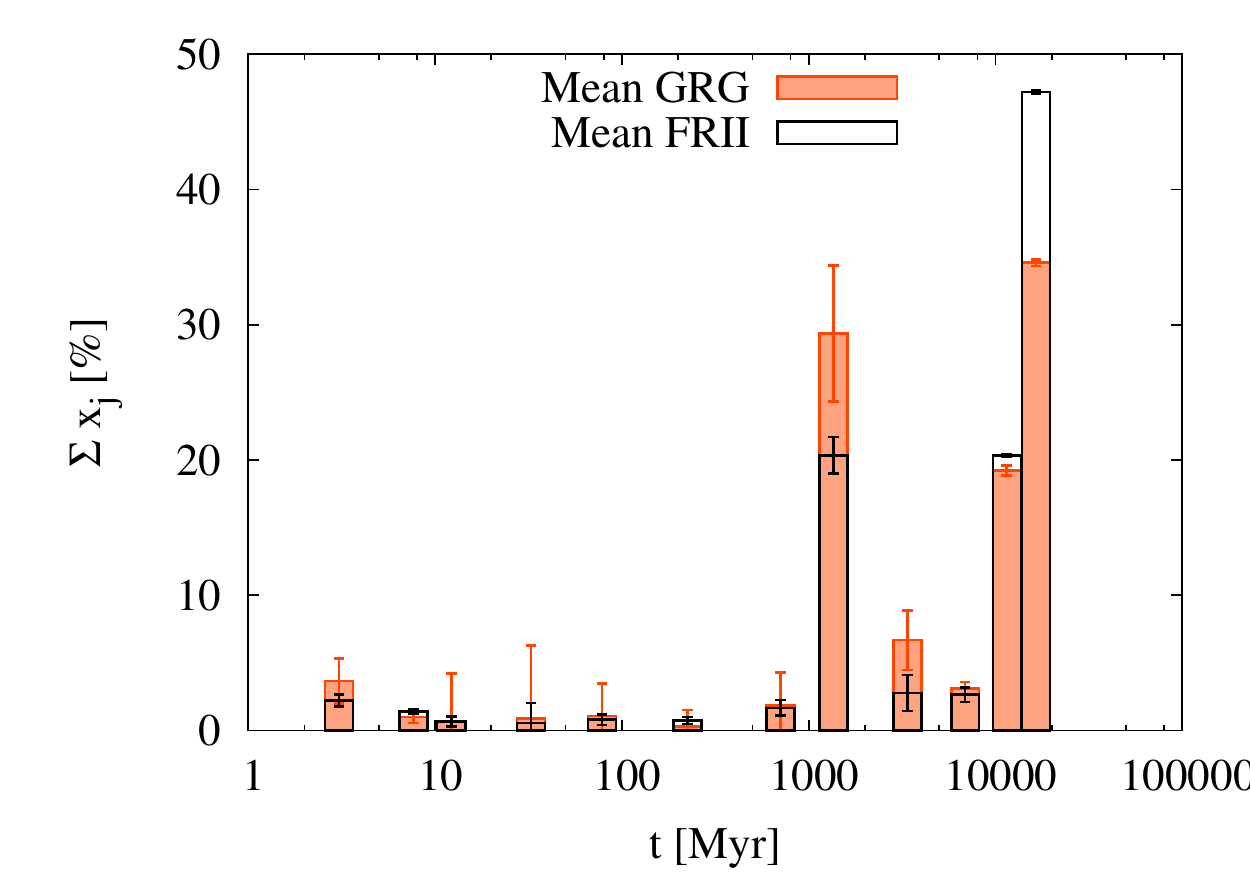}
    \includegraphics[width=0.92\columnwidth]{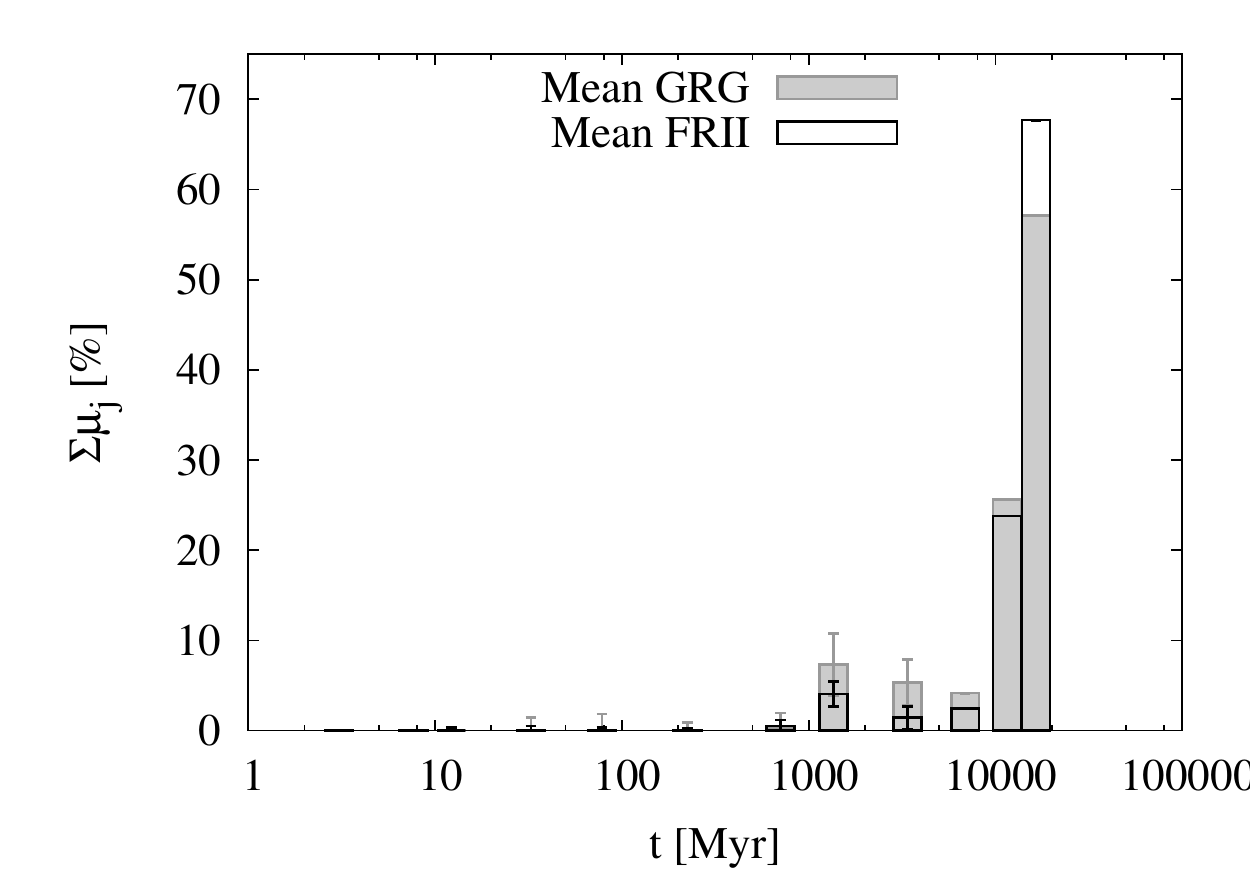}\\
    \includegraphics[width=0.92\columnwidth]{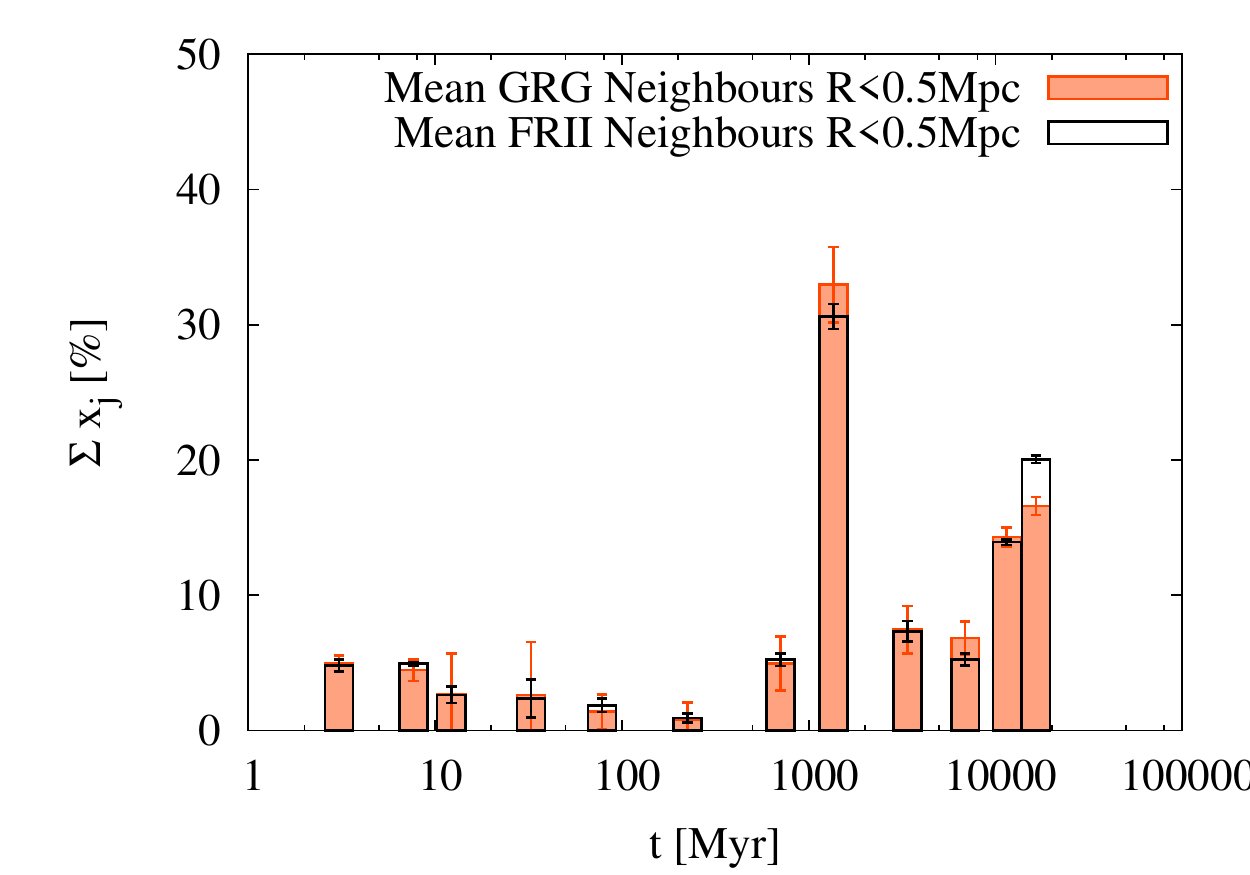}
    \includegraphics[width=0.92\columnwidth]{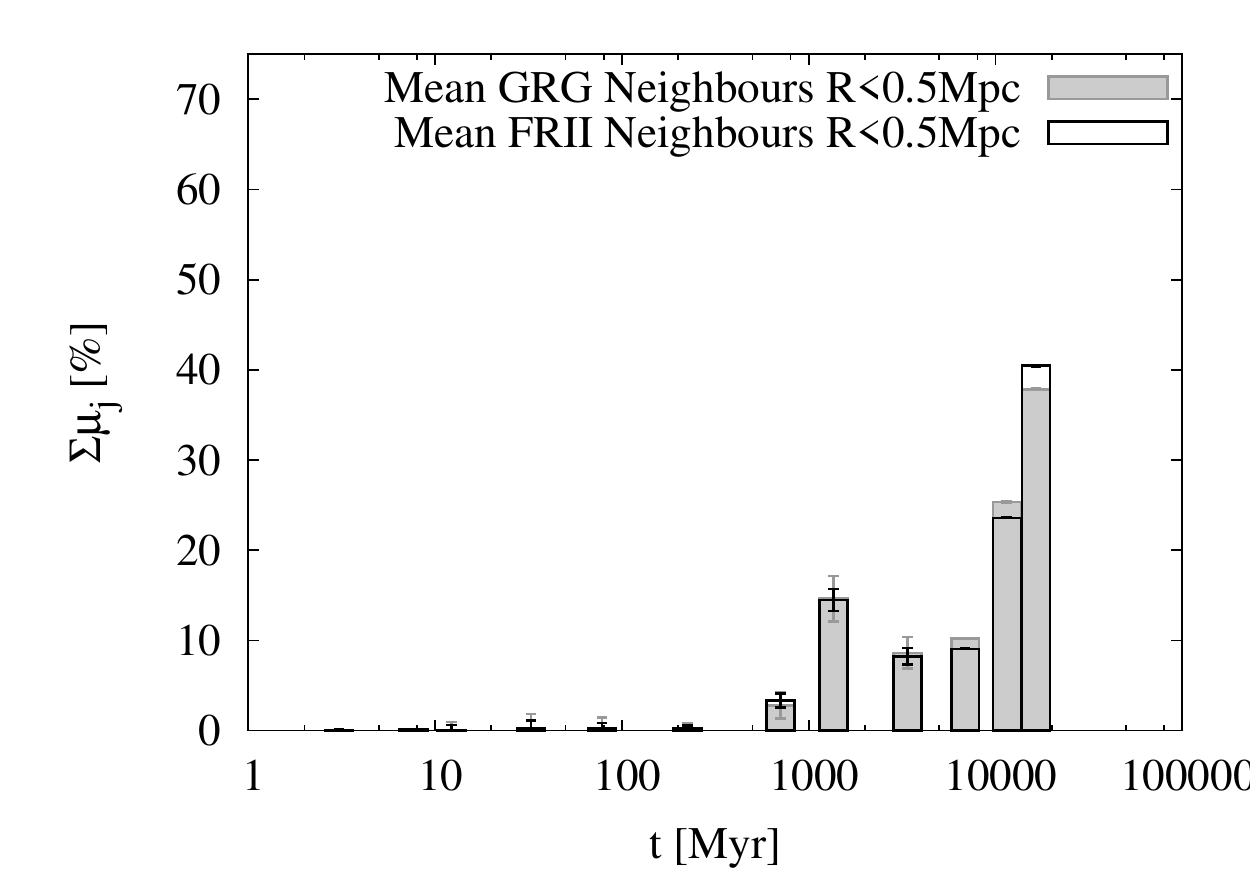}
\caption{Age distribution of the mean light fraction $\Sigma x_j$ population vector (left column) and mean mass fraction $\Sigma \mu_j$ population vector (right column) for GRGs, FRIIs, neighbours of GRGs and neighbours of FRIIs. In the top panels we compare stellar populations of the GRG's hosts with their neighbouring galaxies, in the middle panels -- the hosts of smaller-sized FRII radio galaxies with their neighbouring galaxies, next -- comparison between giants and smaller-sized FRIIs, and in the bottom panels -- between neighbours of giants and neighbours of FRIIs in a radius of 0.5 Mpc from radio galaxy host.}
\label{populations}
\end{figure*}

\begin{figure*}
\centering

    \includegraphics[width=0.92\columnwidth]{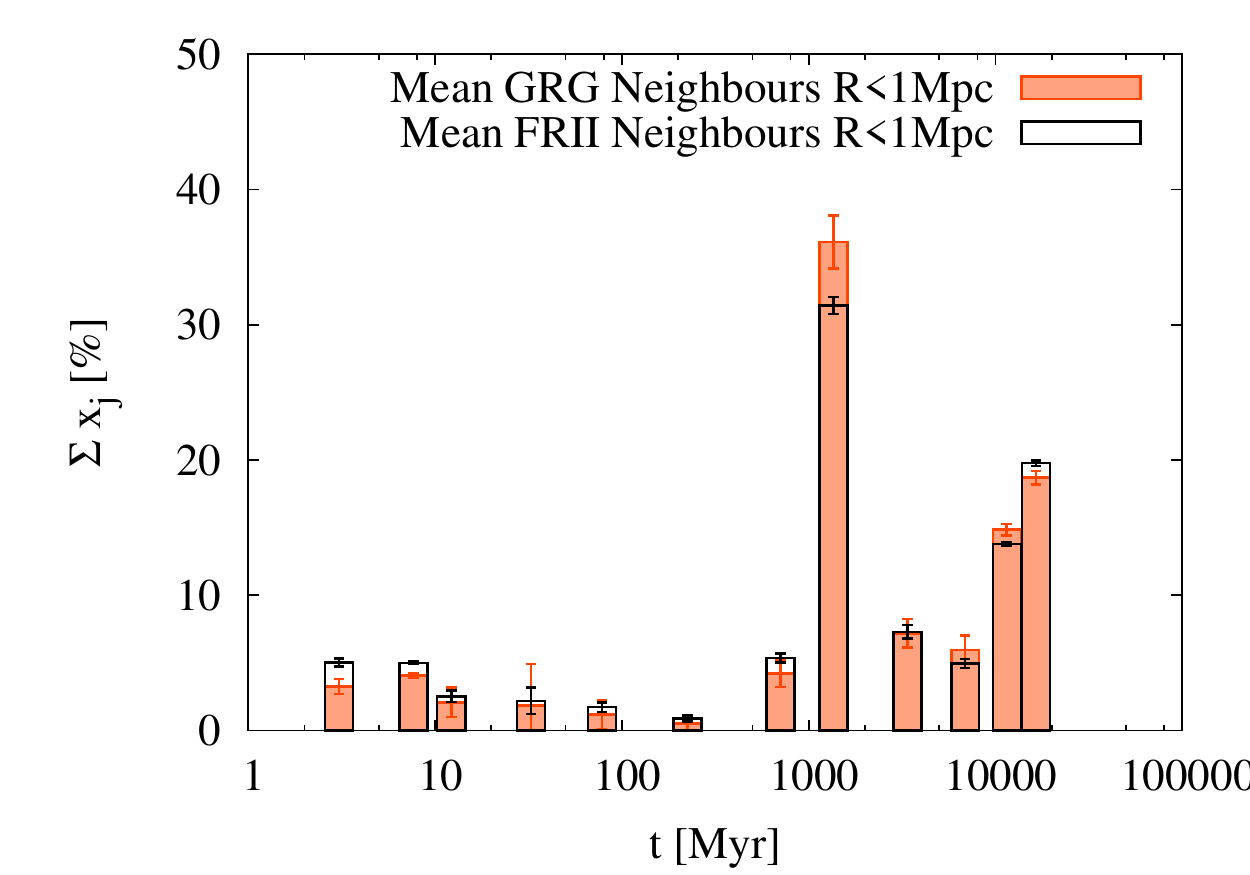}
    \includegraphics[width=0.92\columnwidth]{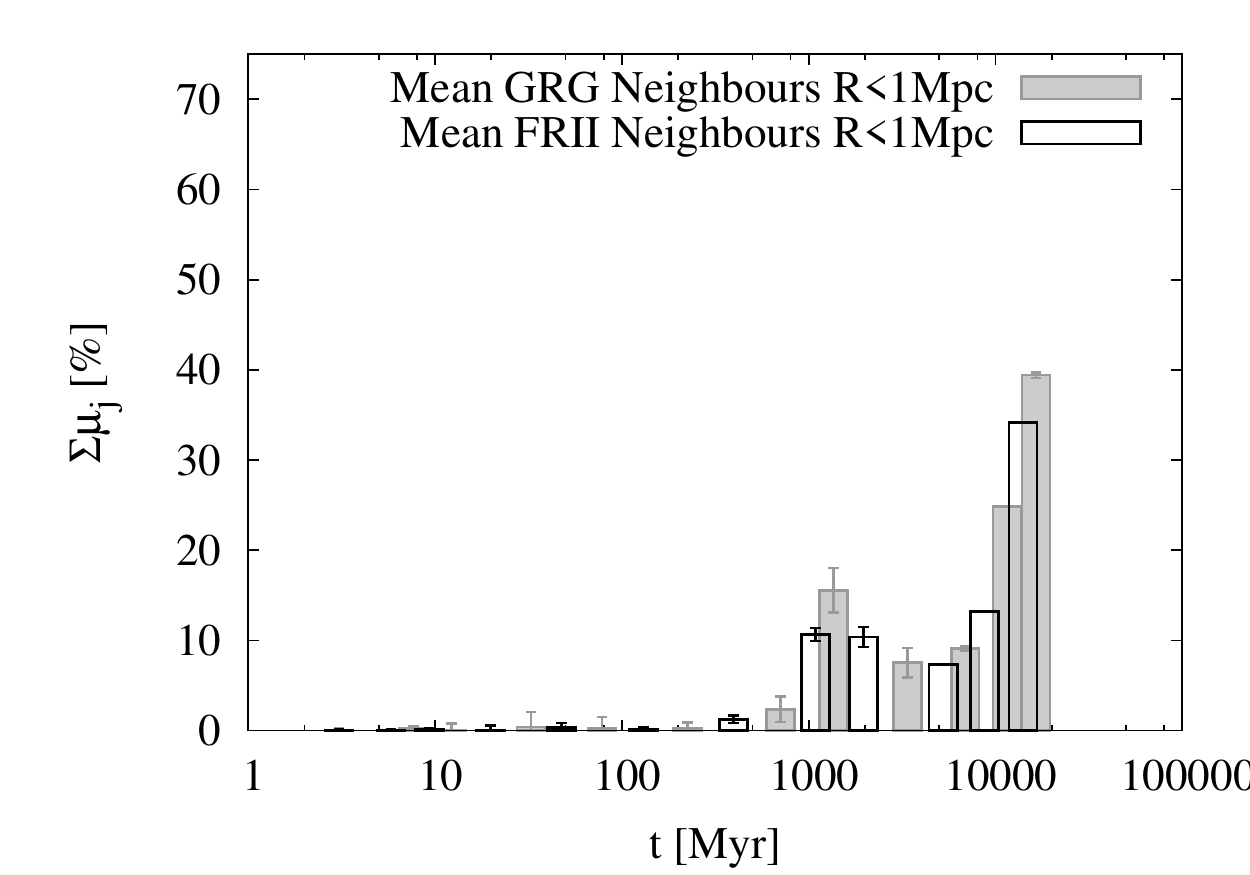}
    \includegraphics[width=0.92\columnwidth]{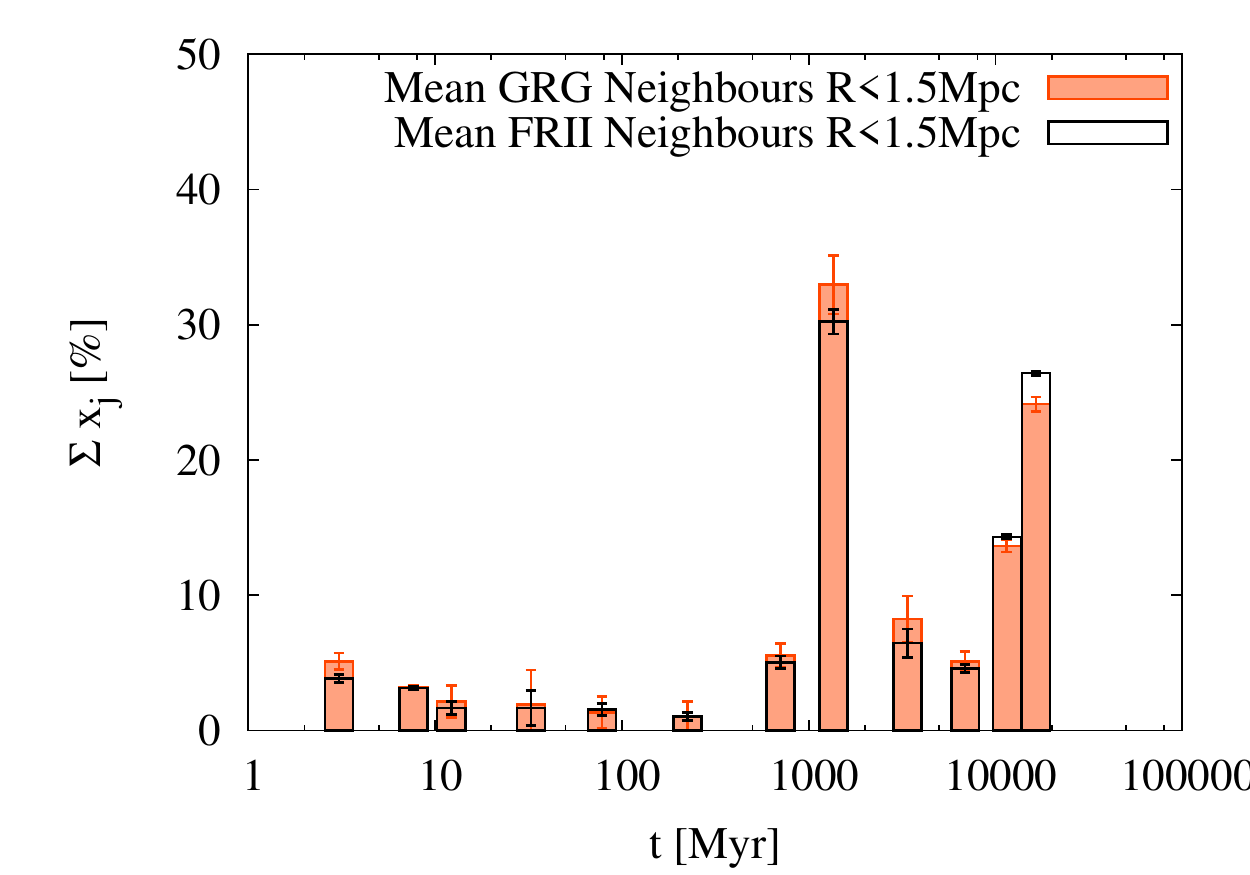}
    \includegraphics[width=0.92\columnwidth]{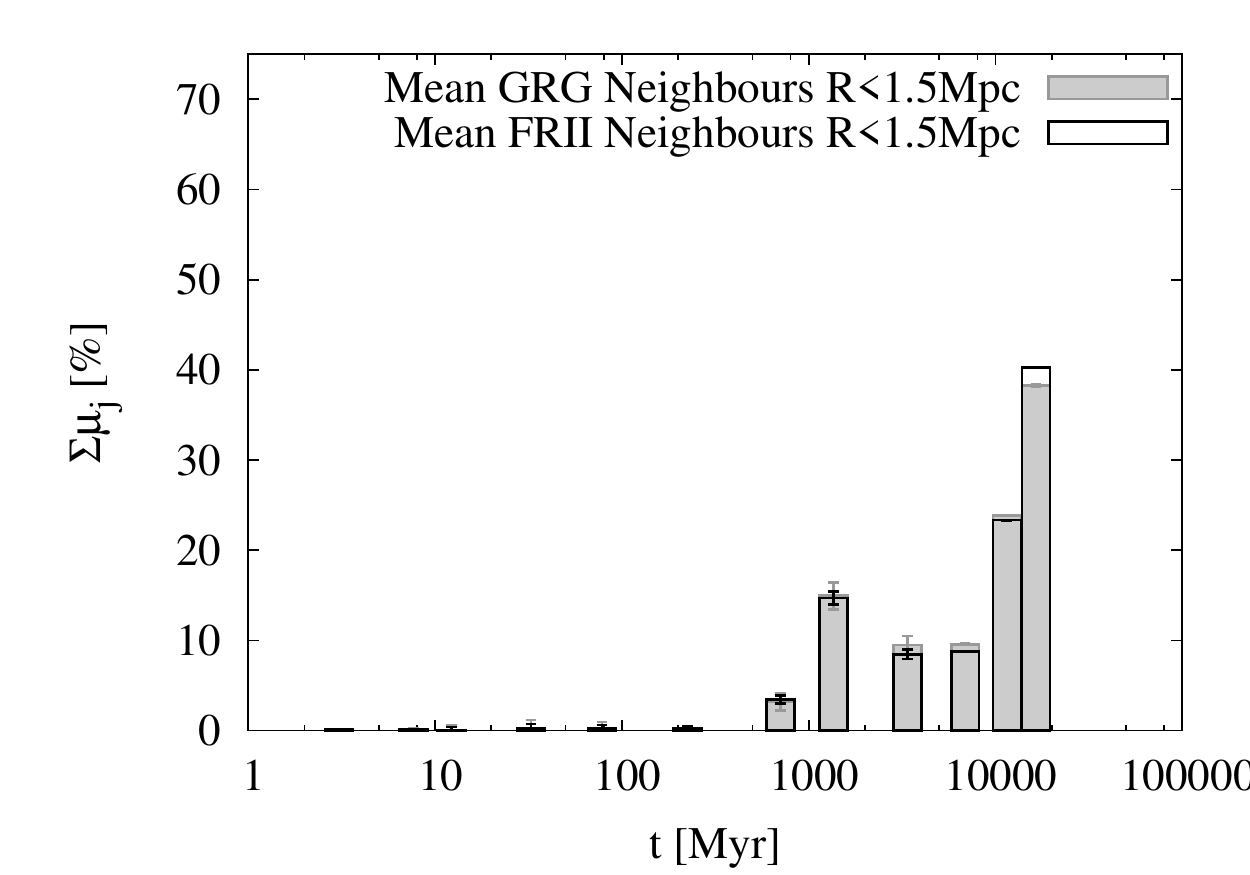}
    \includegraphics[width=0.92\columnwidth]{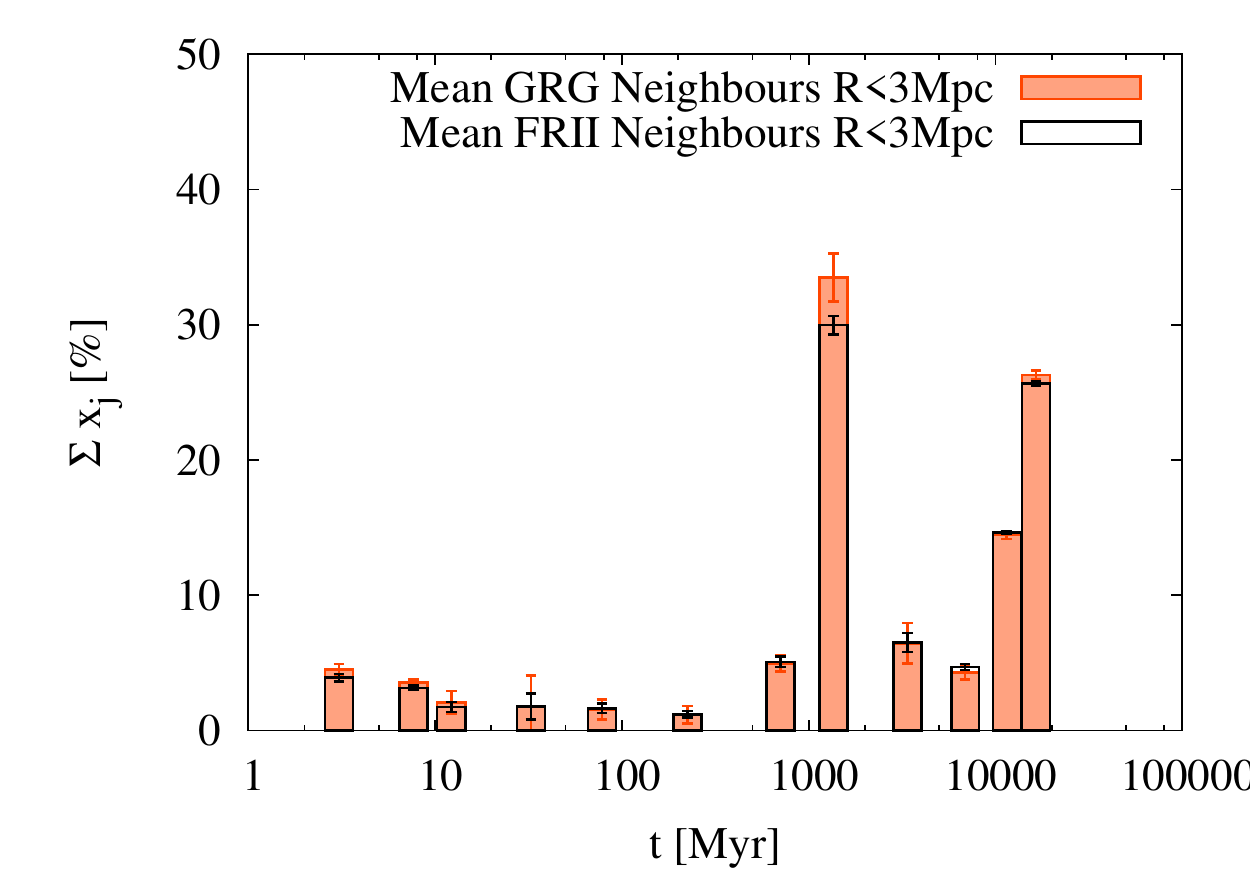}
    \includegraphics[width=0.92\columnwidth]{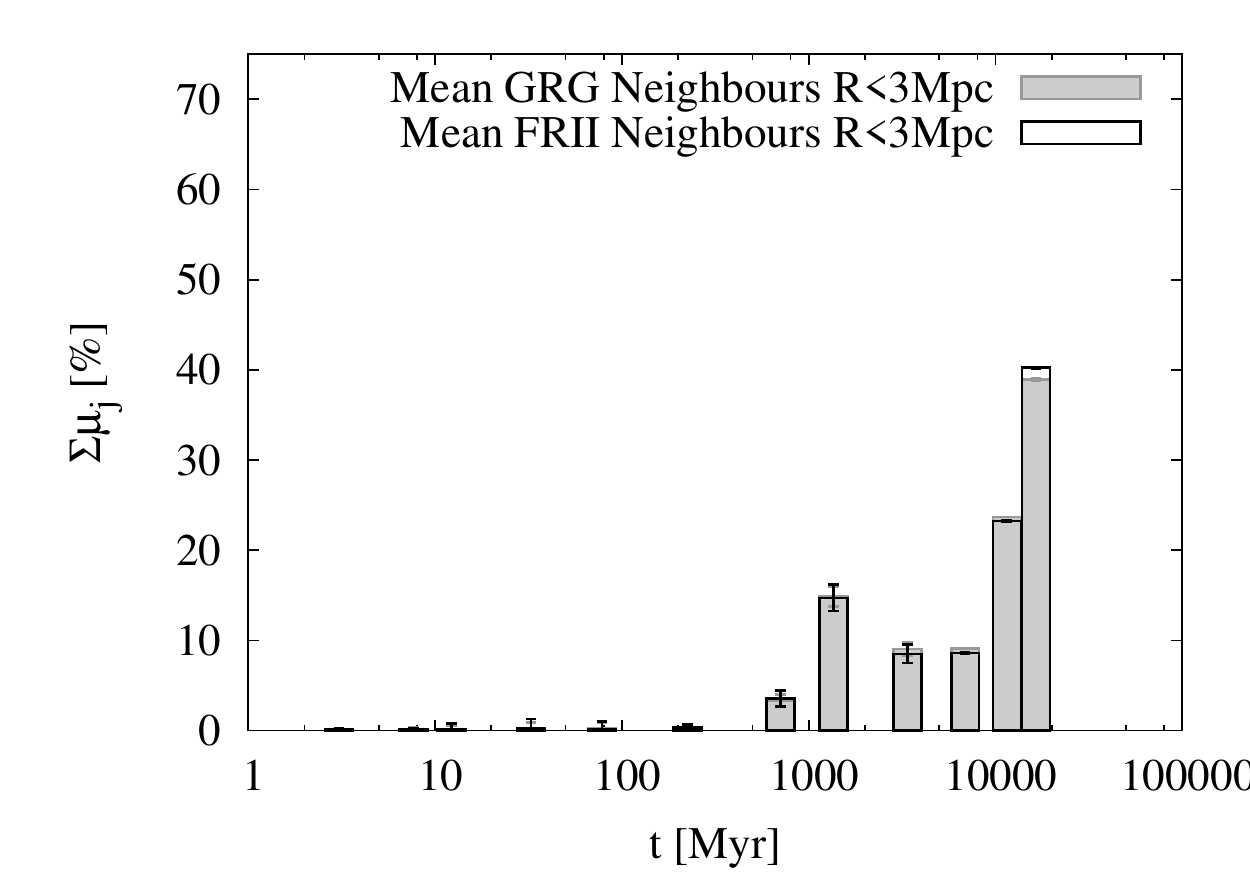}
    \includegraphics[width=0.92\columnwidth]{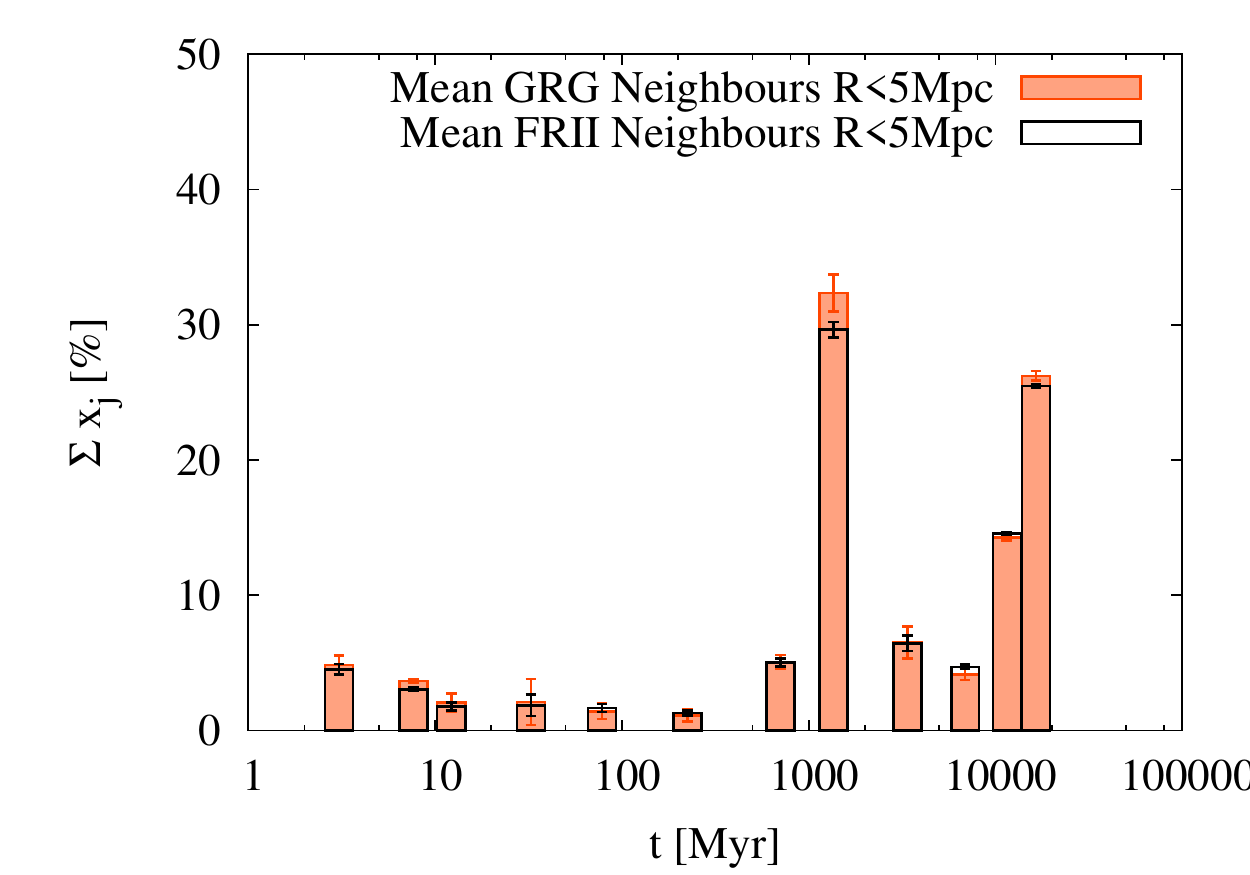}
    \includegraphics[width=0.92\columnwidth]{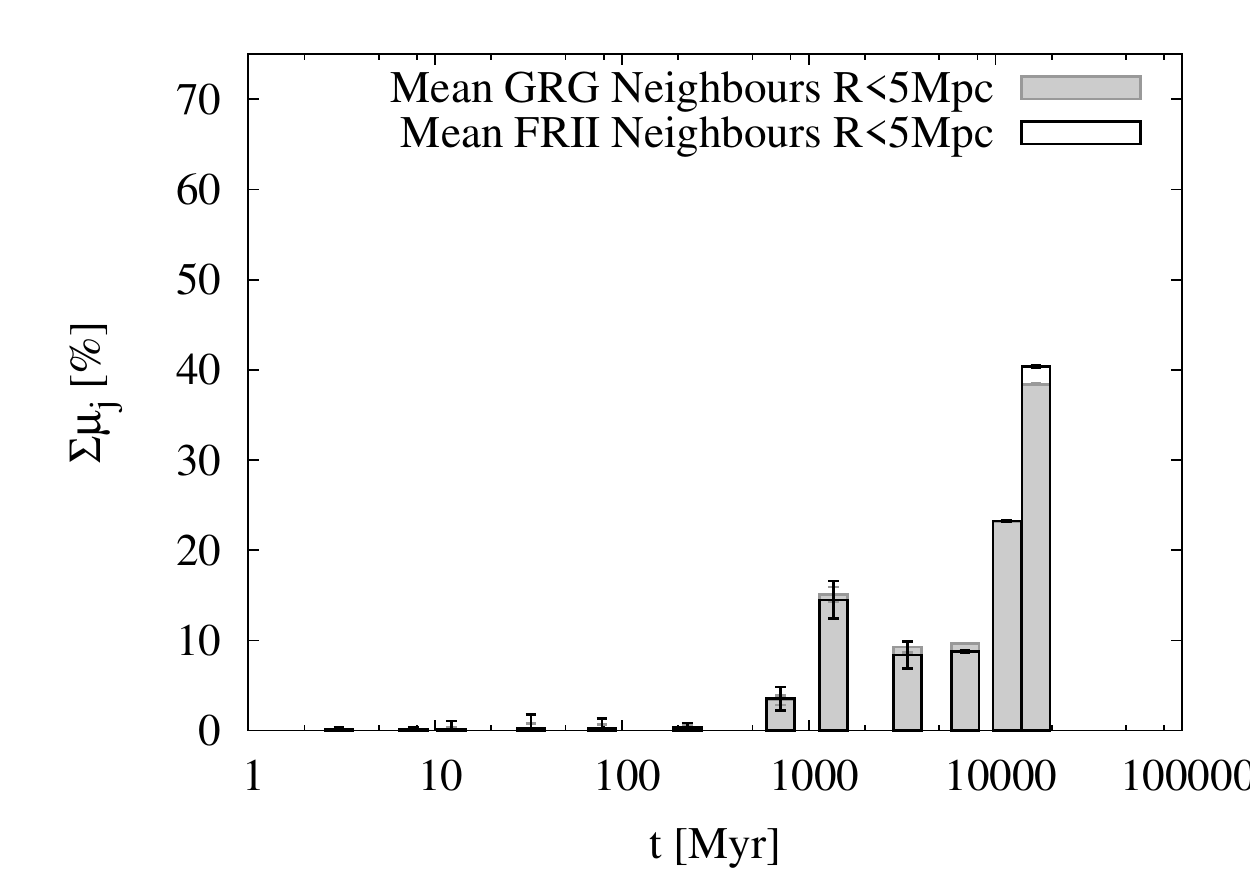}
\caption{Age distribution of the mean $\Sigma x_j$ (left column) and mean $\Sigma \mu_j$ (right column) for neighbours of GRGs and FRIIs in a different radius around radio galaxy host. Top panels -- in a radius of 1 Mpc, upper middle panels -- in a radius of 1.5 Mpc., lower middle panels -- in a radius of 3 Mpc, and in the bottom panel -- in a radius of 5 Mpc.}
\label{populations2}
\end{figure*}

Figures \ref{populations} and \ref{populations2} show the mean light-weighted population vector $\Sigma x_j$ (left column) and mass-weighted $\Sigma \mu_j$ (right column) population vector as a function of stars age $t$. We plotted these figures for particular samples and for five different radii around radio galaxy hosts (0.5, 1, 1.5, 3 and 5 Mpc). The $\Sigma x_j$ and $\Sigma \mu_j$ vectors are summarized by metallicity and then averaged in each galaxy sample. To preserve the clarity of Figures \ref{populations} and \ref{populations2}, we binned the results using 12 age bins instead of 25. \\
The results of stellar population composition in studied galaxies are summarized in Table \ref{tab2}, where the obtained SSPs are divided into three age bins: young populations with stellar ages t$_{\star}<5\times10^8$yr, intermediate populations with $9\times10^8$ yr $<$t$_{\star}$$<7.5\times10^9$ yr, and old populations with t$_{\star}>10^{10}$yr. Contributions of each age bin are given in percentage of $\Sigma$x$_j$ and $\Sigma$$\mu_j$ separately. The uncertainties of $\Sigma$x$_j$ and $\Sigma$$\mu_j$ in each sample of galaxies were calculated as the standard deviation of the mean value.\\
In first two rows of Figure \ref{populations} we show the comparison of stellar populations between hosts of radio galaxies (GRGs and FRIIs) and their neighbours. We can see that both for GRGs and smaller-sized FRIIs the host galaxies are dominated by stellar populations with ages above 1 Gyr. In comparison with their neighbouring galaxies they have a larger fraction of the oldest populations ($\sim$10 Gyr) and a smaller fraction of intermediate age stars ($\sim$1 Gyr). This fact can be explained by the different types of galaxies considered in the samples. The hosts of radio galaxies are old ellipticals located predominantly in the centres of clusters/groups of galaxies, while their neighbours are of various types where the star formation processes are more common than in ellipticals.

In the next step we compare the SSPs of GRGs with smaller-sized FRIIs. As it can be seen in Figure \ref{populations} and Table \ref{tab2} the GRGs have a larger fraction of intermediate age stars ($\sim$1 Gyr) and significantly smaller fraction of the oldest ones ($\sim$10 Gyr). This fact is evidence of the differences in the structure of GRG hosts compared with hosts of smaller-sized counterparts.
The same difference in SSP composition is observed for neighbours of GRGs and FRIIs. Galaxies located around GRGs have a larger fraction of middle age stars compared with galaxies around FRIIs. It is also clearly visible that this effect is more significant when we take into account galaxies located closer to the radio source's host (up to radius of 1.5 Mpc). The difference between SSPs in neighbouring galaxies of GRGs and FRIIs is not as prominent as for radio galaxy hosts, however we observe that the intermediate age populations in neighbours of GRGs are systematically higher than in neighbours of FRIIs, while old stellar populations are systematically lower.\\
The above results show that giants, together with their neighbouring galaxies, could have different formation histories comparing to the groups with smaller sized radio galaxies. It may be evidence of different global properties of the ambient medium where those groups evolve. For example, the scenario of a close interaction or minor merger event in the central part of a galaxy group can indicate larger star formation in the most central galaxies.

We also have carried out simple statistical analyses of the types of neighbouring galaxies in a radius of 5 Mpc from the radio galaxy host. Based on the SDSS classification of galaxies, we found that 31\% of spectroscopic GRG neighbours are star forming galaxies, 6\% are starburst galaxies and 2.4\% are AGNs. In a sample of FRII neighbours we observe 27\% star forming galaxies, 5.7\% starburst galaxies and 2.7\% AGNs. It shows that environments around GRGs and FRIIs statistically consist of similar galaxies and the obtained larger amount of intermediate age populations in GRG neighbours is not due to the larger amount of galaxies which formed stars more recently.

In all graphs which show the distribution of ages represented by the light fraction population vector we observe a large contribution of stars with ages $\sim$1 Gyr. Some authors (e.g. \cite{chen2010}) found that the fraction of $\sim$1 Gyr stars depend on the stellar libraries used in spectral modelling, however qualitatively it does not affect the obtained results, since we use the same fitting procedures for all of our galaxies. Otherwise, the evidence of intermediate age stellar populations was suggested by other authors to be common in local elliptical galaxies (e.g. \cite{huang2009}). 
   

\begin{table}
\caption{Summarized light and mass fraction population vector for samples of GRGs, smaller-sized FRIIs and neighbouring galaxies of GRGs and FRIIs, divided into three age bins: young (t$_{\star}<5\times10^8$yr), intermediate ($9\times10^8$yr$<$t$_{\star}$$<7.5\times10^9$yr), and old (t$_{\star}>10^{10}$yr) stellar populations. }
\label{tab2}
\centering
\begin{tabular}{lccc}
\hline
\hline
                &               &       $\Sigma$x$_j$   &               \\
                &       Young   & Intermediate  & Old      \\
\hline
GRGs            &       8.1$\pm$4.1     &       39.4$\pm$6.2    & 52.5$\pm$0.5    \\
FRIIs           &       7.3$\pm$3.3     &       26.5$\pm$3.6    & 66.2$\pm$0.2    \\
{\bf R$<$0.5 Mpc}       &                       &                       &                   \\
Neighbours of GRGs &    18.7$\pm$6.1    &       50.4$\pm$4.4    & 30.9$\pm$1.1    \\
Neighbours of FRIIs &   19.6$\pm$3.6    &       46.4$\pm$2.4    & 34.0$\pm$0.5    \\
{\bf R$<$1 Mpc} &                       &                       &                  \\
Neighbours of GRGs&     13.9$\pm$6.3    &       52.5$\pm$4.6    & 33.6$\pm$0.9    \\
Neighbours of FRIIs&    19.5$\pm$2.5    &       47.0$\pm$1.7    & 33.5$\pm$0.4    \\
{\bf R$<$1.5 Mpc}       &                       &                       &                   \\
Neighbours of GRGs&     16.2$\pm$6.5    &       50.4$\pm$5.1    & 37.8$\pm$1.0    \\
Neighbours of FRIIs&    14.9$\pm$3.1    &       44.3$\pm$2.5    & 40.8$\pm$0.3    \\
{\bf R$<$3 Mpc} &                       &                       &                  \\
Neighbours of GRGs&     15.4$\pm$5.3    &       46.2$\pm$4.1    & 38.4$\pm$0.6    \\
Neighbours of FRIIs&    15.4$\pm$2.5    &       44.3$\pm$1.8    & 40.3$\pm$0.2    \\
{\bf R$<$5 Mpc} &                       &                       &                  \\
Neighbours of GRGs&     15.2$\pm$4.4    &       45.3$\pm$3.3    & 39.5$\pm$0.6    \\
Neighbours of FRIIs&    16.1$\pm$2.2    &       43.8$\pm$1.5    & 40.1$\pm$0.2    \\

\hline
\hline
                &       &$\Sigma$$\mu_j$        &       \\
                &       Young   & Intermediate  & Old      \\
\hline
GRGs            &        0.5$\pm$2.6    & 17.1$\pm$7.5          & 82.4$\pm$0.1  \\
FRIIs           &        0.1$\pm$1.1    & 8.5$\pm$3.5           & 91.5$\pm$0.2  \\
{\bf R$<$0.5 Mpc}       &                       &                       &                   \\
Neighbours of GRGs &     0.4$\pm$4.6    & 36.4$\pm$6.1          & 63.2$\pm$0.1  \\
Neighbours of FRIIs &    0.5$\pm$1.9    & 35.4$\pm$3.5          & 64.1$\pm$0.1  \\
{\bf R$<$1 Mpc} &                       &                       &                  \\
Neighbours of GRGs &     1.0$\pm$3.5    & 34.7$\pm$6.9          & 64.3$\pm$0.3  \\
Neighbours of FRIIs &    0.6$\pm$1.3    & 34.8$\pm$2.5          & 64.6$\pm$0.1  \\
{\bf R$<$1.5 Mpc}       &                       &                       &                   \\
Neighbours of GRGs &     0.6$\pm$2.1    & 37.3$\pm$4.2          & 62.1$\pm$0.1  \\
Neighbours of FRIIs &    0.7$\pm$1.1    & 35.7$\pm$2.0          & 63.6$\pm$0.1  \\
{\bf R$<$3 Mpc} &                       &                       &                  \\
Neighbours of GRGs &     0.6$\pm$1.9    & 36.8$\pm$3.2          & 62.6$\pm$0.1  \\
Neighbours of FRIIs &    0.7$\pm$1.3    & 35.8$\pm$2.1          & 63.5$\pm$0.1  \\
{\bf R$<$5 Mpc} &                       &                       &                  \\
Neighbours of GRGs &     0.6$\pm$1.3    & 37.8$\pm$2.5          & 61.6$\pm$0.1  \\
Neighbours of FRIIs &    0.8$\pm$1.2    & 35.6$\pm$1.9          & 63.6$\pm$0.1  \\

\hline
\end{tabular} 
\end{table}

\subsection{Uncertainty of STARLIGHT fitting}
 
In our studies we obtain relatively small differences between resultant SSPs for studied samples of galaxies, therefore it is important to consider the uncertainties of fitting procedure. \cite{cidfernandes2005} check the recovery of spectral parameters modelled by STARLIGHT based on mock spectra with an assumed star formation history. The synthetic spectra were perturbed to obtain different signal-to-noise ratios (S/N) and the error spectrum was adopted to reconstruct real observed spectrum. The authors find that individual components of x$_j$ are very uncertain but the binning x$_j$ on to young, intermediate and old components gives the robust description of the star formation. Other output parameters, for example $\left<\log t^{\star}\right>_L$, $\left <\log t^{\star}\right >_M$ also recover the input parameters well.  
On average our spectra have S/N equal to 15. According to parameter uncertainties obtained by \cite{cidfernandes2005} (listed in their Table 1.) the mean and the dispersion between input and output values of xj in an individual objects are equal to $0.62 \pm 4.04$\%, $0.01 \pm 7.88 $\%, and $-0.63 \pm 7.61$\% for young, intermediate and old stellar populations respectively. Thus for an individual galaxy the dispersion is large. On the other hand, we finally compare a sample of 41 giant radio galaxies, so these uncertainties reduce to  $0.62 \pm 0.63$\%, $0.01 \pm 1.23 $\%, and $-0.63 \pm 1.19$\%, respectively, due to the reduction of the dispersion. For $\mu_j$ the uncertainties of parameter recovery for a single galaxy are equal to $0.16 \pm 1.18$\%, $0.93 \pm 6.10$\% and $-1.09 \pm 6.54$\%, reducing to $0.16 \pm 0.18$\%, $0.93 \pm 0.95$\% and $-1.09 \pm 1.02$\%. The uncertainties are much lower than the differences we report in Table 2.

\subsection{The ages and masses}
Based on stellar continuum fits for individual galaxies we determine the stellar mass, black hole mass, light and mass weighted mean stellar age $\left <\log t^{\star}\right >$ and metallicity $\left <\log Z^{\star}\right >$ defined by \citet{cidfernandes2005} as

\begin{equation}
\left <\log t^{\star}\right >_L=\sum_{j=1}^{N} x_j \log t_j \hspace{0.5cm} and \hspace{0.5cm} \left <\log Z^{\star}\right>_L=\sum_{j=1}^{N} x_j \log Z_j\\
\label{eq1}
\end{equation}
\begin{equation}
\left <\log t^{\star}\right >_M=\sum_{j=1}^{N} \mu_j \log t_j \hspace{0.5cm} and \hspace{0.5cm} \left <\log Z^{\star}\right>_M=\sum_{j=1}^{N} \mu_j \log Z_j
\label{eq2}
.\end{equation}

\begin{figure}
\centering
    \includegraphics[width=0.99\columnwidth]{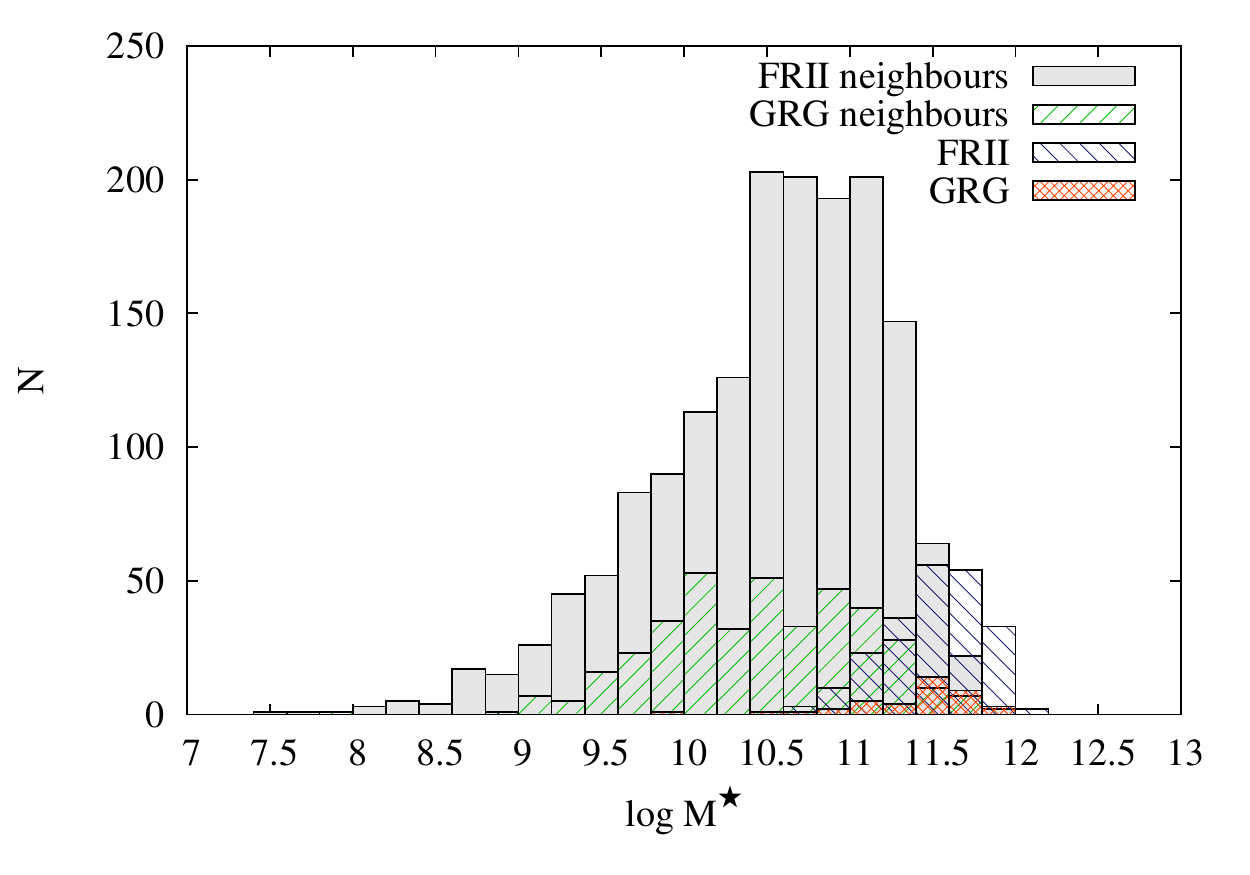}
    \includegraphics[width=0.99\columnwidth]{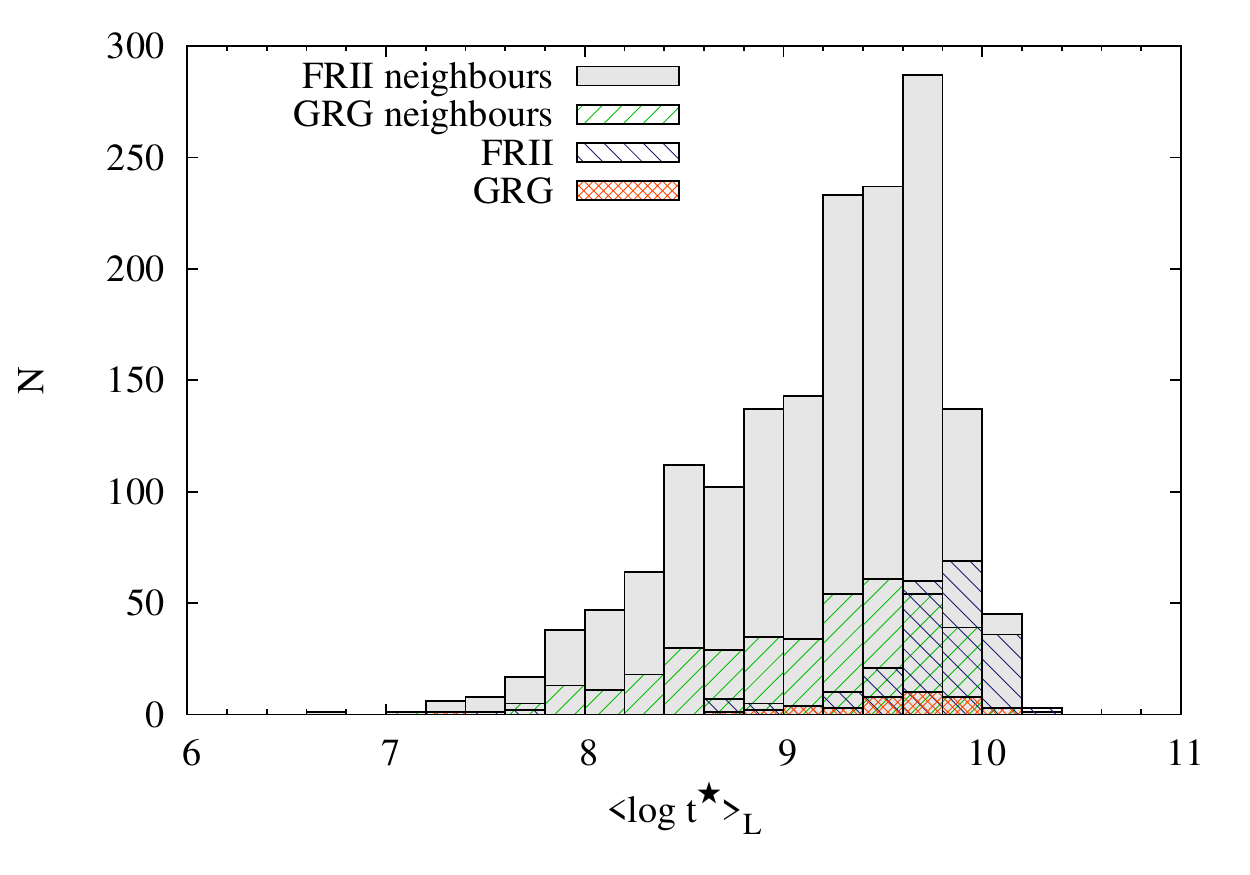}
    \includegraphics[width=0.99\columnwidth]{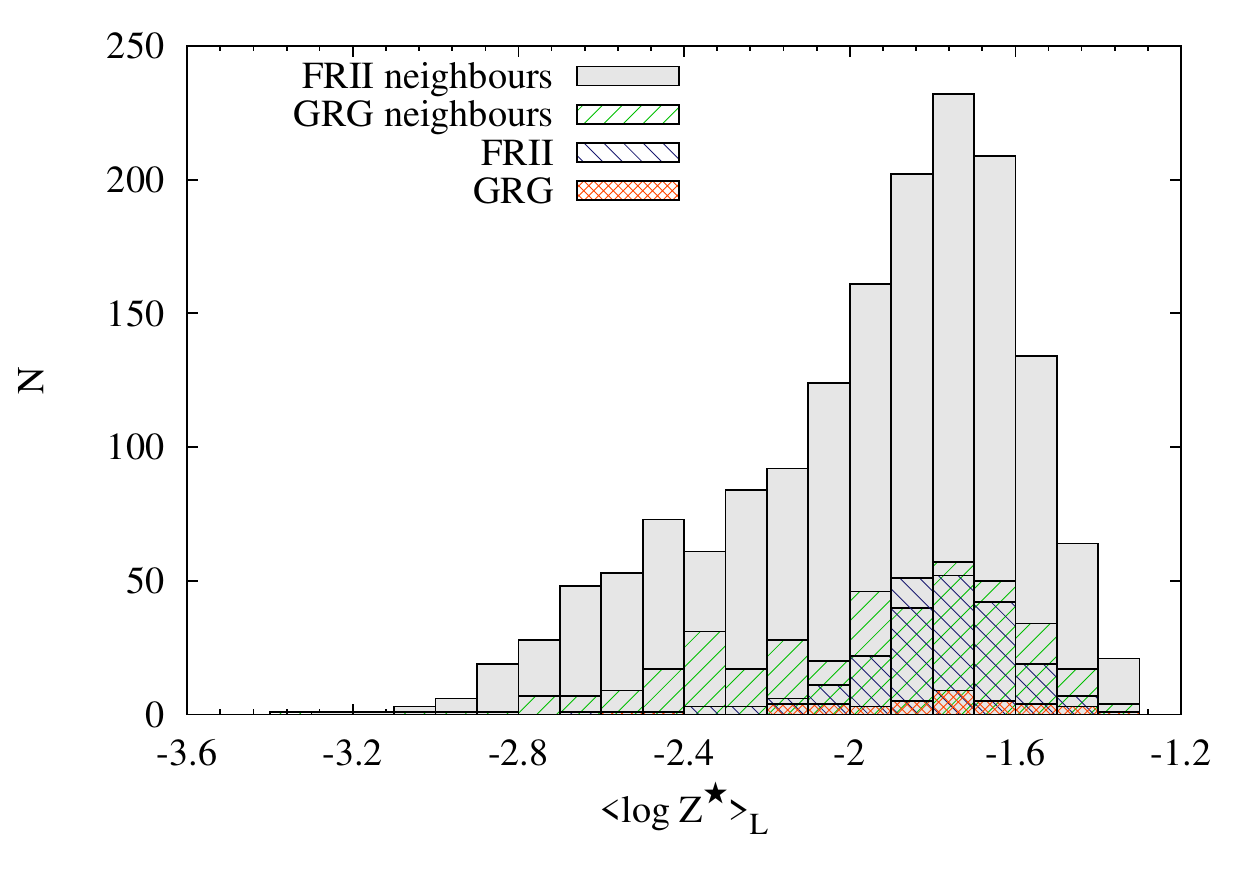}
\caption{Distributions of mean stellar mass, mean stellar age and mean stellar metallicity for samples of GRGs, smaller-sized FRII radio sources, and their neighbouring galaxies.}
\label{distrib}
\end{figure}

In Figure \ref{distrib} we plot the distributions of stellar mass, $\left <\log t^{\star}\right >_L$, and $\left <\log Z^{\star}\right >_L$ obtained for individual galaxies in each sample. In the first graph, where we present the distribution of mean stellar mass, it can be seen that the stellar masses in hosts of radio sources are mostly higher than masses of neighbouring galaxies, but in both samples of neighbours we also observe galaxies with stellar masses as high as in radio sources. 

In the next two panels we plot the distributions of light weighted stellar ages and metallicities. Similarly to the graphs with mass distributions, the plotted parameters obtained for neighbouring galaxies span a wide range of values and the hosts of radio sources from GRG and FRII samples are concentrated in the right end of these ranges.\\
        
\begin{figure}
\centering
    \includegraphics[width=0.99\columnwidth]{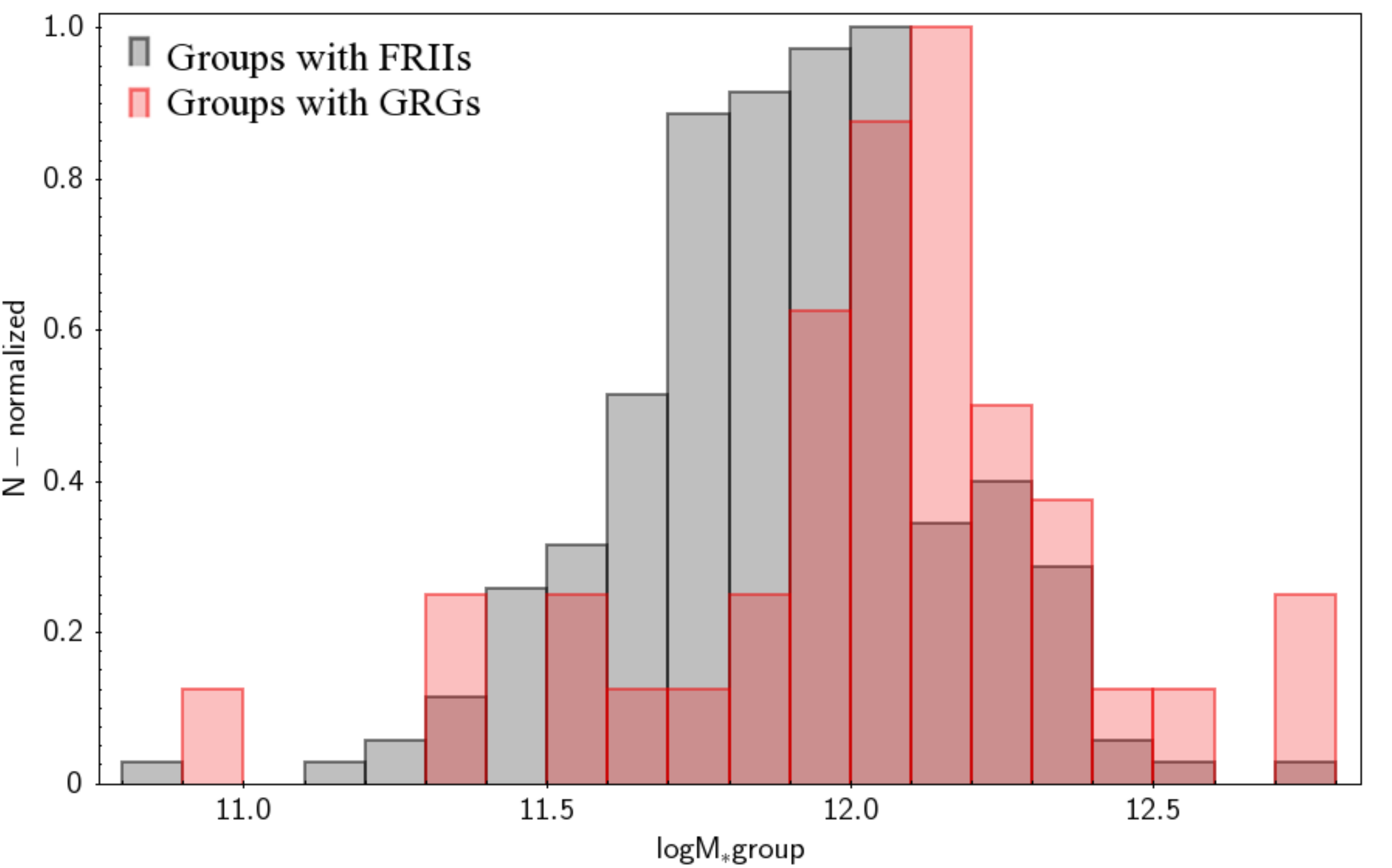}
    \includegraphics[width=0.99\columnwidth]{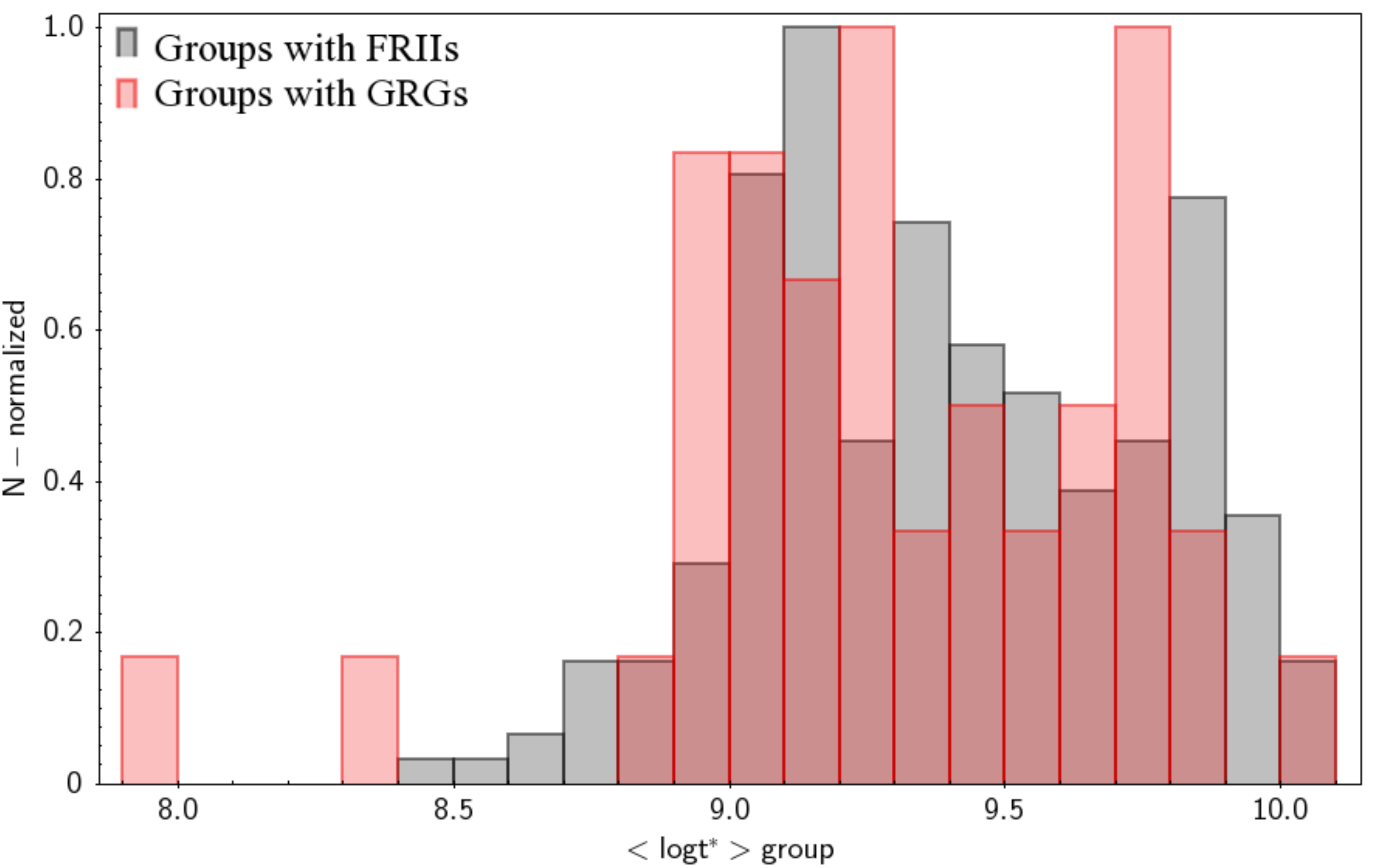}
    \includegraphics[width=0.99\columnwidth]{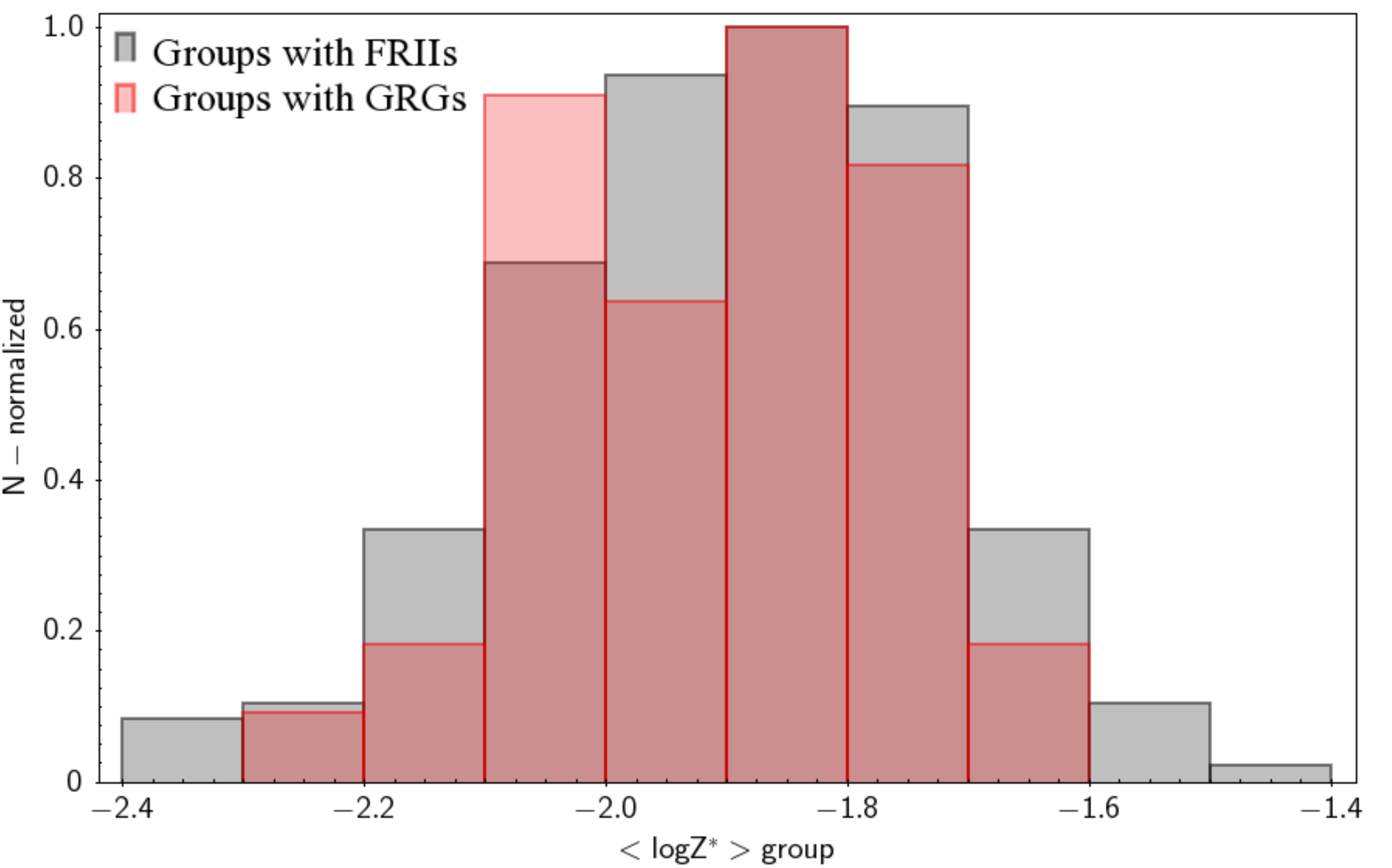}
\caption{Normalized distributions of mean stellar masses, ages and metallicities for groups of galaxies with GRGs and smaller-sized FRIIs.}
\label{group}
\end{figure}

To find any difference in properties between the groups concentrated around GRGs and FRIIs, we determine the parameters characterizing groups as a whole (the host of a radio galaxy with its neighbours). We summed up the stellar masses, $\left <\log t^{\star}\right >_L$, and $\left <\log Z^{\star}\right >_L$ of individual galaxies in each galaxy group separately. 
In Figure \ref{group} we plot the normalized distributions of above parameters. We obtained that both for groups with GRGs and FRIIs, all of the parameters span similar ranges of values and have similar distribution shapes. However, we observe that the summarized stellar mass in groups with giants have slightly higher values than in groups with smaller FRIIs. This is evidence that the larger amount of stellar mass is cumulated in galaxies around giants.  

In Table \ref{tab3} we summarize the average values of light and mass-weighted stellar ages, metallicities and stellar masses obtained for all galaxies in each sample. However, it can be seen that there are no statistically significant differences between considered samples. Also when we compare the values characterizing the whole groups with GRGs and FRIIs, we do not see the differences in mean stellar ages, metallicity and stellar masses. 
The obtained parameters for groups with giants and smaller-sized FRIIs are nearly the same, so all considered groups look very similar. However, the effects of environmental influences on internal properties of cluster members can be small enough to be visible. It can only be well recognized in studies of individual groups for which we have good quality spectroscopic data for all cluster members. Any subtle differences are usually not visible when we average the large number of values, because the uncertainties of these quantities become larger than presumed differences.\\

\begin{table*}
\caption{Mean values of light and mass weighted stellar ages, metallicities, and mean stellar mass for each sample of galaxies. At the bottom of the table we give the mean values for whole groups with GRGs and FRIIs. }

\label{tab3}
\centering
\begin{tabular}{r c c c c c }
\hline
                &  Mean                              & Mean                              &  Mean                       & Mean                              &  Mean         \\      
                &  $\left <\log t^{\star}\right >_L$ & $\left <\log t^{\star}\right >_M$ &  $\left <\log Z^{\star}\right >_L$ & $\left <\log Z^{\star}\right >_M$ &logM$_{\star}$  \\ 
                &  [yr]                              &  [yr]                             &                                     &                                   & [$\odot$]     \\     
\hline                                                                                  
GRGs               &  9.54$\pm$0.49 & 9.96$\pm$0.34  & -1.82$\pm$0.26 & -1.65$\pm$0.15   & 11.38$\pm$0.39        \\
FRIIs              &  9.71$\pm$0.41 & 10.07$\pm$0.29 & -1.79$\pm$0.19 & -1.64$\pm$0.11   & 11.48$\pm$0.34        \\
Neighbours of GRGs &  9.11$\pm$0.58 & 9.80$\pm$0.32  & -1.95$\pm$0.34 & -1.85$\pm$0.33   & 10.46$\pm$0.61        \\
Neighbours of FRIIs&  9.12$\pm$0.59 & 9.80$\pm$0.46  & -1.95$\pm$0.35 & -1.86$\pm$0.35   & 10.51$\pm$0.72        \\
\hline                                                  
Groups with GRGs   & 9.13$\pm$0.29 & 9.89$\pm$0.17   & -1.84$\pm$0.15 & -1.72$\pm$0.15   & 11.92$\pm$0.24                 \\
Groups with FRIIs  & 9.45$\pm$0.35 & 9.94$\pm$0.23   & -1.86$\pm$0.18 & -1.73$\pm$0.16   & 11.89$\pm$0.27                 \\
\hline
\end{tabular} 
\end{table*}

\subsection{Star formation rate} 

\begin{figure}
\centering
    \includegraphics[width=0.99\columnwidth]{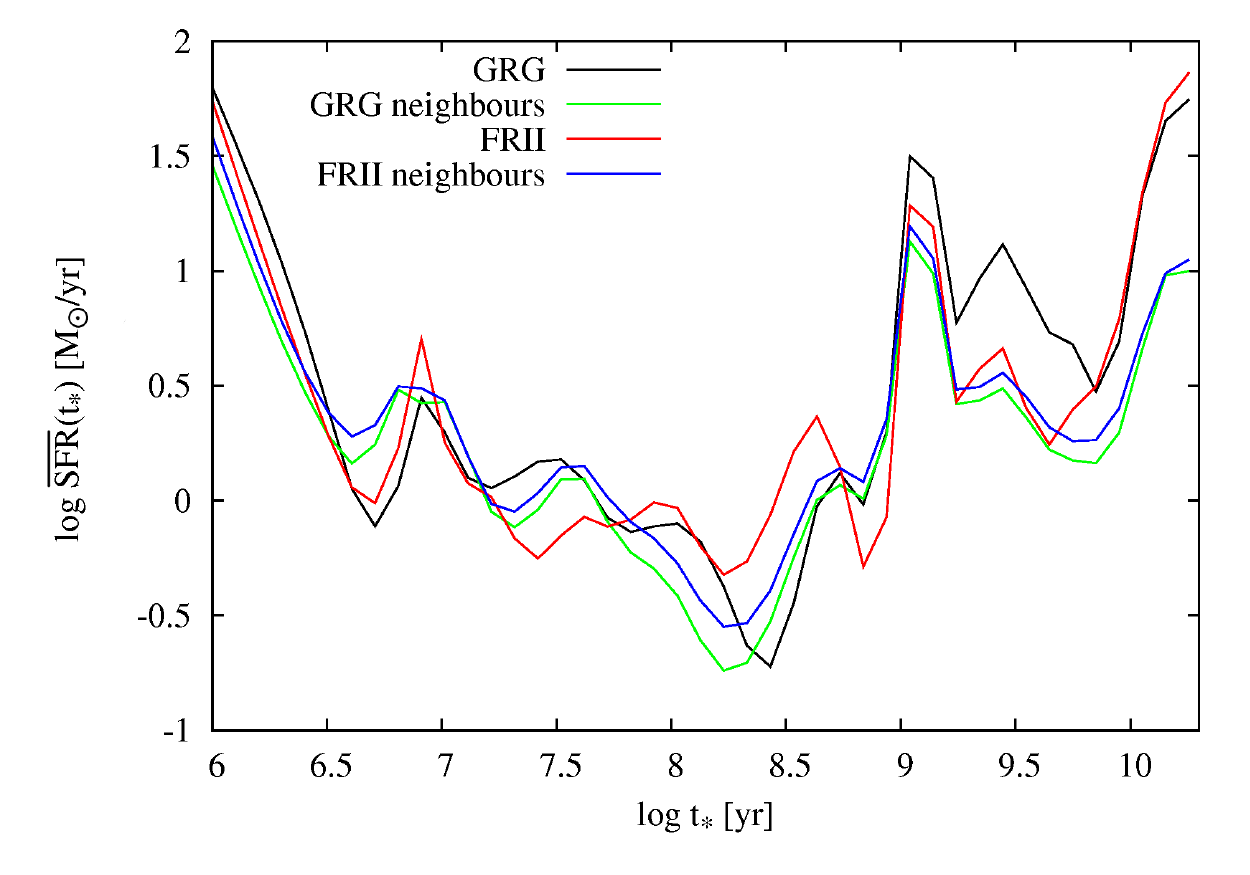}
    \includegraphics[width=0.99\columnwidth]{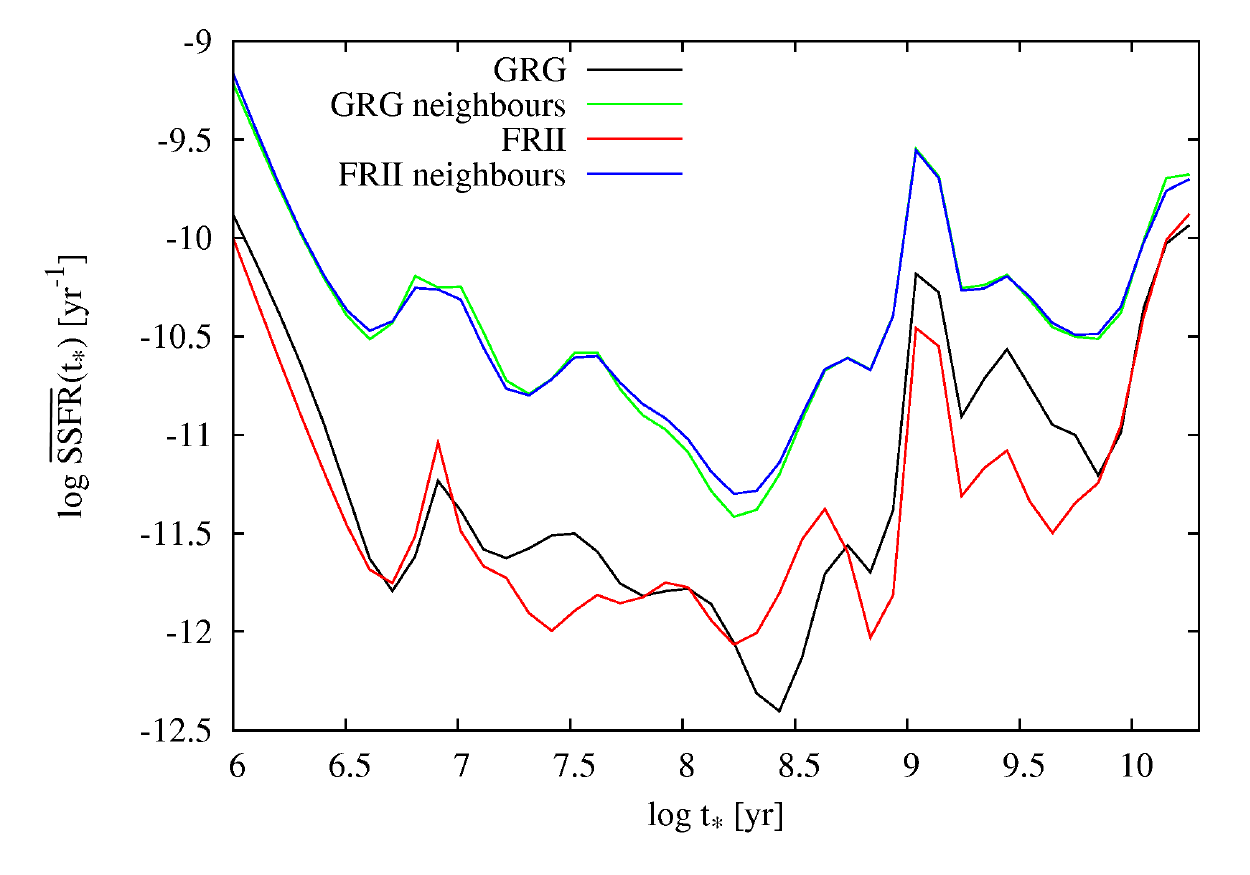}
\caption{ Average time-dependent star formation rate ({\bf top graph}) and specific star formation rate ({\bf bottom graph}) for sample of GRGs, FRIIs, neighbours of GRGs and neighbours of FRIIs.}
\label{SFR}
\end{figure}

For each galaxy we derived the star formation rate (SFR) and specific star formation rate (SSFR) using the definition of \cite{asari2007}:

\begin{equation}
SFR(t_*)=\frac{M_{*}^c {\rm log} e}{t_*}\frac{\mu_{s}^c(t_*)}{\Delta {\rm log} t_*}
\label{eq3}
\end{equation}
where $M_{*}^c$ is the total mass converted to stars throughout the galaxy life, and $\mu_{s}^c(t_*)$ is the fraction of this mass in the $t_*$ bin.
\begin{equation}
SSFR(t_*)=\frac{ {\rm log} e}{t_*}\frac{\mu_{s}^c(t_*)}{\Delta {\rm log} t_*}
\label{eq4}
,\end{equation}
which measures the star formation rate with respect to the mass already converted into stars. Time-dependent star formation rates can be derived from the stellar population synthesis and they are in a good agreement with SFR estimations from H$\alpha$ line \citep{asari2007}.  
In Figure \ref{SFR} we present the averaged SFR(t) and SSFR(t) for each sample of galaxies considered in this paper. It can be seen that star formation occurred $\sim$1 Gyr ago in GRGs started earlier and was higher than in FRII galaxies. However, there are no differences between neighbours of giants and neighbours of FRIIs. 

\begin{figure}
\centering
    \includegraphics[width=0.99\columnwidth]{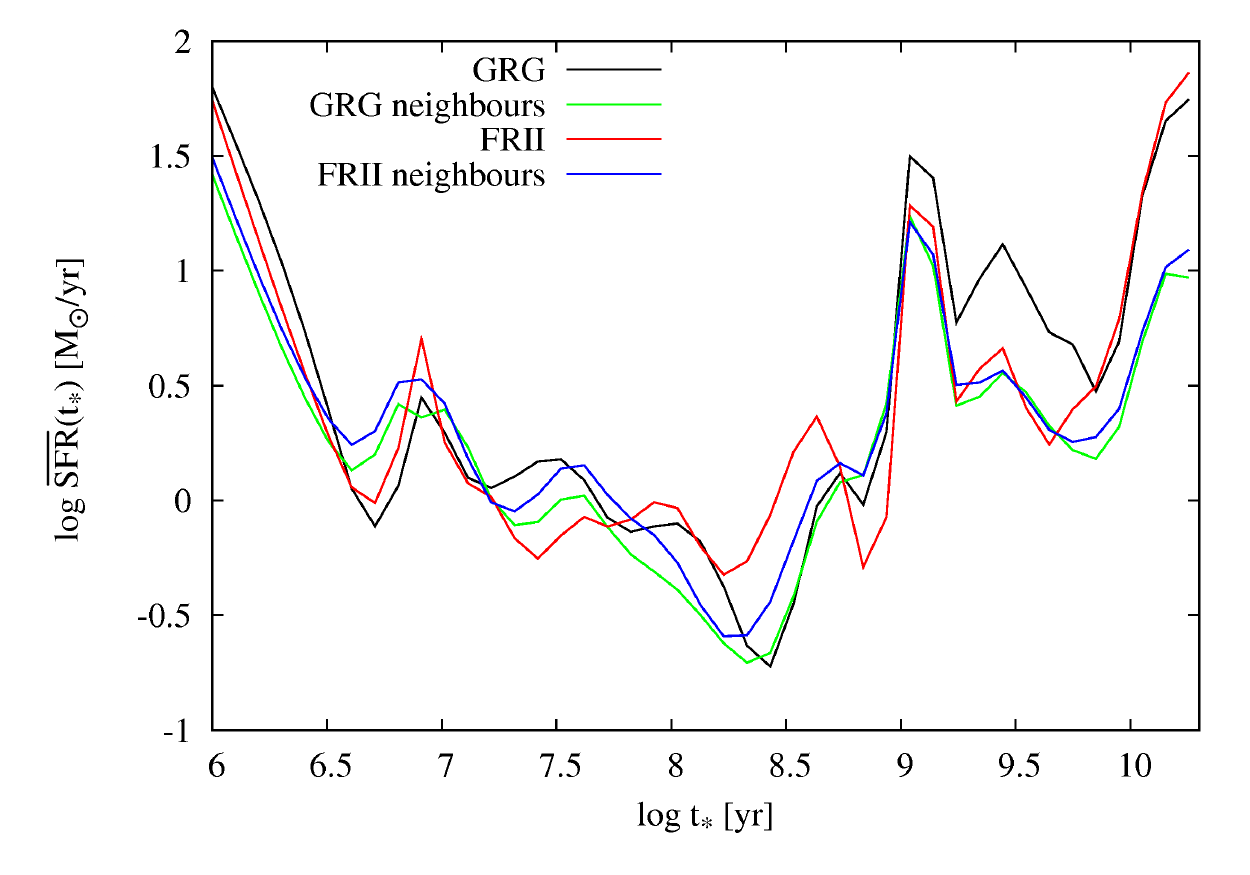}
    \includegraphics[width=0.99\columnwidth]{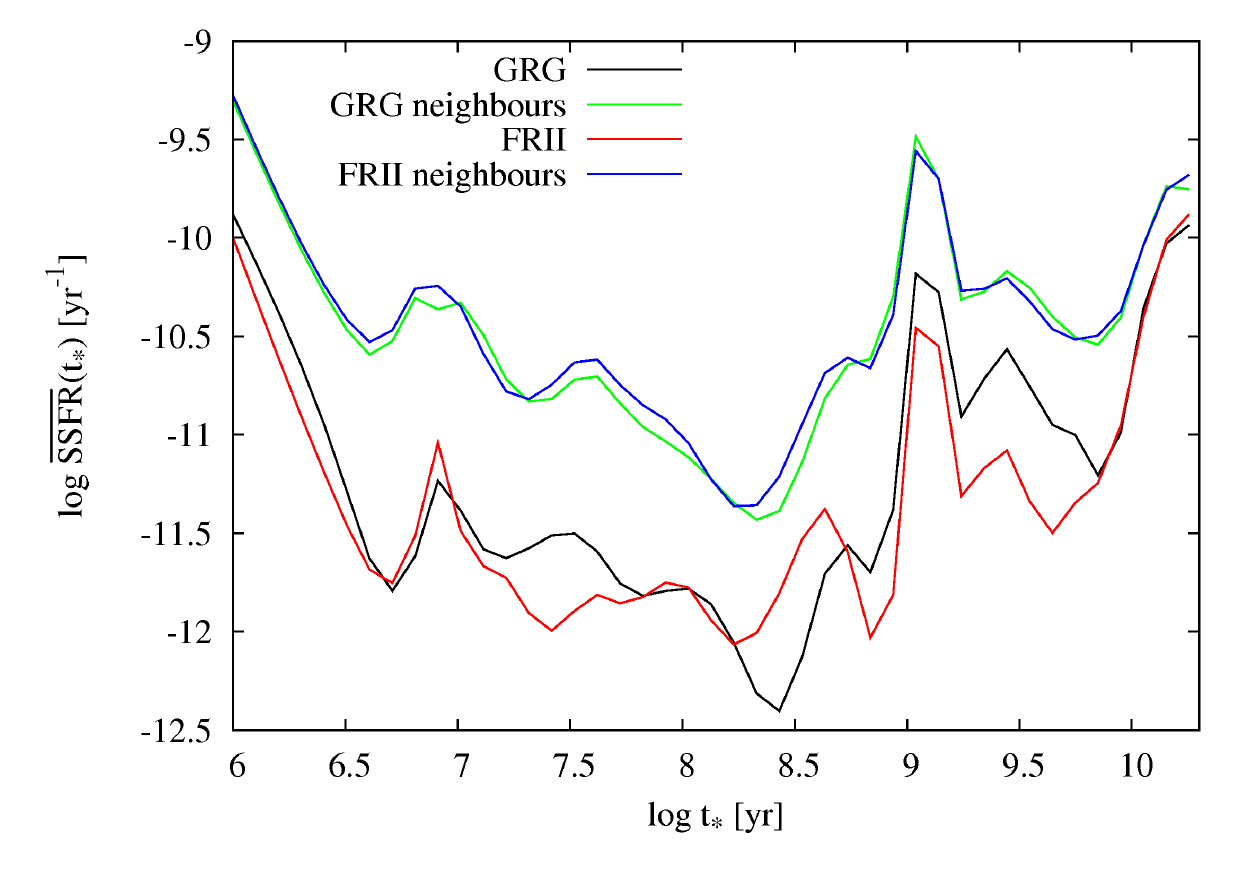}
\caption{Average time-dependent star formation rate ({\bf top graph}) and specific star formation rate ({\bf bottom graph}) for sample of GRGs, FRIIs, neighbours of GRGs within a radius of 1.5 Mpc and neighbours of FRIIs within a radius of 1.5 Mpc. }
\label{SFR2}
\end{figure}

According to our results from Section \ref{Stellar populations}, the neighbours of giants have larger fraction of intermediate age populations compared to their counterparts around FRII galaxies, but this effect is visible for galaxies located up to 1.5 Mpc from the host. Therefore, we plotted the same figure as Figure \ref{SFR} but for neighbouring galaxies located within a radius of 1.5 Mpc. In Figure \ref{SFR2} it is clearly visible that in case of neighbours of giants the $\sim$1 Gyr starburst also started earlier. It confirms that in groups with giants (within a radius of 1.5 Mpc) star formation processes were triggered at almost the same time indicating the specific global conditions occurred in the intergalactic medium of a group.        

\subsection{Galaxy distribution around GRGs}

The distribution of galaxies around GRGs was previously studied by several authors (e.g. \cite{malarecki2015, pirya2012, chen2011a, chen2011b, chen2012a, chen2012b}). They used GRGs to probe the properties of the ambient IGM. As a result they find evidence of radio jet interaction with the group of galaxies around some GRGs. They also find that in most giants, the shorter jet is brighter, suggesting asymmetries in the IGM which may not be apparent in the distribution of neighbouring galaxies \citep{pirya2012}. Also the asymmetries and deformations of radio lobes indicate the influence of environment on to the radio source. \citet{malarecki2015} find that there is a tendency for GRG's lobes to grow in directions that avoid dense regions that have large number of galaxies (i.e. perpendicular to filaments) on both small and large scales. Hence they state that GRGs can grow to large sizes due to their specific location in large scale structure of Universe.\\
In Appendix A we plot the maps of all GRGs from our sample and marked all spectroscopically confirmed neighbours as well as photometric candidates for galaxy group members. Eight of the GRGs from our sample was also studied by \citet{pirya2012} (J0926+6519, J1006+3454, J1147+3501, J1247+6723, J1311+4059, J1328-0307, J1400+3019, J1428+2918, J1635+3608). Based on the distribution of spectroscopically identified galaxies that we study in this paper, we can see that many GRGs have radio lobes directed towards the less dense regions in the cluster (as it was shown by \cite{malarecki2015}), but there are some evident examples of GRGs where the radio jets are directed to the denser regions (e.g. J0134-0107, J0918+3151, J1004+5434, J1021+1217, J1032+5644, J1311+4058 and J1429+0715). In these radio sources we observe neighbouring galaxies located along the radio lobes with few or no galaxies in the orthogonal direction. Another example of such a giant is DA240 studied by \citet{chen2011b} who show that neighbouring galaxies lie along the major axis of the radio source. It shows that GRGs are located in relatively various environments and future studies are needed to verify if their orientation in large scale structure of Universe can be a significant factor responsible for their sizes. We also note that a large fraction (45\%) of GRGs from our sample have at least one relatively close neighbouring galaxy 0.2 Mpc away from GRG's host, while in a comparison sample of smaller sized FRIIs we observe close neighbours in 36\% of radio galaxies, but this tendency also have to be tested in future studies. 

\section{Discussion}

There are many factors which are closely related to star formation in galaxies. We can distinguish two groups of them: the factors related to physical properties of individual galaxies (e.g. mass, luminosity, morphological type, gas richness, etc.), and the environmental factors (galaxy interactions and mergers, tidal forces, cold streams, gas stripping, strangulation, density of IGM, etc.) Some of them can trigger a starburst, and some of them can suppress star formation. The results obtained for groups with GRGs and FRIIs indicate the importance of environmental factors because we observe higher fractions of $\sim$1 Gyr aged star formation not only in hosts of GRGs but also in their neighbouring galaxies.

\subsection{Merger events}

Galaxy interactions and mergers are thought to be a major process driving galaxy formation and they may be responsible for triggering the star formation in interacting galaxies. However, it is well known that major galaxy mergers are rare in the nearby Universe (e.g. \cite{patton2000}). We also know that the majority of galaxy stellar mass was reached at the cosmic time corresponding to z$\sim$2--3 (e.g. \cite{stott2013}). This star formation epoch is visible as the peak of light and mass fraction population vectors in Figures \ref{populations} and \ref{populations2} near 10 Gyrs.\\      
A significant fraction of intermediate age stars ($\sim$1 Gyr) are visible in all samples considered in this paper. This was also found by \citet{raimannl2005} for a sample of 24 radio galaxies. They suggest that there is a connection between starburst episodes occurring 1 Gyr ago and the radio activity at the present time. They also state that the starbust was a consequence of, for example, interaction with a passing external galaxy, or merger. This scenario is also proposed by \citet{huang2009} although they do not find any obvious evidence of morphological disturbance in a sample of low redshift elliptical galaxies. They also state that the merger events leading to star formation are relatively minor and that the morphological disturbances could not already be visible. It is also confirmed by the observations that massive galaxies ($>$10$^{10} M\odot$) passed one or two major merger events within z$<$1.2 \citep{conselice2009}. According to these results it is possible that star formation which happened $\sim$1 Gyr ago is a result of mergers at z$\sim$1.  \\
Merger events could be a good explanation of $\sim$1 Gyr starbursts in GRGs and smaller sized radio galaxies. However, it is more significant for central galaxies of the group. The galaxies located at larger distances from the centre are not disturbed by the central merger, so the $\sim$1 Gyr star formation visible in these galaxies is likely to have other origin.

\subsection{Environment}

Global star formation may be a result of environmental factors. The environmental influence on star formation has been studied by many authors. For example the studies of \citet{hoyle2005a, hoyle2005b} show that galaxies located inside voids have higher star formation rates than galaxies in denser regions and that they are still forming stars at the same rate as in the past. However the GRSs are mostly located out of cosmic voids \citep{kuzmicz2018}.\\
Also \citet{ceccarelli2008} find that bluer galaxies with a wide range of luminosity and local density, which are located at the void peripheries, show  increased star formation. They explain this effect as a consequence of lower accretion and the merger history of galaxies arriving at void walls from the emptier inner void regions.\\

\subsection{Cold streams}
The other possible explanation of global star formation in groups of galaxies is the interaction of group galaxies with cold (10$^4$K -- 10$^5$K) intergalactic gas which penetrates the galaxies. Such cold flows are filamentary and clumpy \citep{keres2005}, particularly in the low density environment. The star formation caused by cold streams occurs only in low-mass galaxy halos ($<$10$^{12}\odot$). For more massive halos the cold stream is preheated in a standing shock to nearly virial temperature of 10$^6$K and star formation does not follow \citep{dekel2006}. Therefore, when we consider cold streams as an explanation of higher star formation occurring $\sim$1 Gyr ago, it can only be the case for low mass galaxy halos.\\
The evidence of cold streams passing through the galaxy group should be visible in the ages of stellar populations. The typical velocities of cold streams are $\sim$10$^4$ km/s \citep{zinger2018} and to pass the distance equal to the assumed diameter of the group ($\sim$10 Mpc) it needs about 1 Gyr. The intermediate-age stellar populations have ages in the range of $9\times10^8$yr $<$t$_{\star}$$<7.5\times10^9$ yr, so we should be able to see evidence of higher star formation in the whole group along the cold stream. For our sample of GRGs we do not see any evidence that the starburst occurred along any preferred direction which could correspond to the cold stream direction. In some cases the $\sim$1 Gyr starburst is initially visible in central galaxies of the group, and in some cases it initially occurs at the edges. However, we do not know the sizes and geometries of these supposed cold streams and it is possible that they could pass through the galaxy group in more complicated ways. 

\section{Conclusions}

In this paper we studied the stellar populations of 41 GRGs and galaxies which belong to the groups around them. We compare our results with a sample of 217 smaller-sized FRII radio galaxies and their neighbours in order to find systematic differences in properties of the GRG's hosts and their environment, which can be responsible for the origin of large scale radio lobes. The main conclusions of this work can be summarized as follows:

The average stellar populations in samples of galaxies -- the GRG hosts, FRII hosts, neighbours of GRGs and neighbours of FRIIs -- are dominated by old stellar populations (t$_{\star}>10^{10}$yr) but they also comprise significant fraction of intermediate age populations with ages $9\times10^8$yr$<$t$_{\star}$$<7.5\times10^9$yr. 

The GRG's hosts have larger intermediate age stellar populations compared to smaller-sized FRIIs, in which the larger fraction of the oldest populations with ages above t$_{\star}>10^{10}$yr can be observed. The same effect can be seen for neighbouring galaxies located up to 1.5 Mpc from radio galaxy host -- the neighbours of giants have larger fractions of intermediate age populations compared to their counterparts around FRII galaxies.  

We do not find differences in the mean values of stellar mass, $\left <\log t^{\star}\right >_L$, and $\left <\log Z^{\star}\right >_L$ obtained for each sample of galaxies. Also, the differences in these parameters derived for individual groups of galaxies are statistically insignificant, indicating that groups with GRGs and groups with smaller-sized FRIIs are similar. 

Based on the distribution of neighbouring galaxies around GRGs, we found that radio jets are usually oriented towards the regions with smaller numbers of surrounding galaxies, however there are also a fraction of giants with jets oriented towards the dense regions. Therefore, future detailed studies are needed to confirm the scenario of specific orientation of GRGs in large scale structure of Universe, postulated as a possible explanation of large sizes of giants.\\

The larger fraction of intermediate age stellar populations in GRGs and their neighbouring galaxies can be explained as, for example, a result of past merger events or cold streams penetrating the group of galaxies, which can trigger star formation. The smaller radio sources also have a large fraction of intermediate age stellar populations but this number is lower than in GRGs. This means that in groups with GRGs, the processes responsible for star formation could be globally more efficient and they not only occurred in the central elliptical galaxy, but also in surrounding members of the group. These processes have larger significance on spatial scales of 1.5 Mpc around the radio source. Therefore, either the global properties of the intergalactic medium or past events that happened in the galaxy groups can be responsible for the giant sizes of radio structures. Both mergers and cold streams may also supply the central AGN of the radio source. This indicates that in such galaxies the central black hole is fed by new material and the radio activity mode may persist for a longer time, or it occurs more frequently than in smaller radio sources. This scenario may support the idea that the longer activity phase of central AGN in GRGs may be responsible for giant radio source sizes. The obtained results show that future studies of larger samples of GRGs with accompanying multi-object spectroscopy can be very helpful in investigations of GRGs origin and evolution in cluster environments.

\begin{acknowledgements}
This project was supported by the Polish National Center of Science under decision UMO-2016/20/S/ST9/00142.
\end{acknowledgements}

\bibliographystyle{aa} 
\bibliography{aa.bib}

\begin{appendix}
\begin{onecolumn}
\newpage
\section{Radio maps of giant radio galaxies}
\vspace{0.1cm}

\begin{figure}[hb!]
\caption{Distribution of galaxies around giant radio galaxies studied in this paper. All of the 1.4 GHz radio maps were taken from NVSS survey. The red plus symbols denote the neighbouring galaxies within the radius of 5 Mpc from GRGs host, for which spectroscopic redshifts were available and have $\Delta z\leqslant$0.003. The host of GRGs are marked as red pluses inside the red boxes. With the black crosses we indicate the galaxies with available spectroscopic redshifts but with low signal to noise ratio - this are not used to our studies. The blue boxes denote galaxies that are five magnitudes fainter than the host galaxy of GRG with $\Delta z_{phot}\leqslant$0.02. In each map we plot the circles of 0.5 Mpc, 1 Mpc and 5 Mpc radius around GRGs host.}
\vspace{1cm}
    \includegraphics[width=0.45\columnwidth]{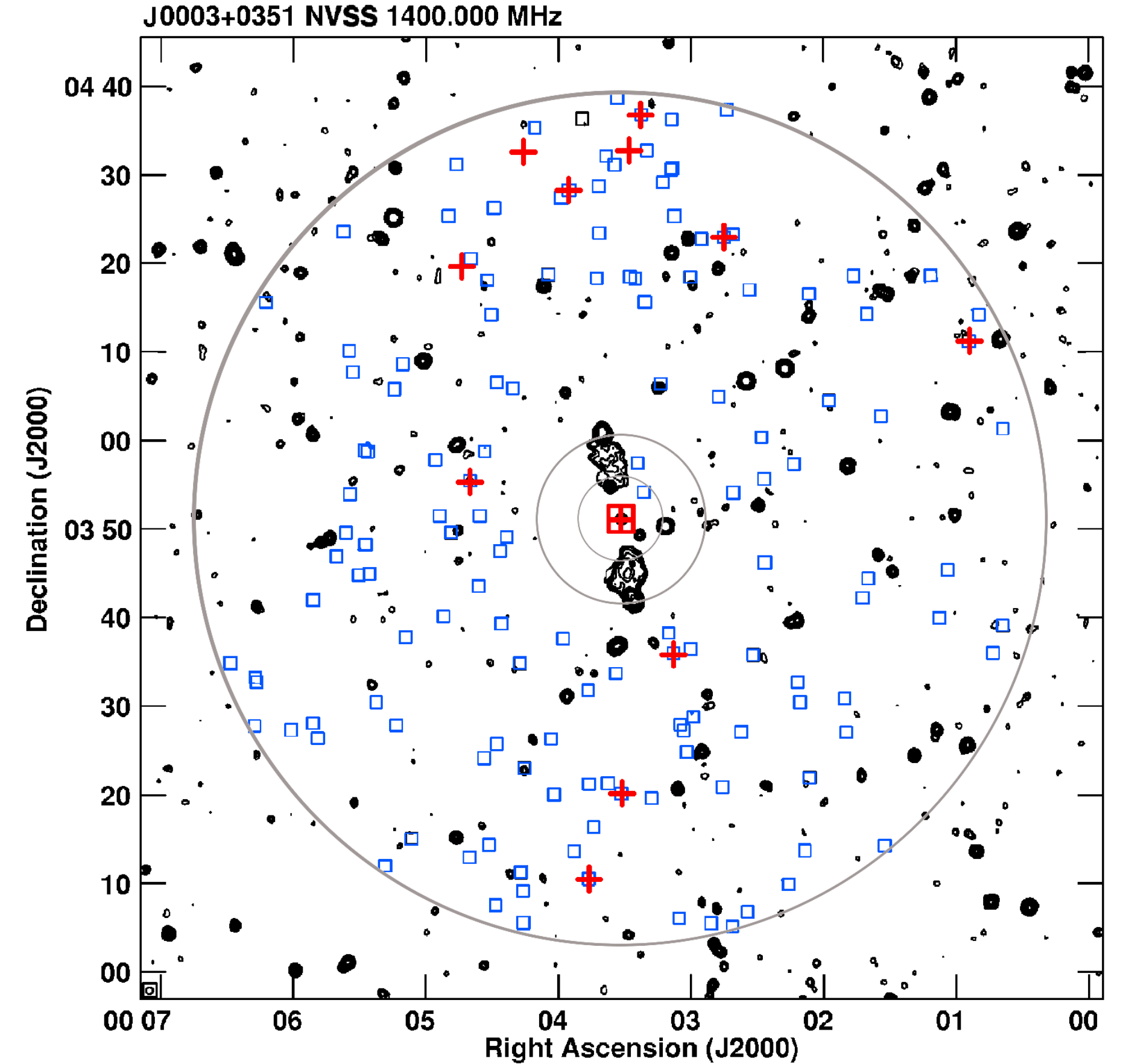} 
    \includegraphics[width=0.45\columnwidth]{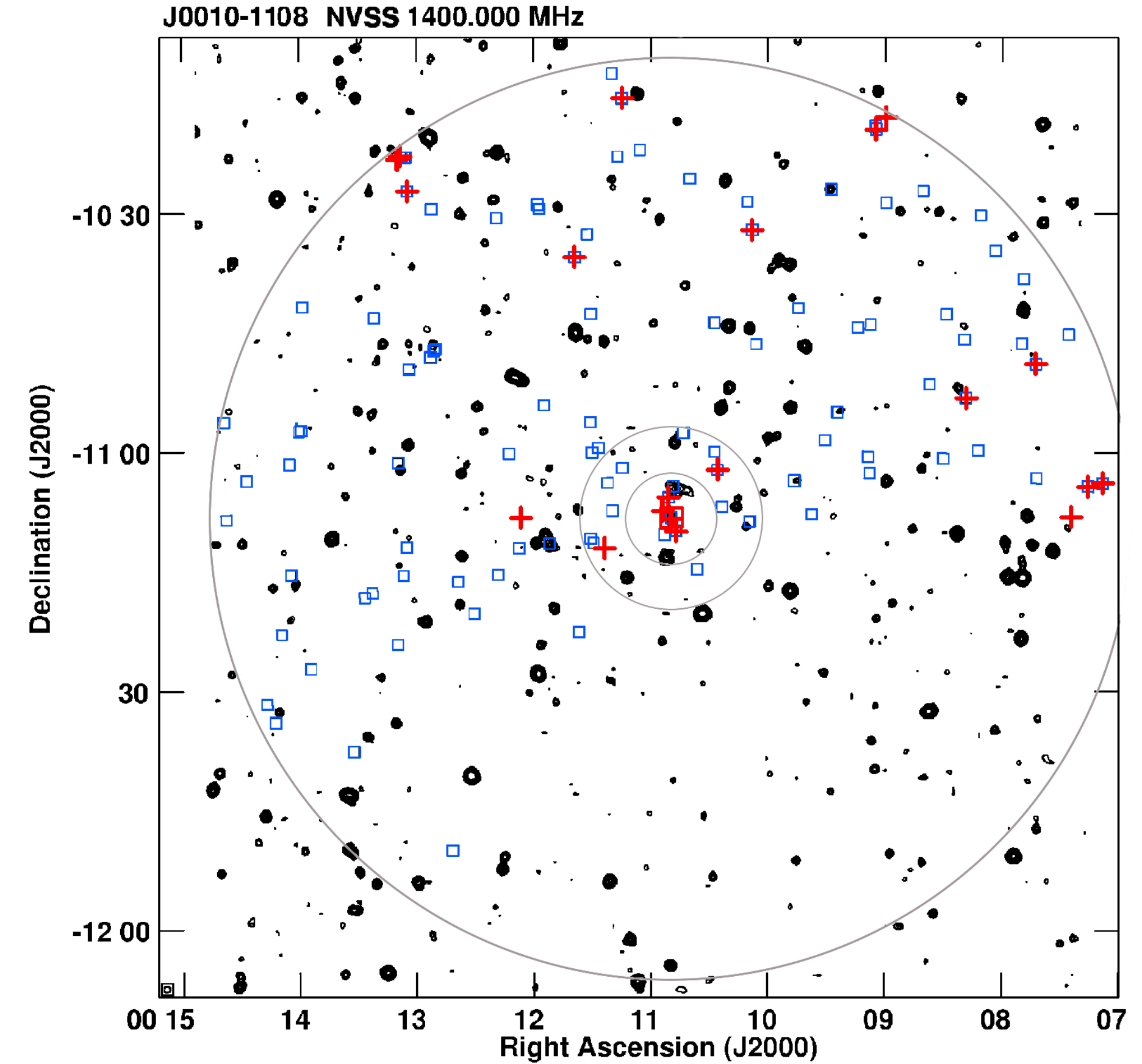} \\\\
    \includegraphics[width=0.45\columnwidth]{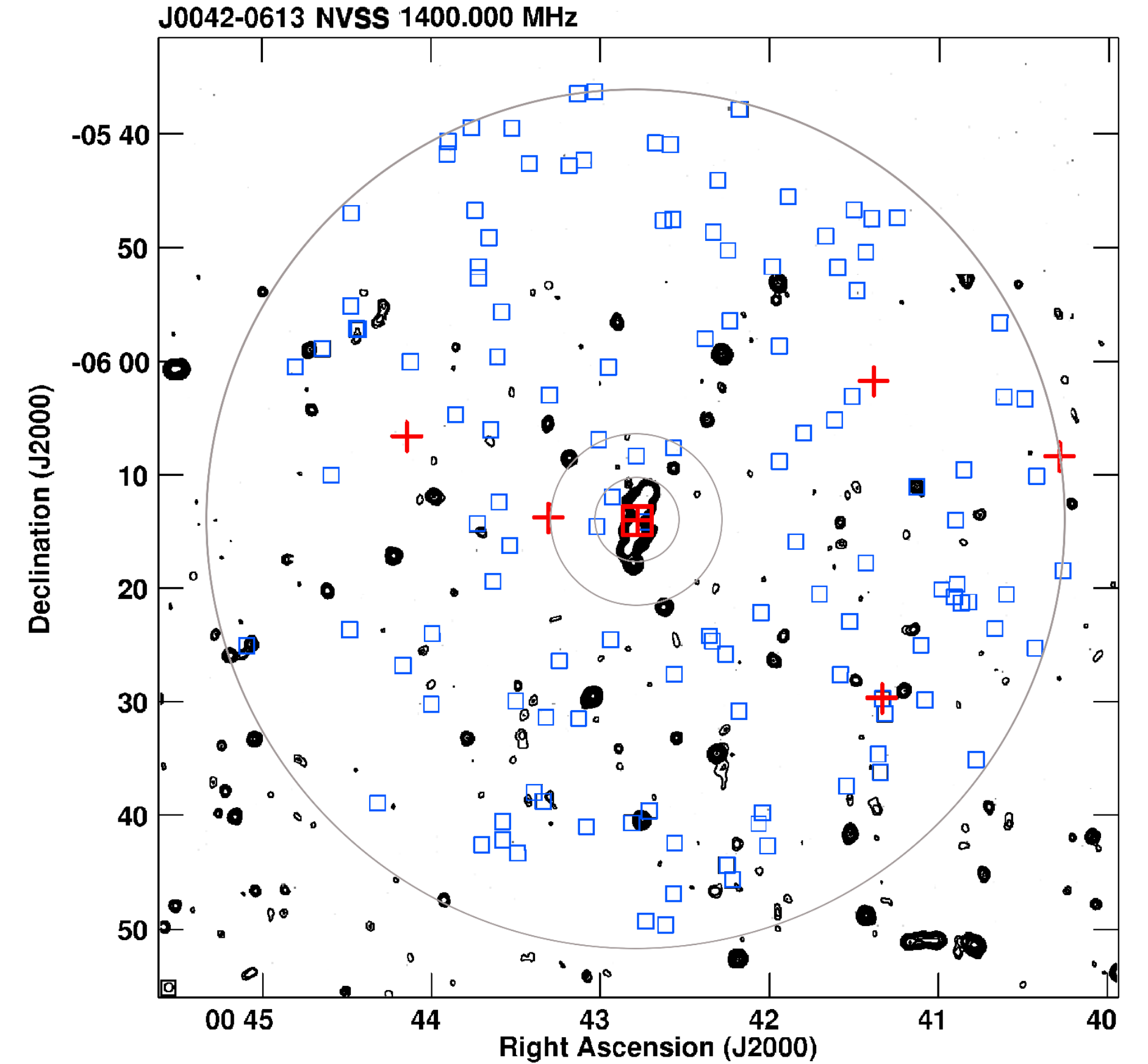}
    \includegraphics[width=0.45\columnwidth]{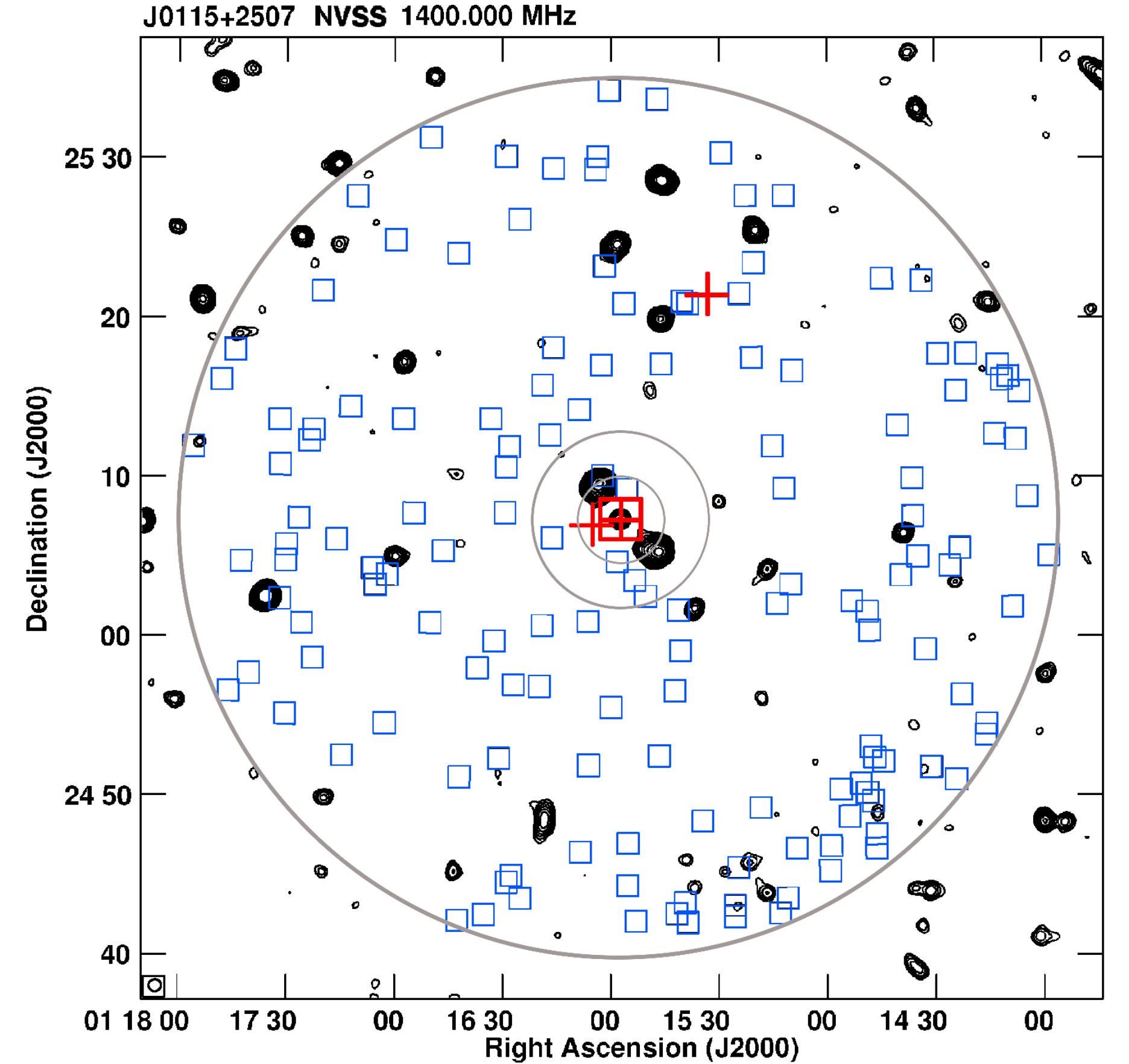}\\

\end{figure}

\begin{figure}
    \includegraphics[width=0.45\columnwidth]{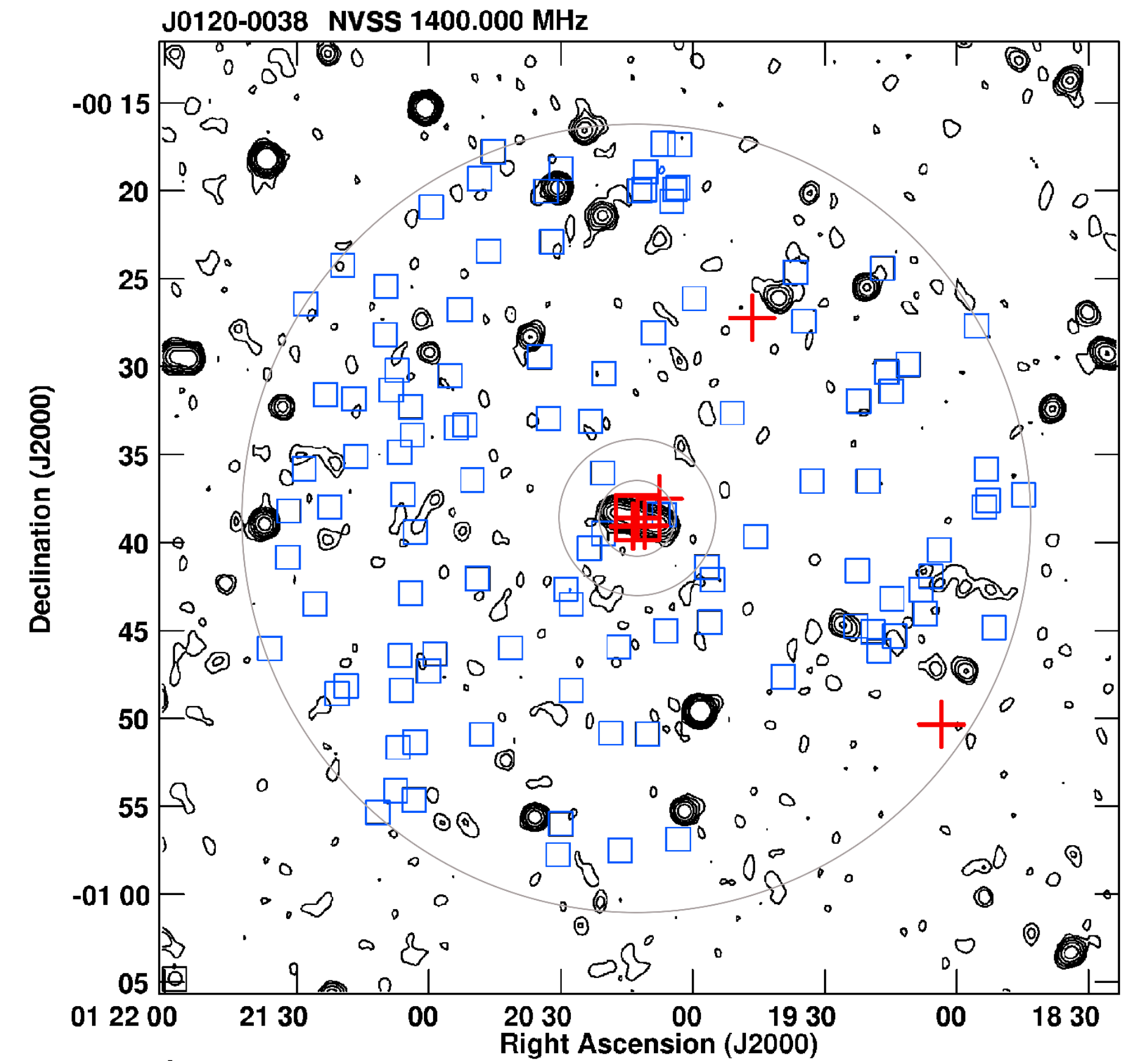}
    \includegraphics[width=0.45\columnwidth]{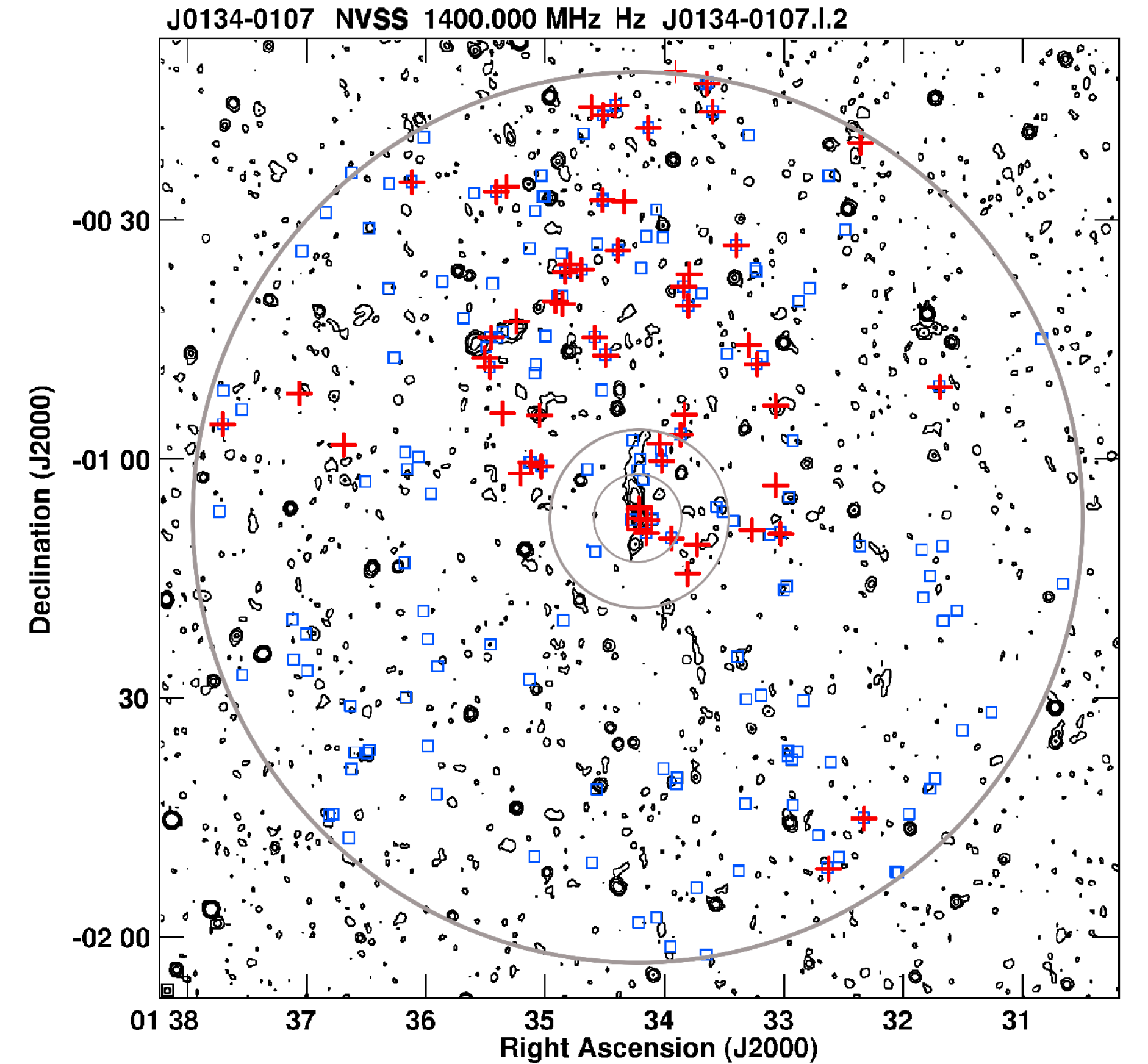} \\\\
    \includegraphics[width=0.45\columnwidth]{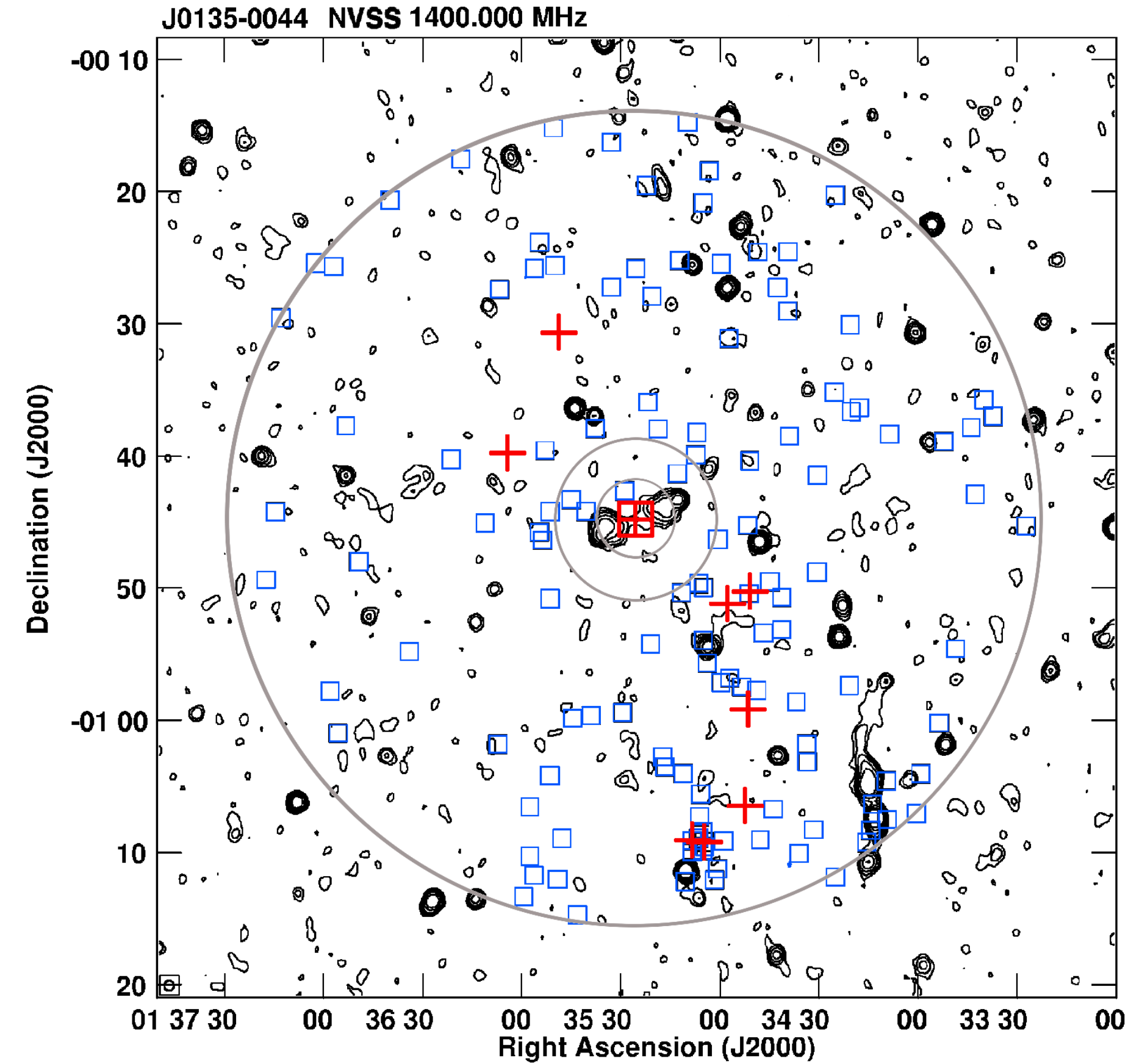}
    \includegraphics[width=0.45\columnwidth]{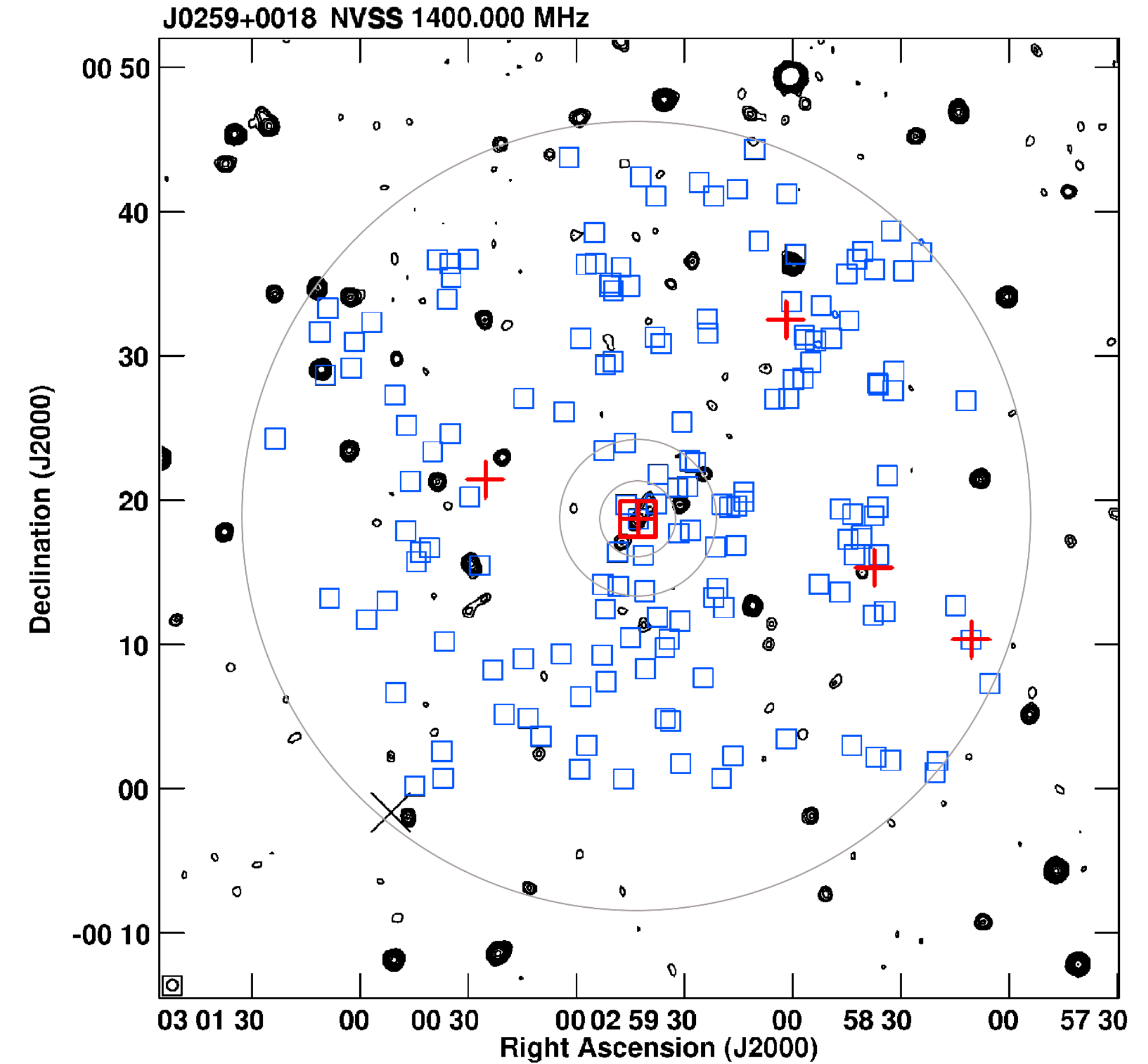}\\\\
    \includegraphics[width=0.45\columnwidth]{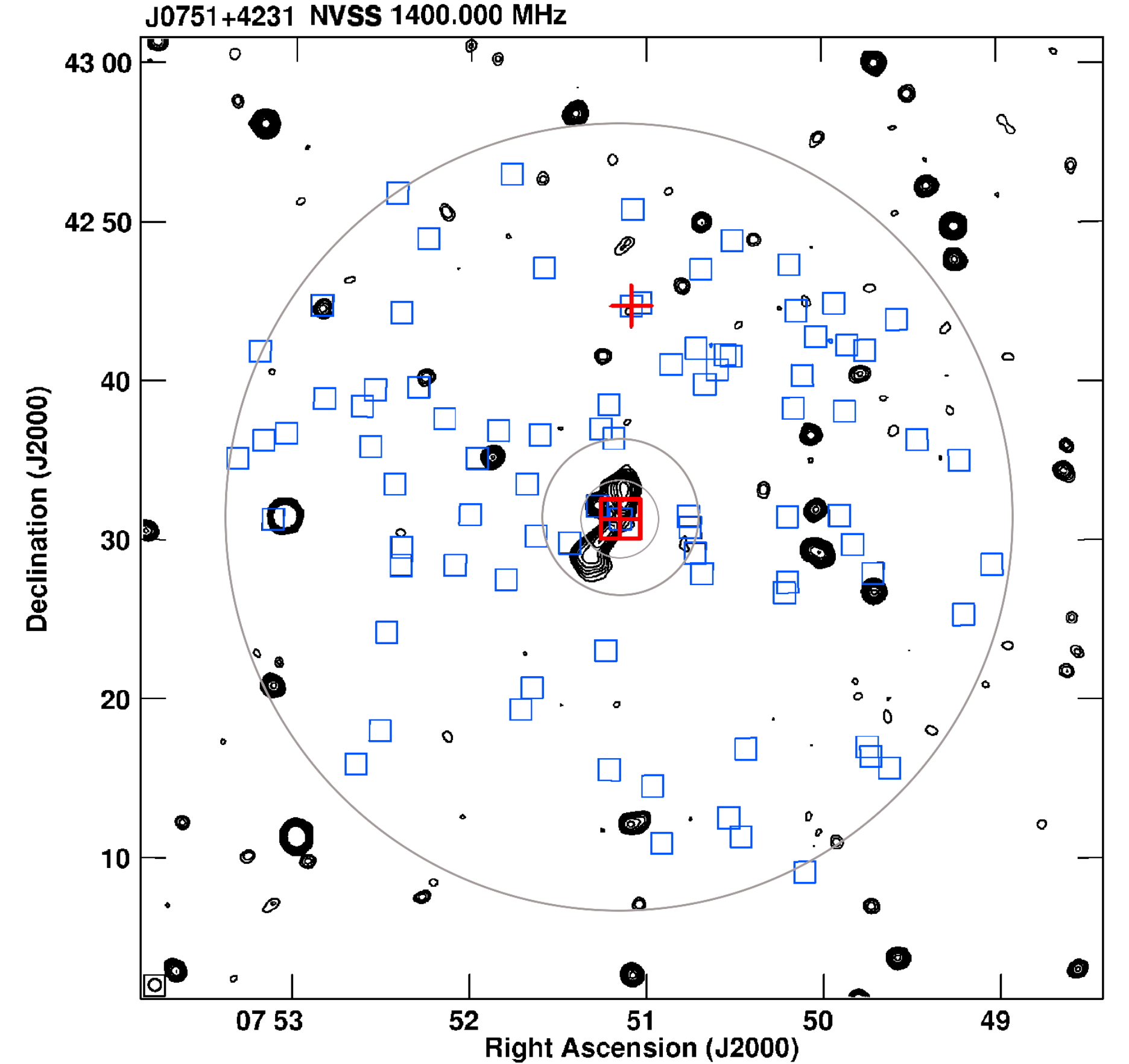} 
    \includegraphics[width=0.45\columnwidth]{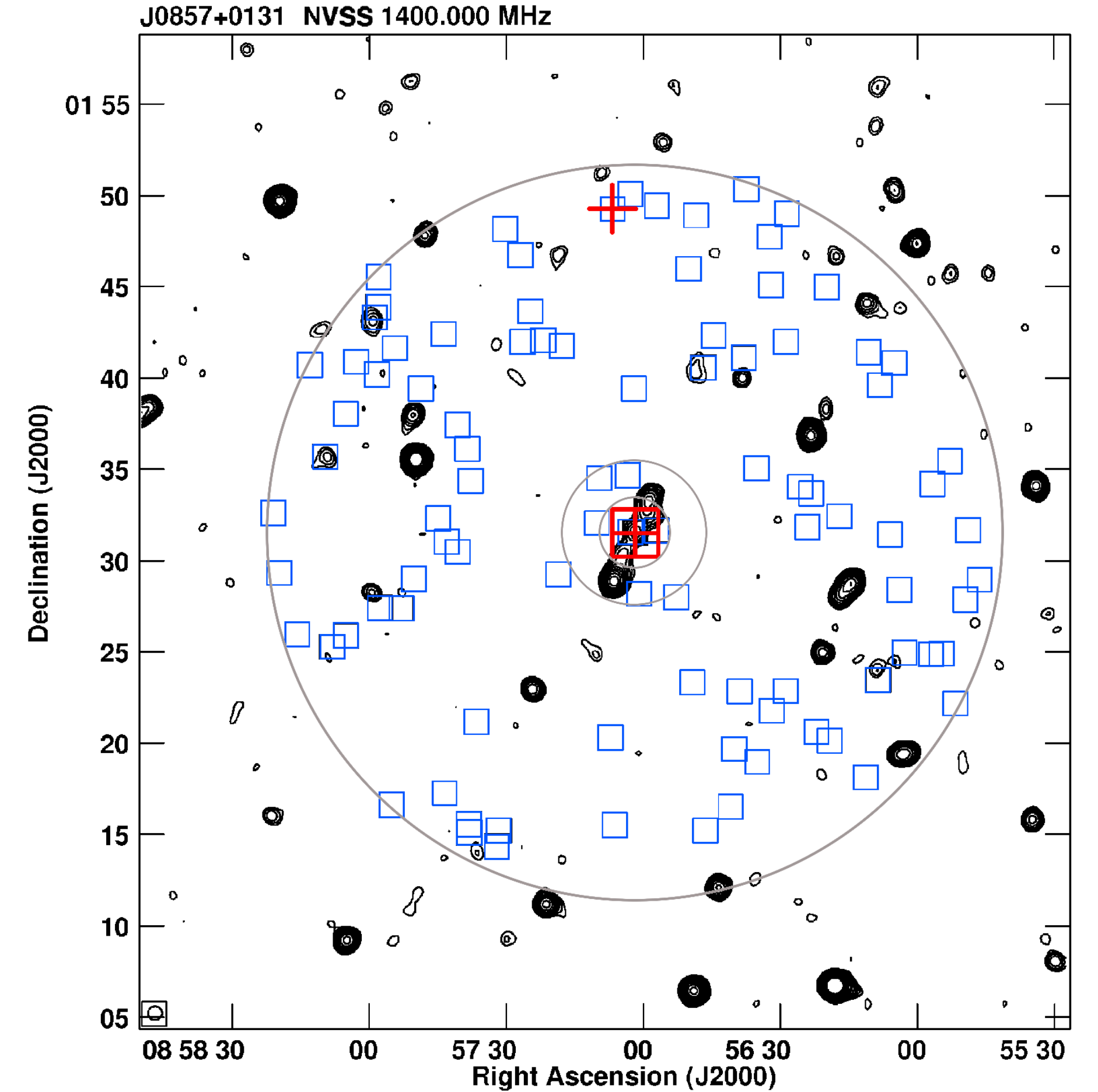}\\\\
\newpage
\end{figure}

\begin{figure}
    \includegraphics[width=0.45\columnwidth]{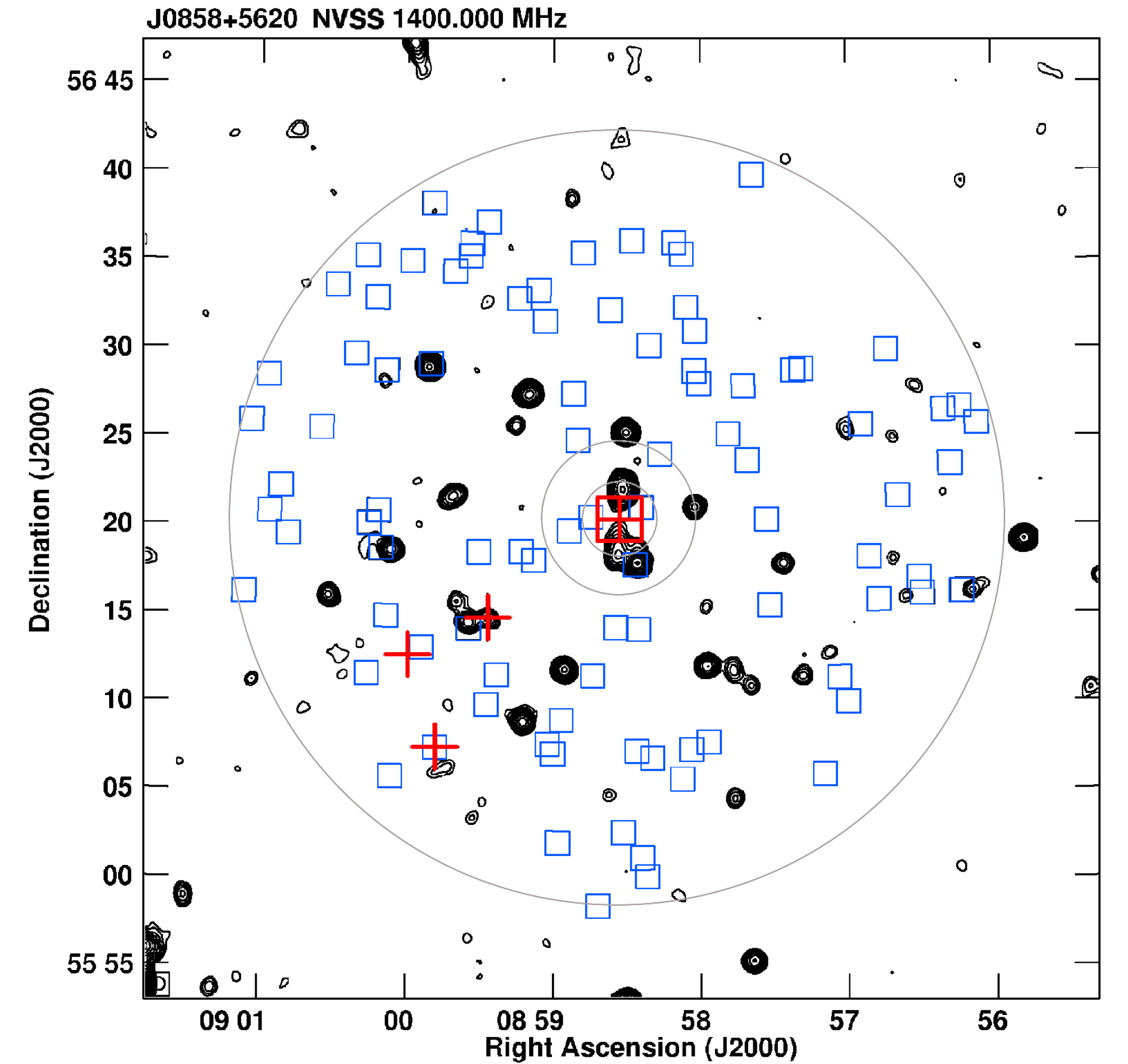}
    \includegraphics[width=0.45\columnwidth]{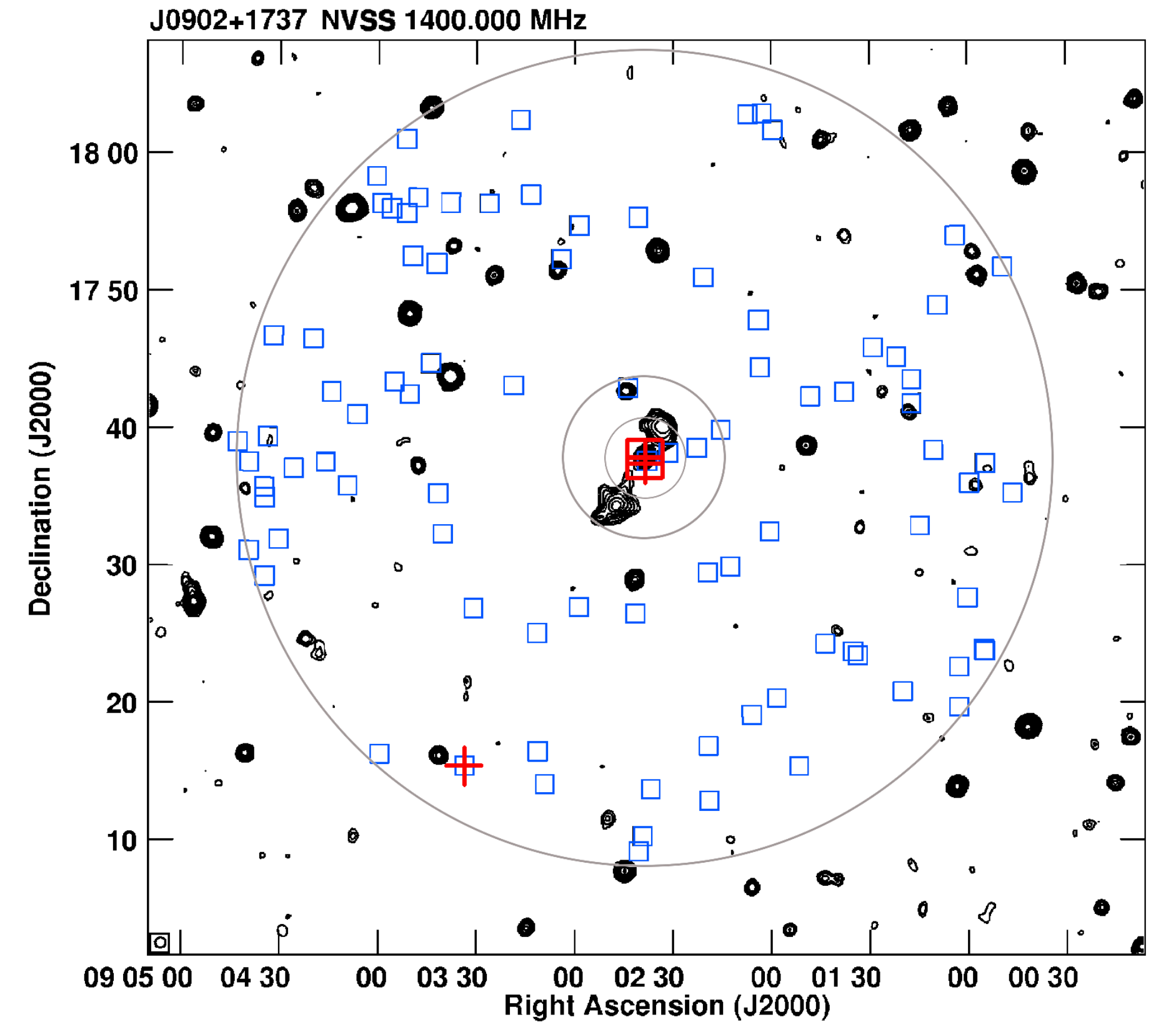}\\\\ 
    \includegraphics[width=0.45\columnwidth]{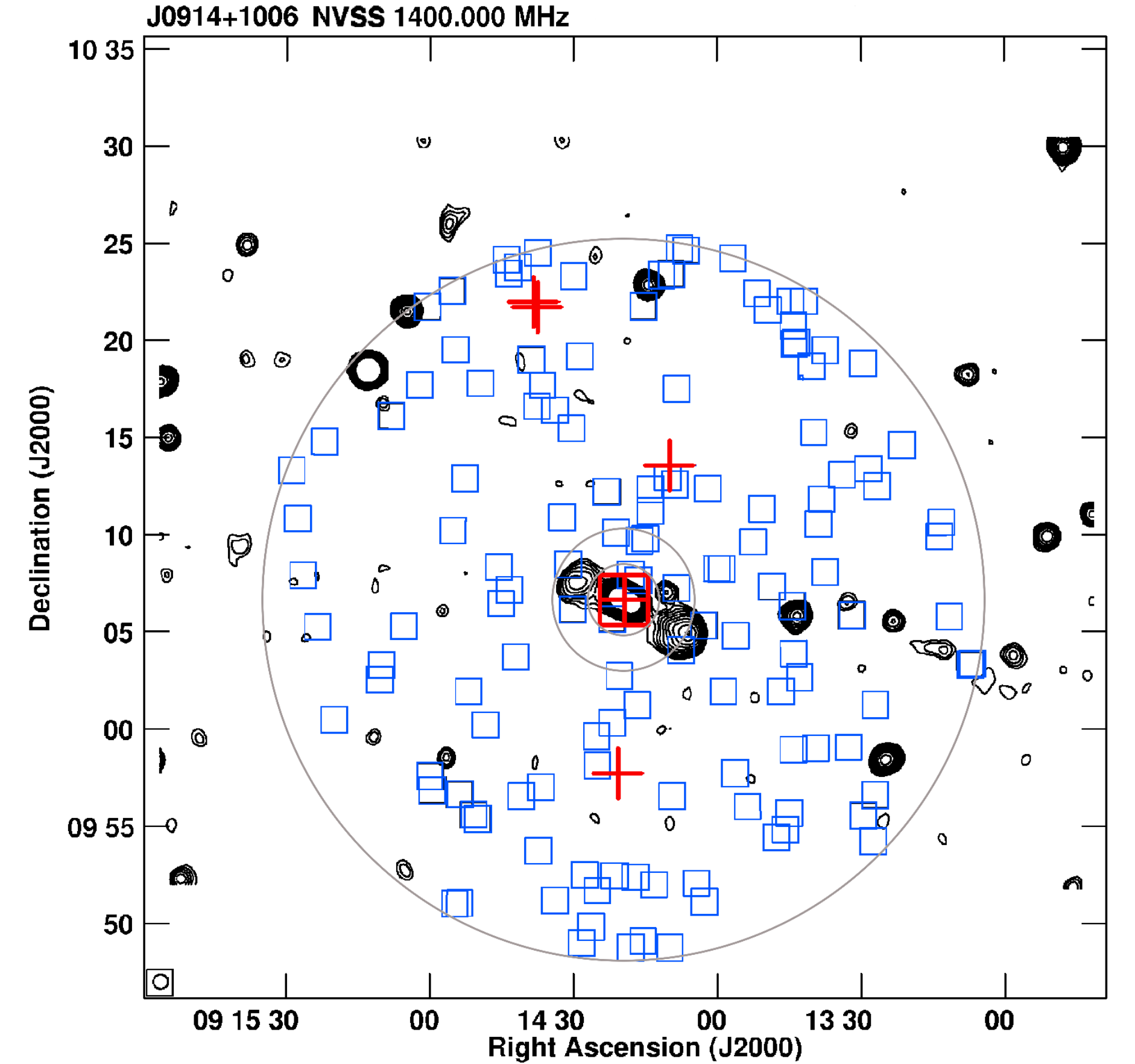}
    \includegraphics[width=0.45\columnwidth]{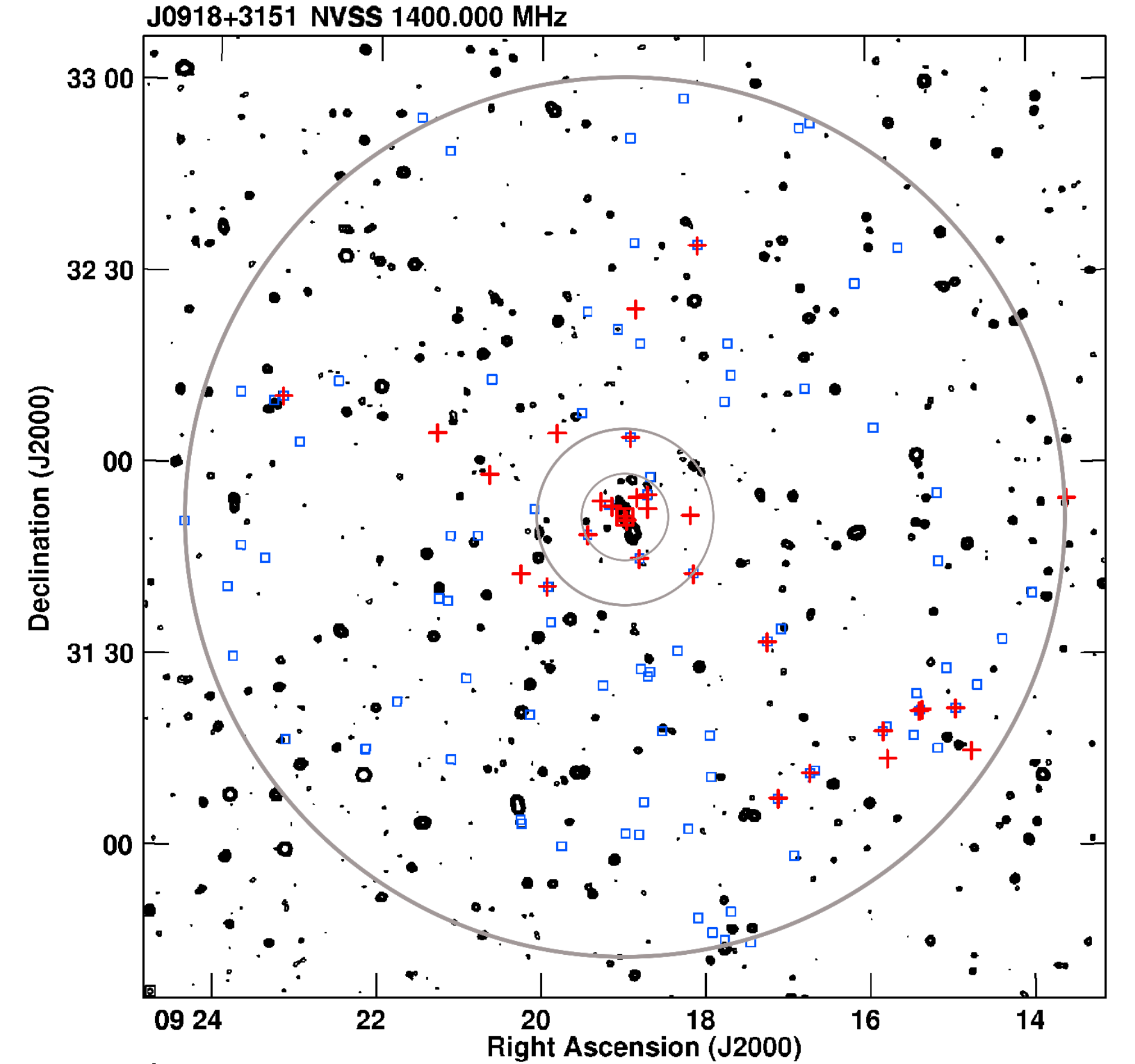}\\\\
    \includegraphics[width=0.45\columnwidth]{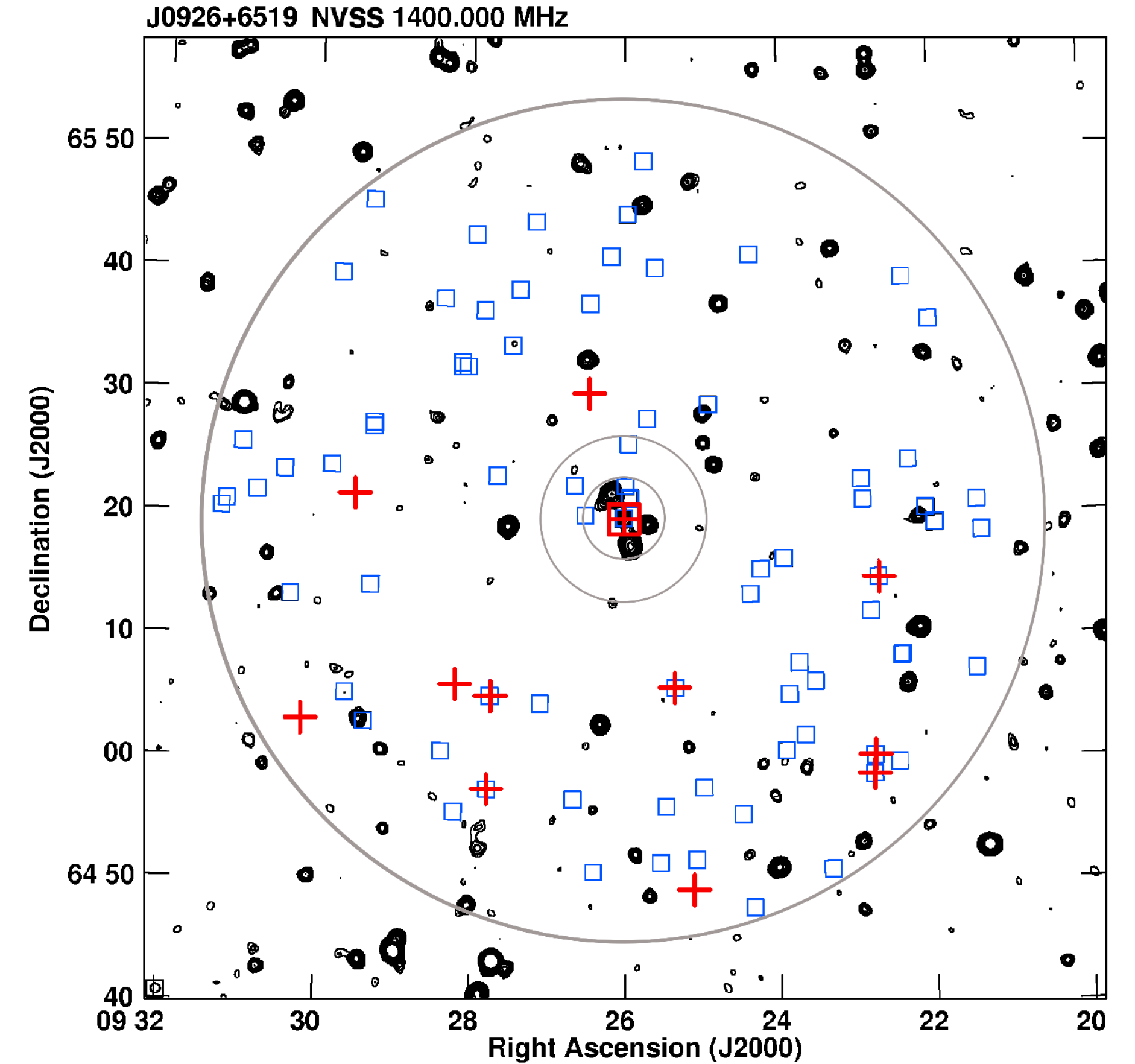} 
    \includegraphics[width=0.45\columnwidth]{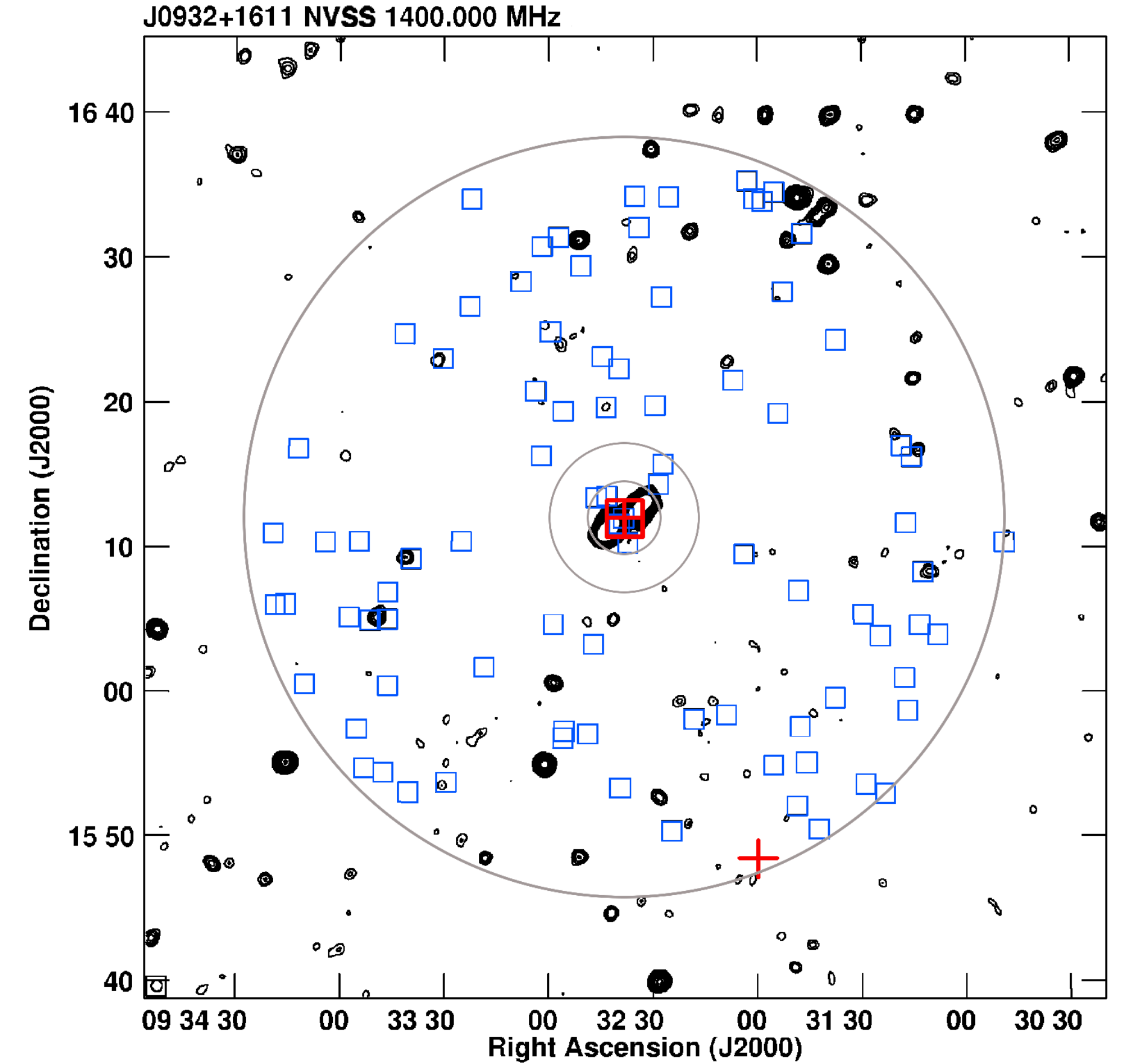}\\\\
\end{figure}

\begin{figure}
    \includegraphics[width=0.45\columnwidth]{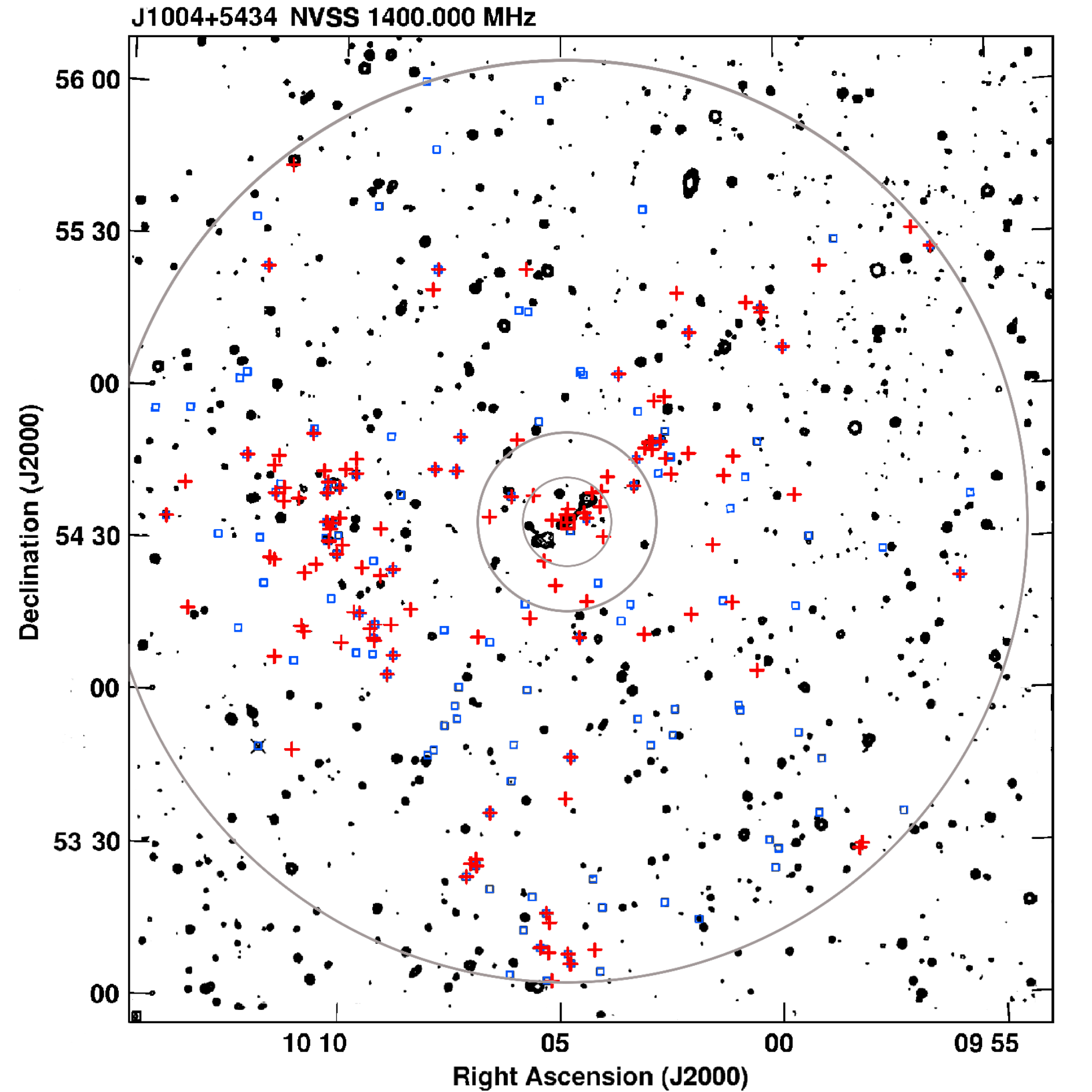}
    \includegraphics[width=0.45\columnwidth]{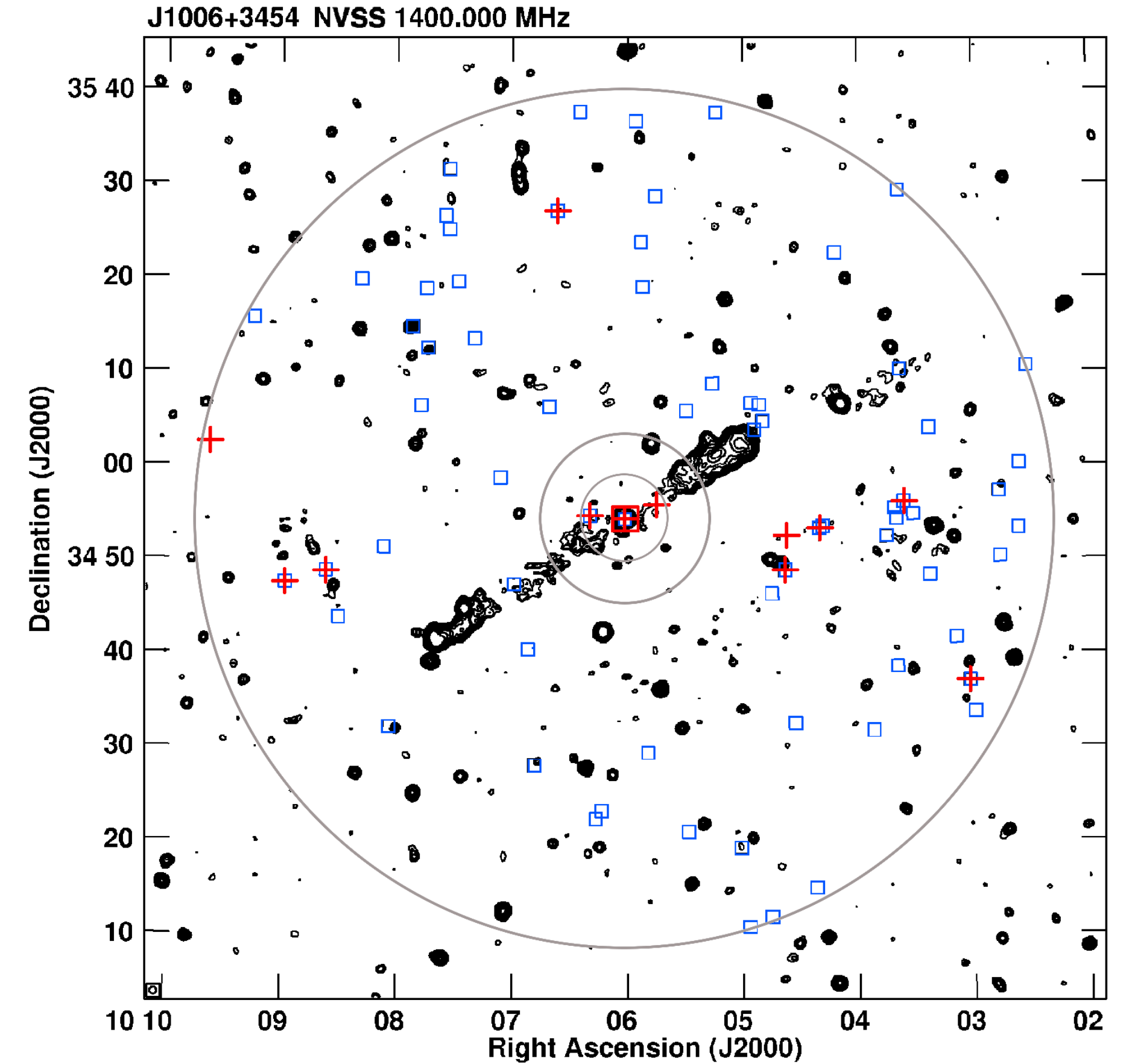} \\\\
    \includegraphics[width=0.45\columnwidth]{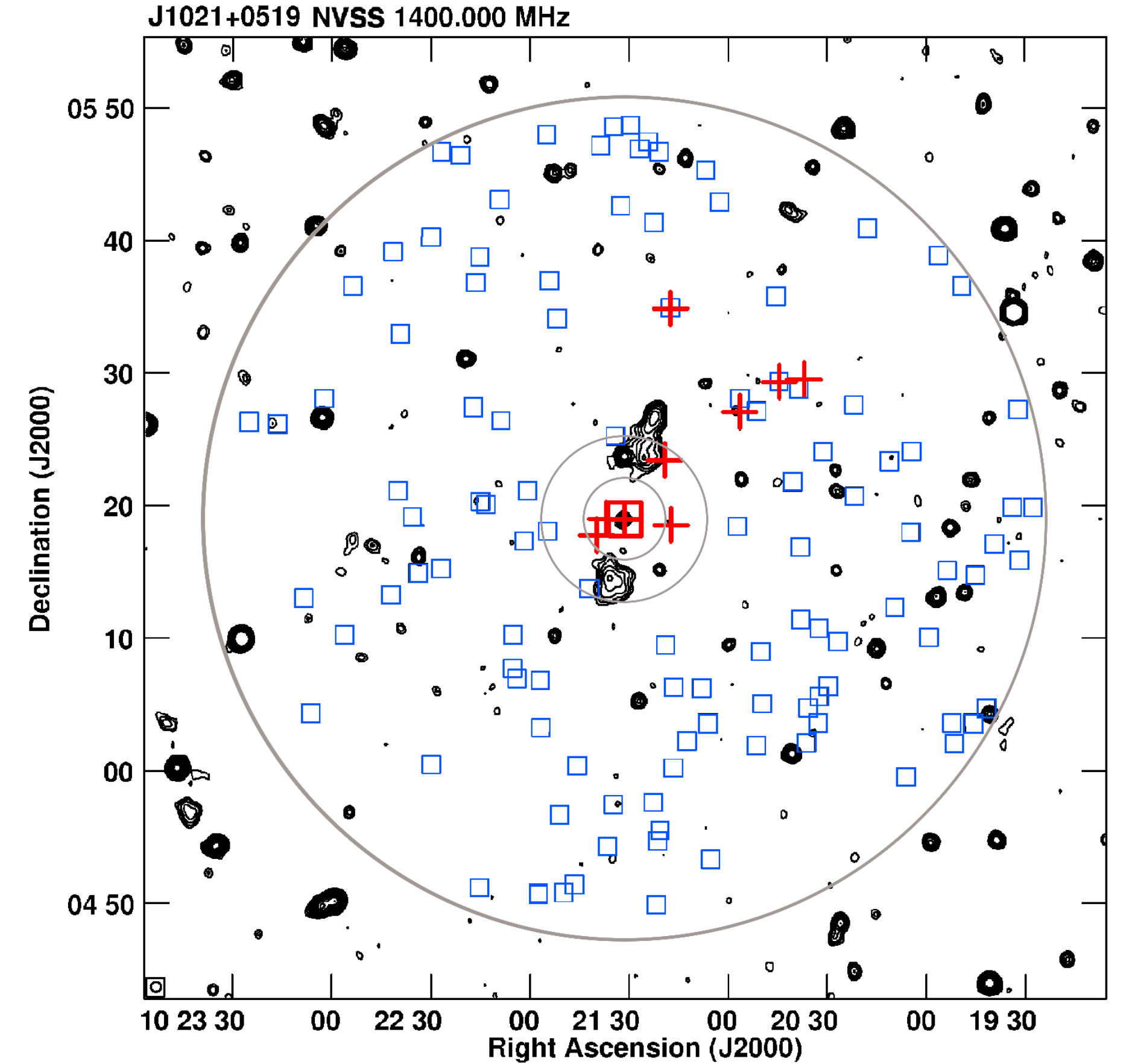} 
    \includegraphics[width=0.45\columnwidth]{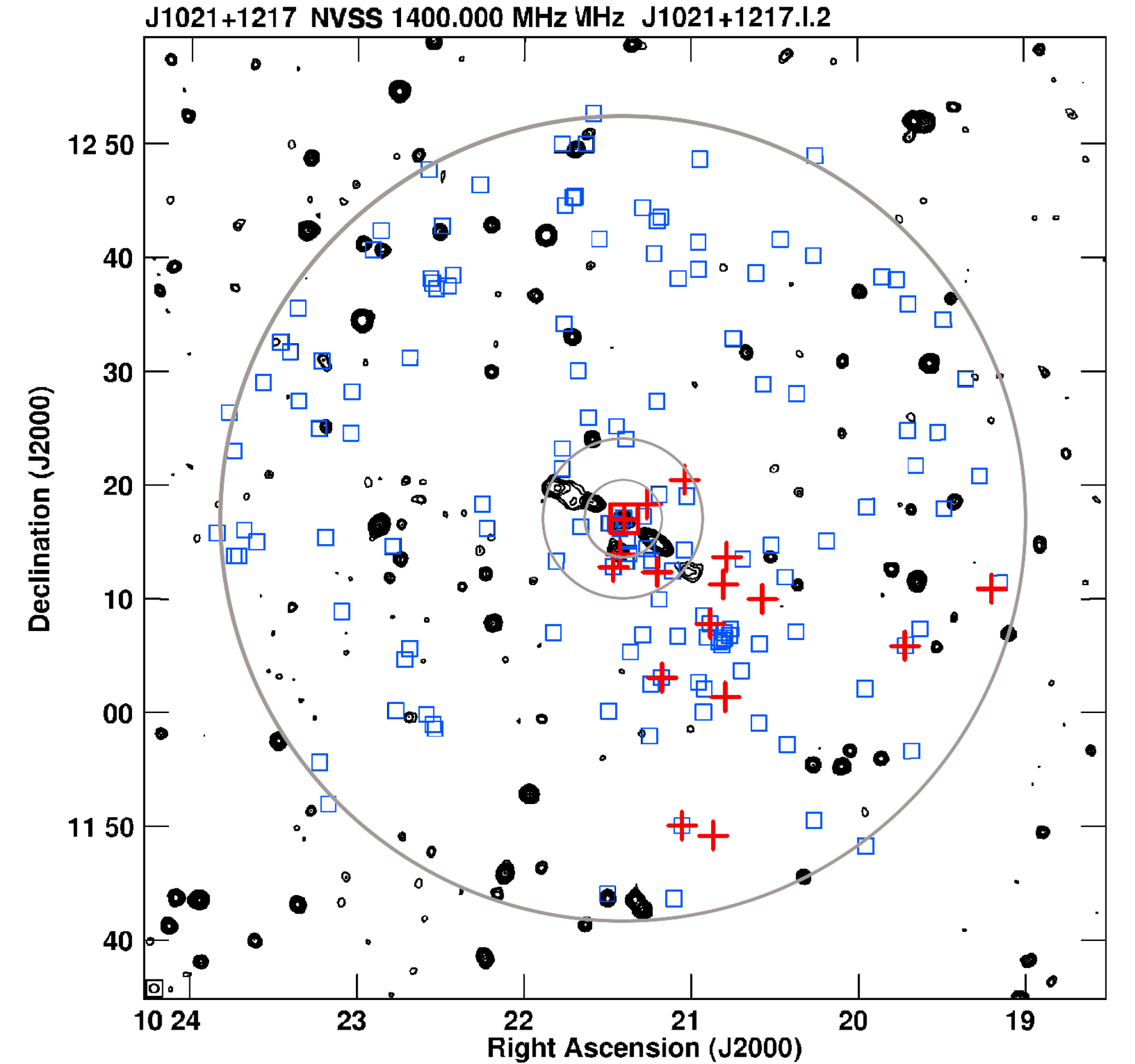}\\\\
    \includegraphics[width=0.45\columnwidth]{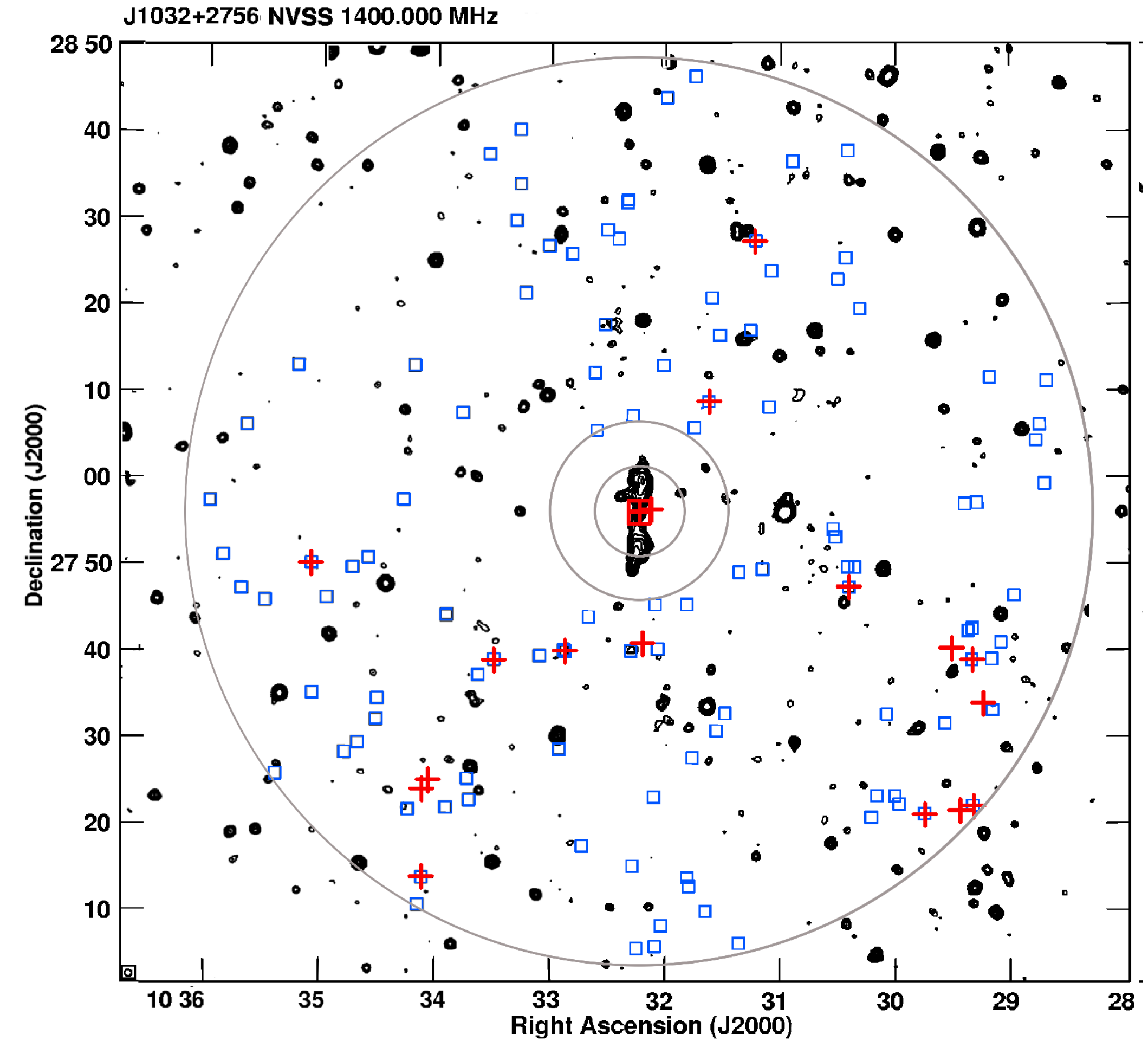} 
    \includegraphics[width=0.45\columnwidth]{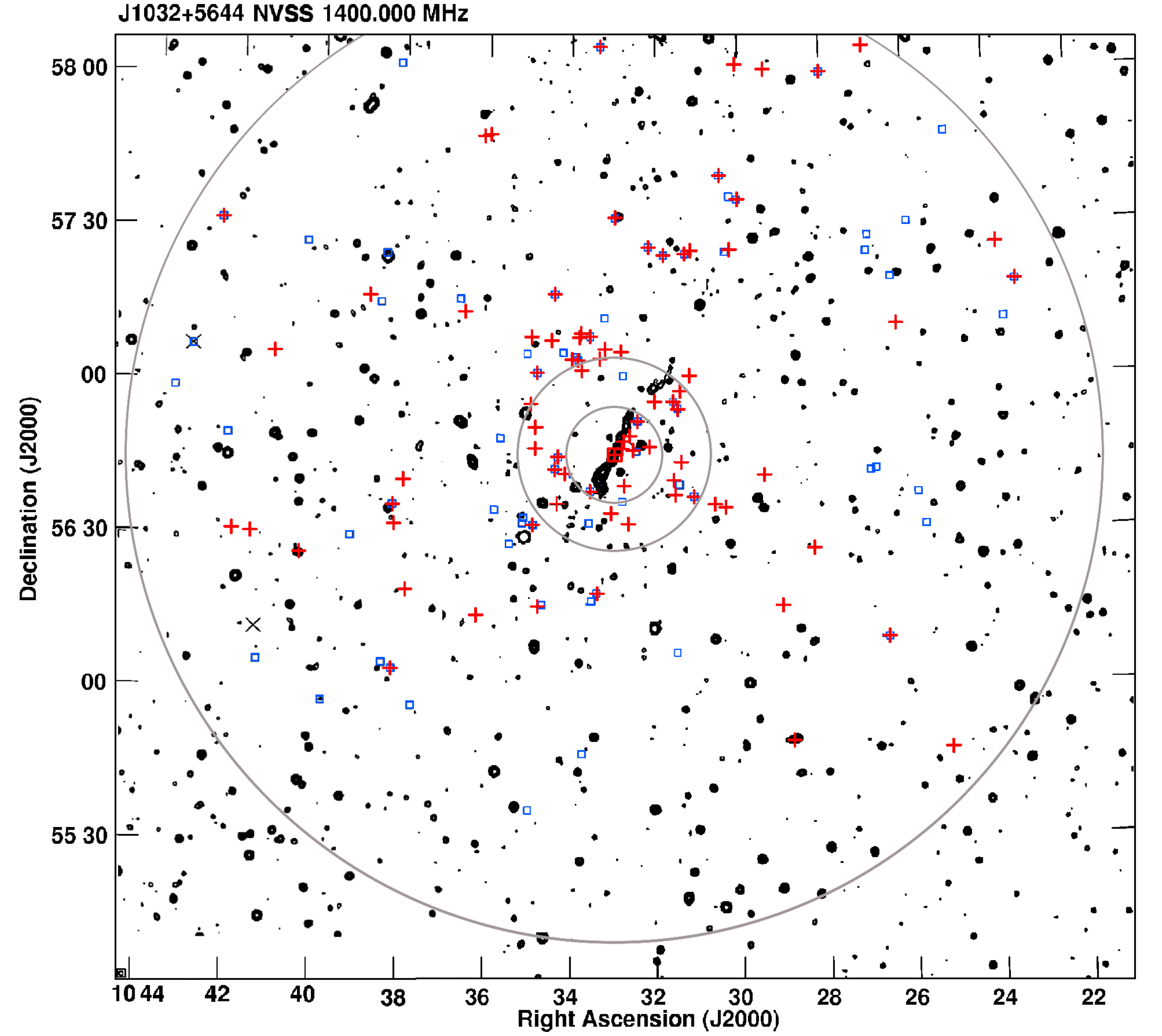}\\\\
\end{figure}
 
\begin{figure}
    \includegraphics[width=0.45\columnwidth]{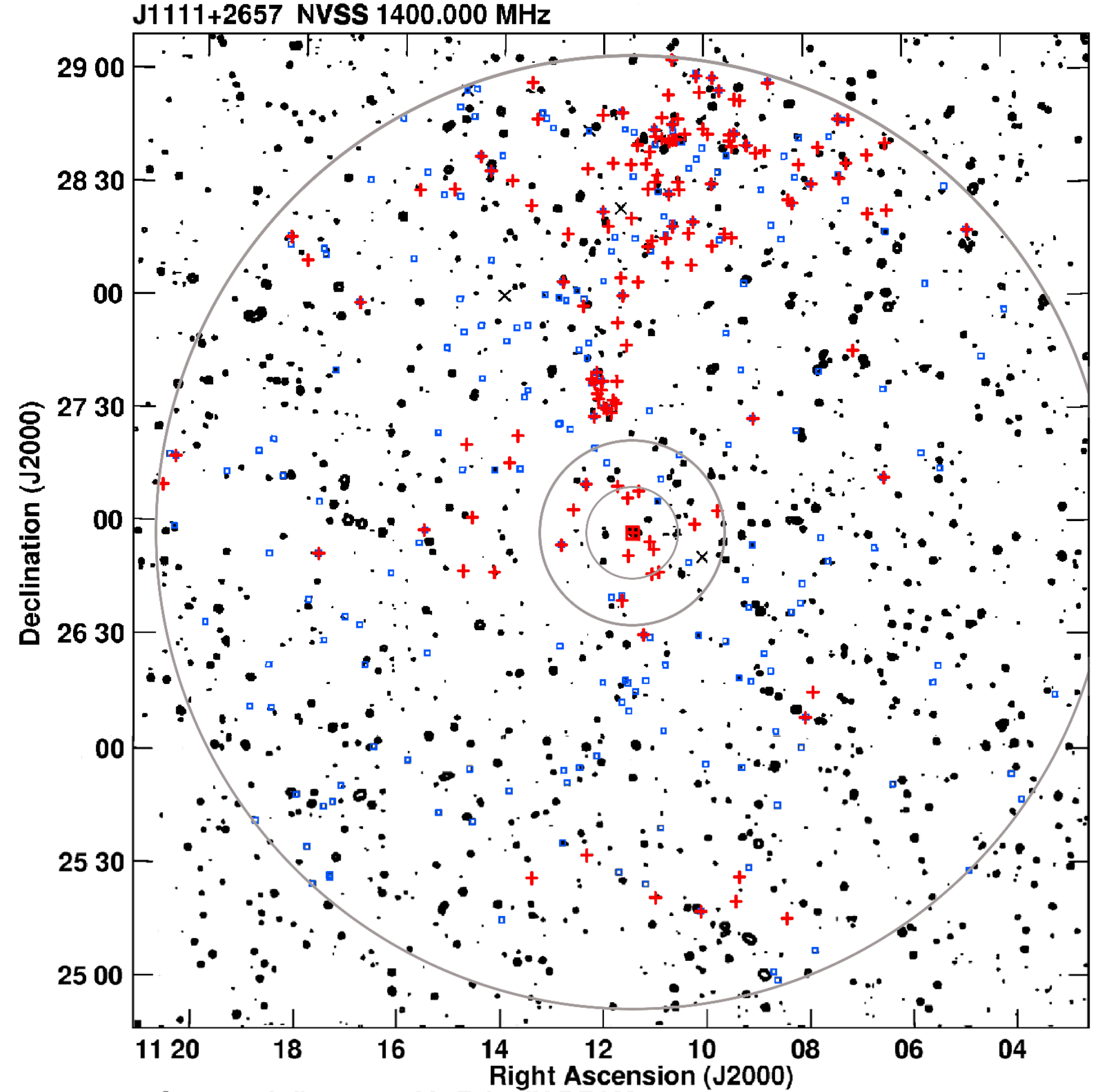}
    \includegraphics[width=0.45\columnwidth]{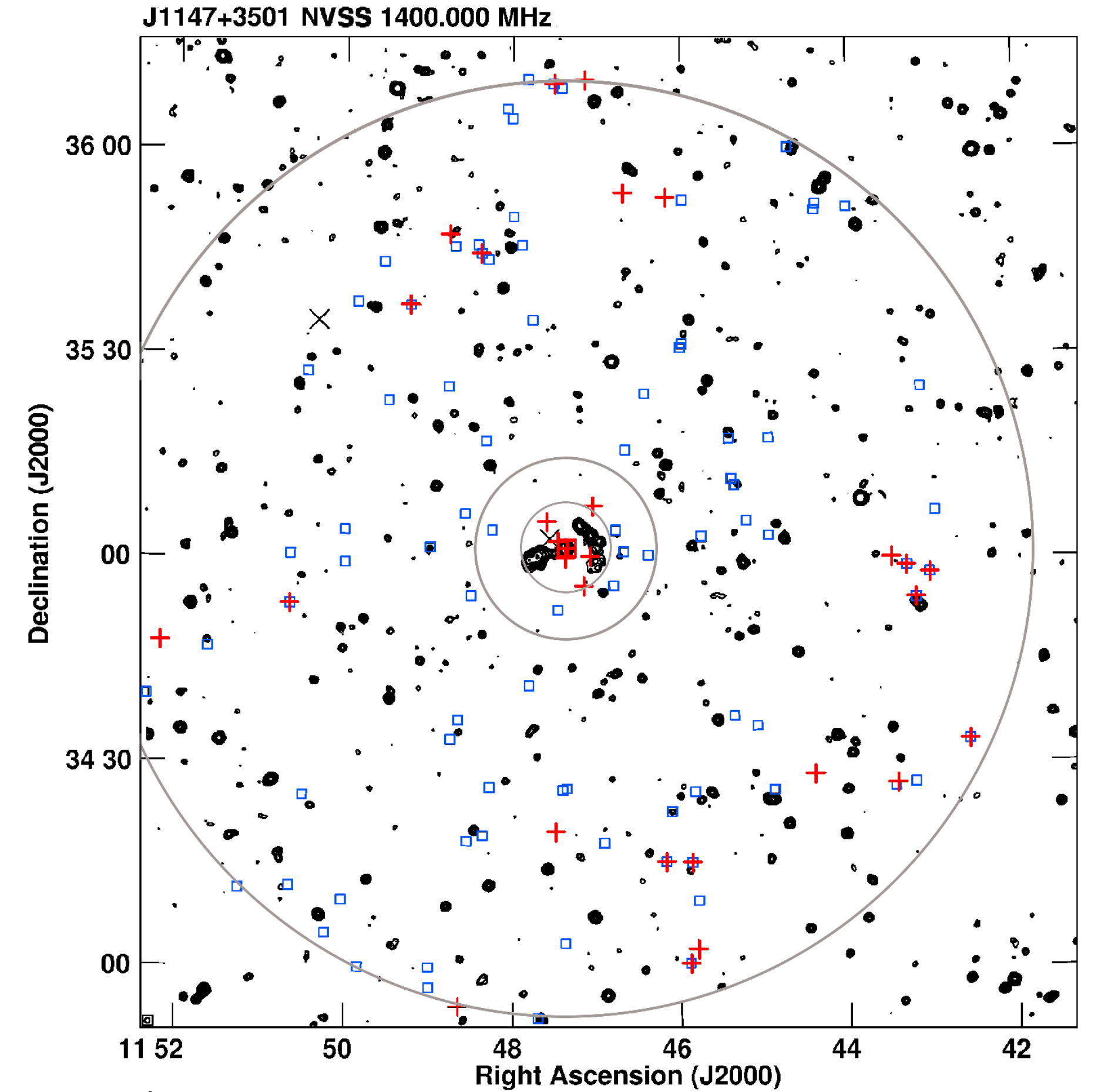}\\\\
    \includegraphics[width=0.45\columnwidth]{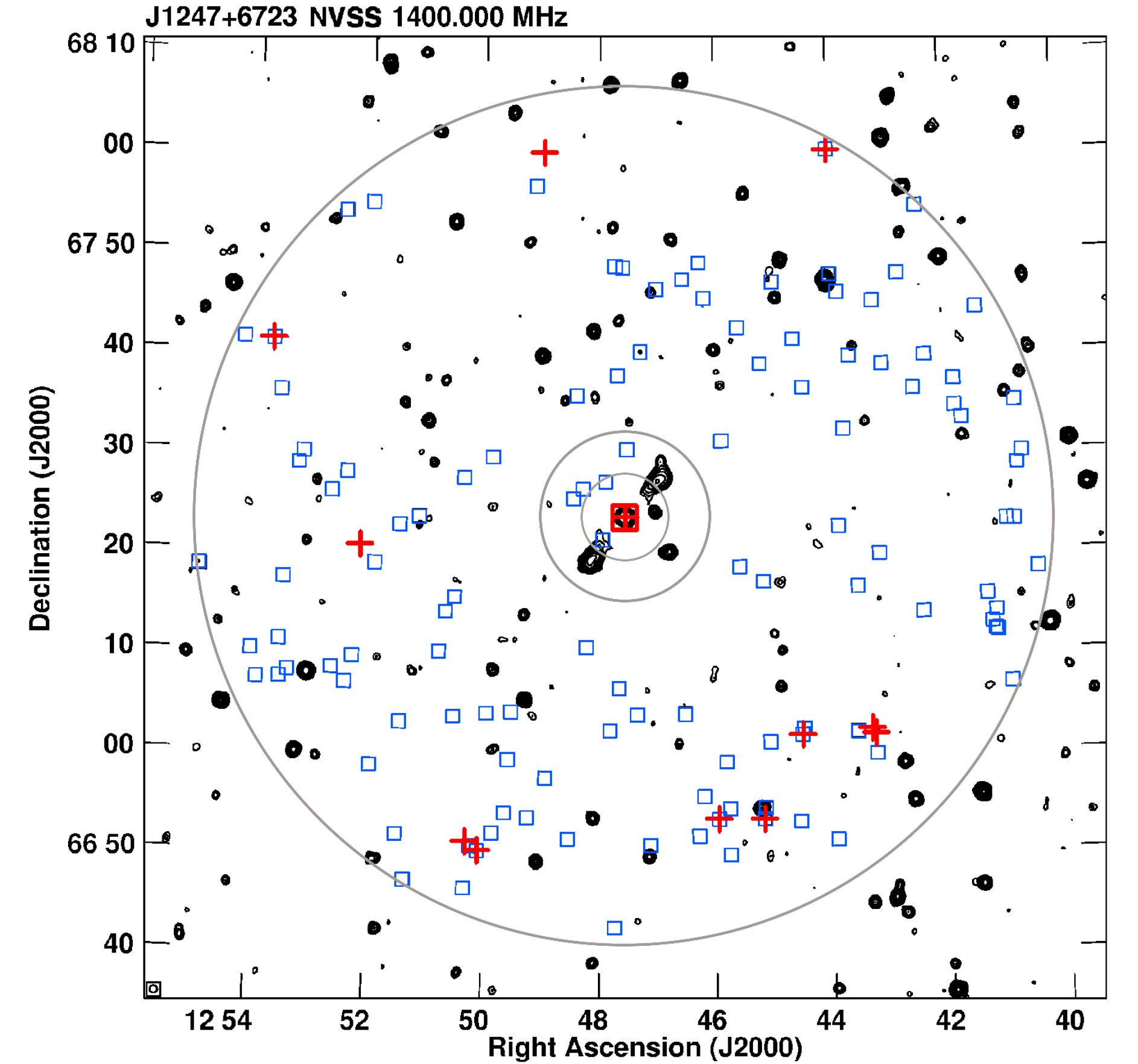}
    \includegraphics[width=0.45\columnwidth]{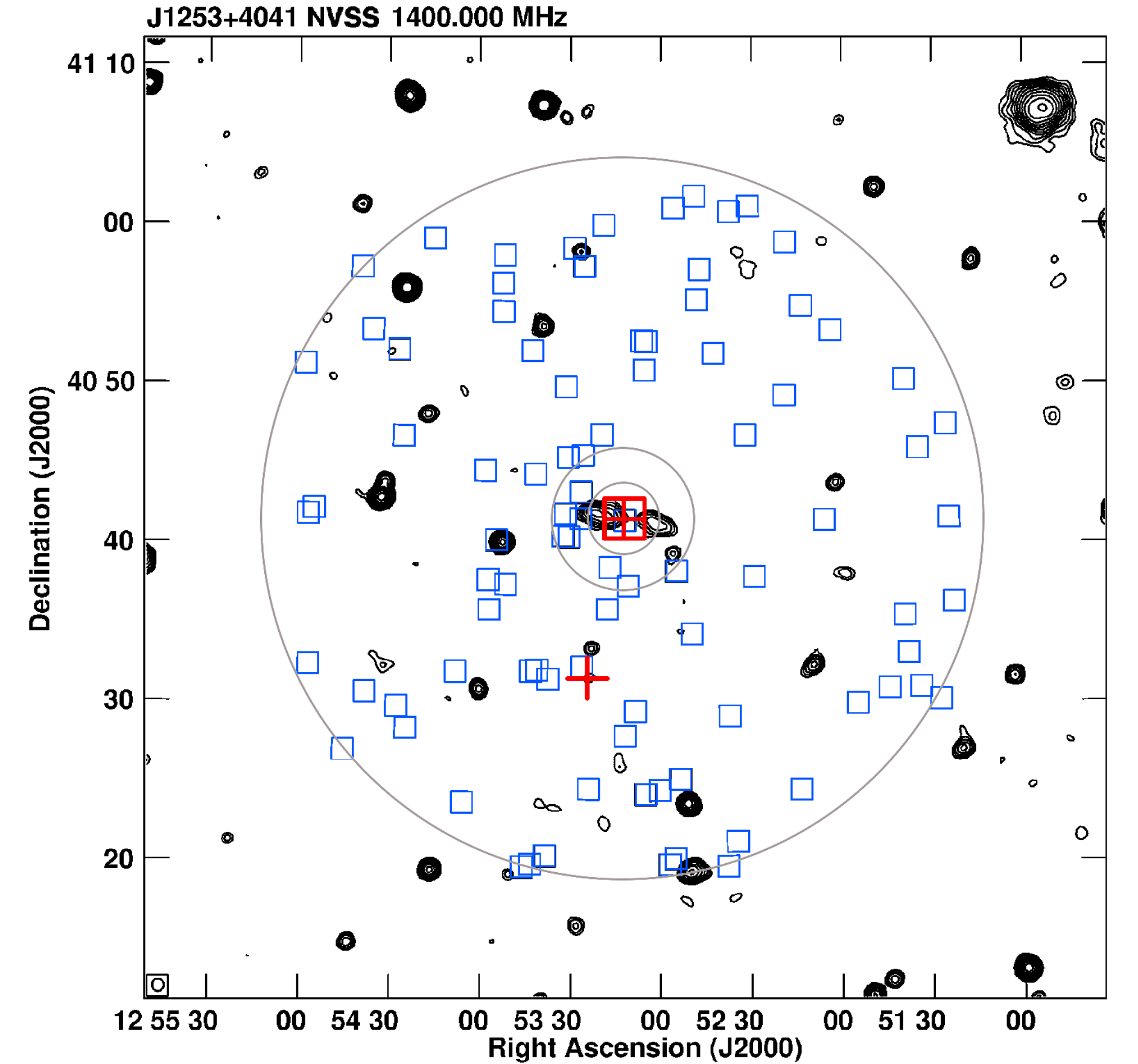}\\\\
    \includegraphics[width=0.45\columnwidth]{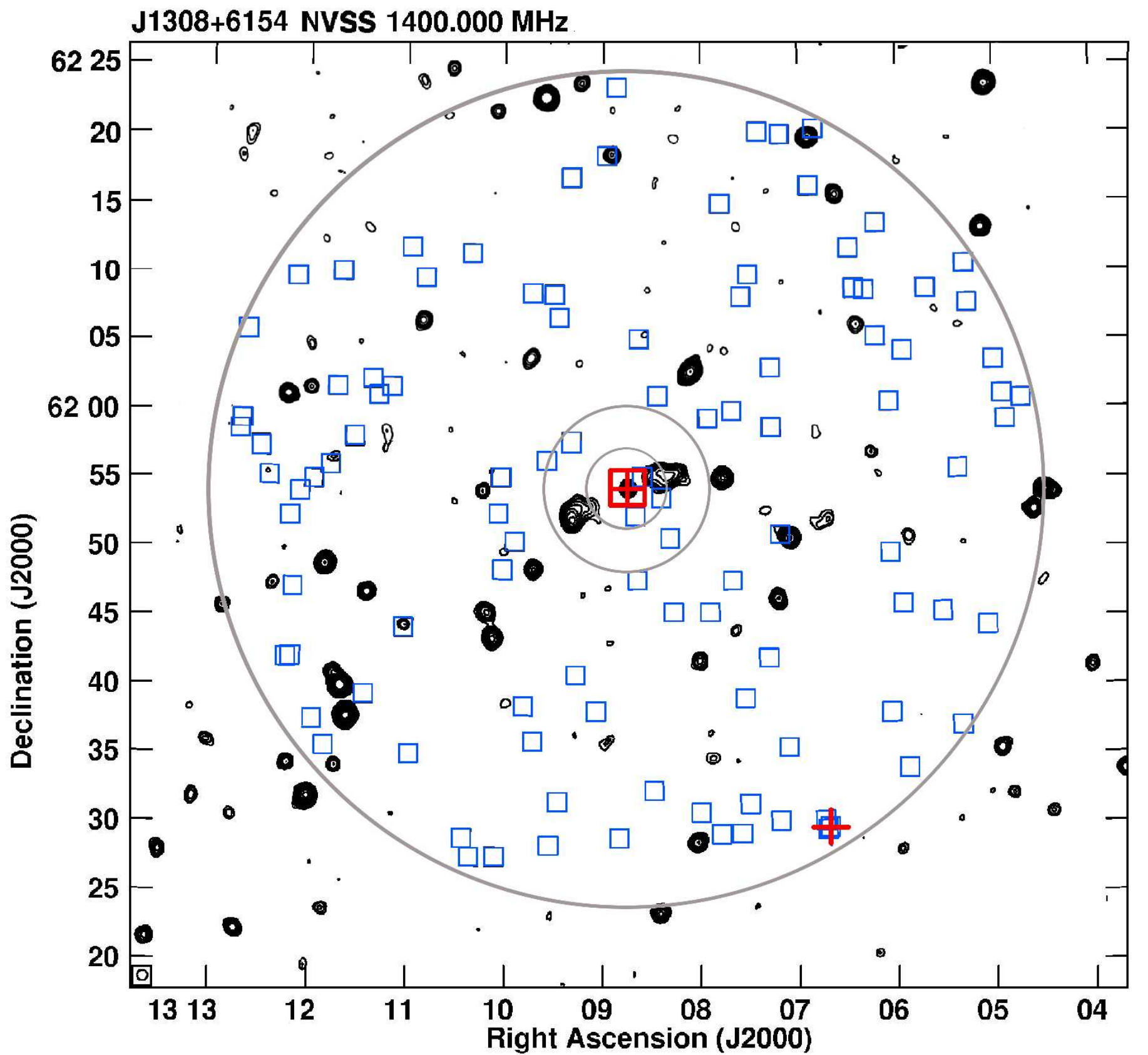}
    \includegraphics[width=0.45\columnwidth]{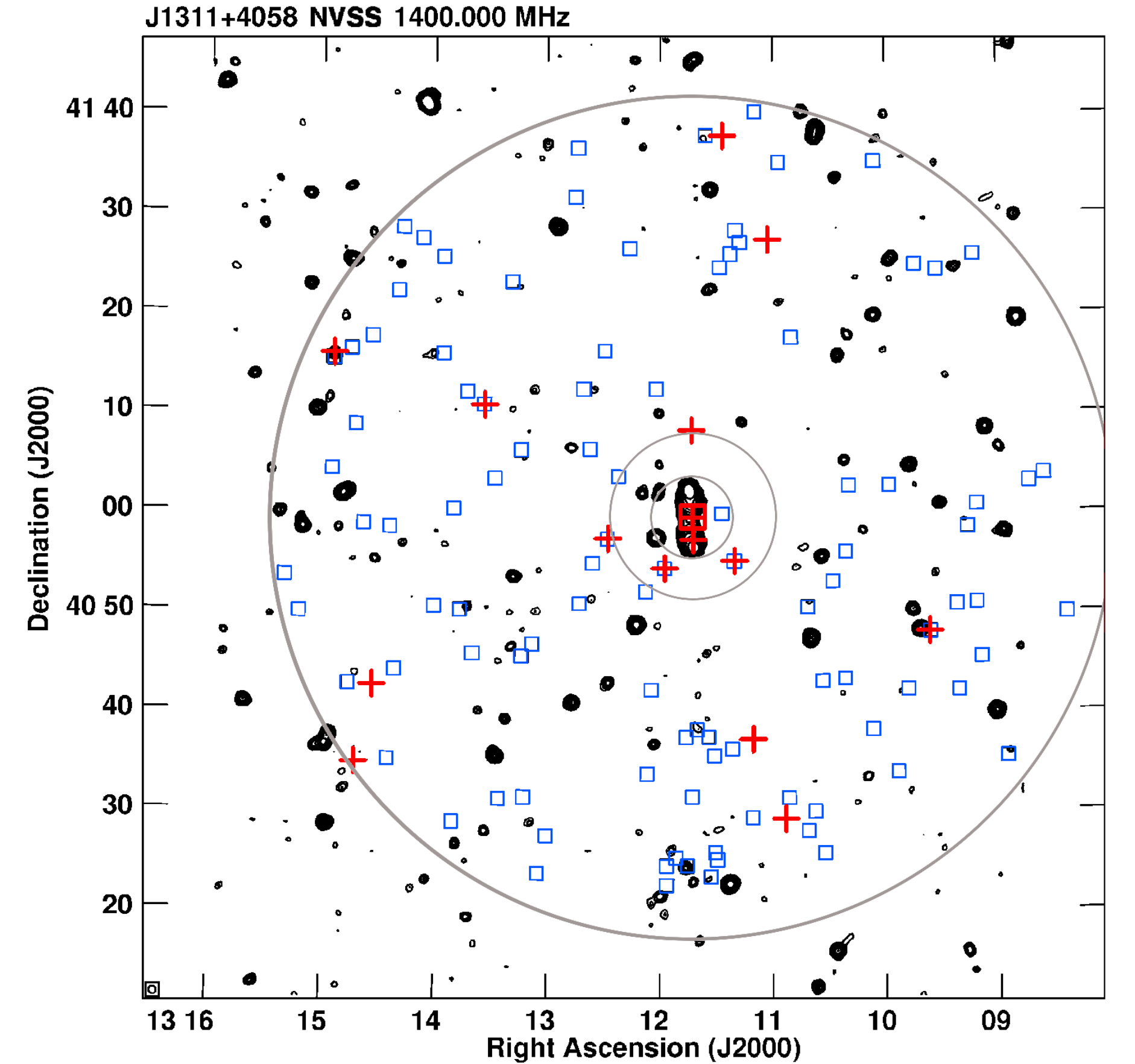}\\\\
\end{figure}

\begin{figure}
    \includegraphics[width=0.45\columnwidth]{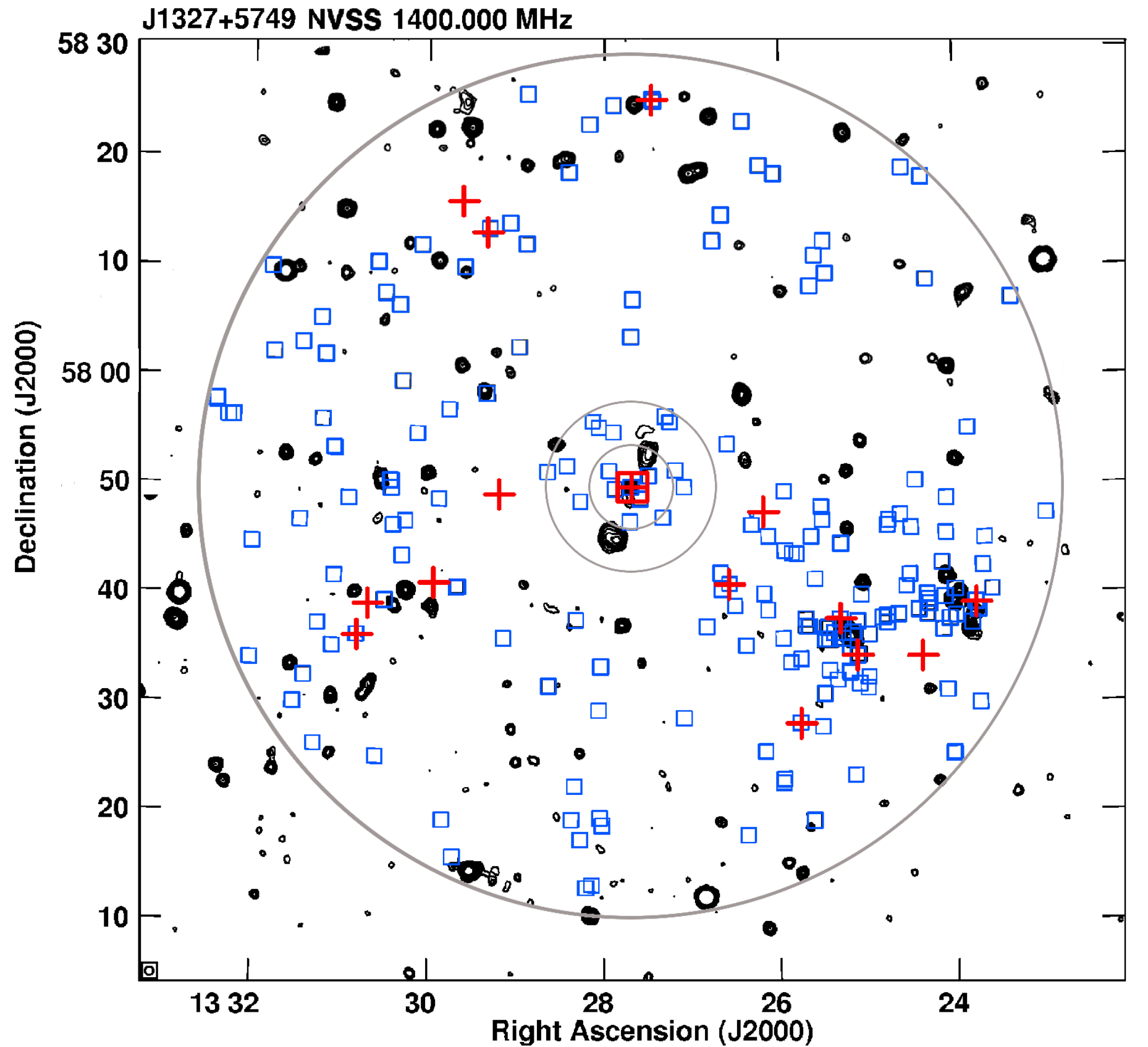}
    \includegraphics[width=0.45\columnwidth]{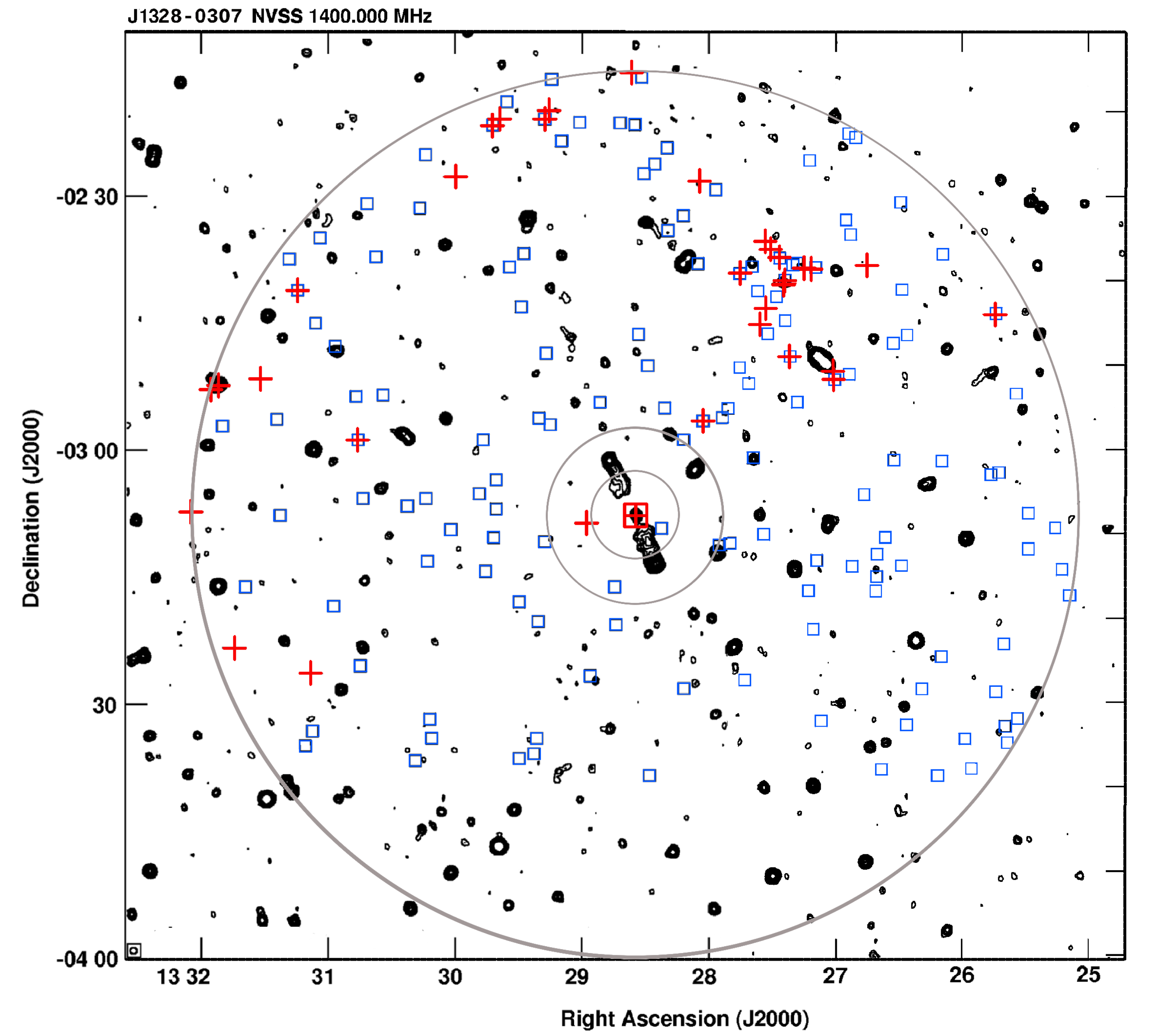}\\\\
    \includegraphics[width=0.45\columnwidth]{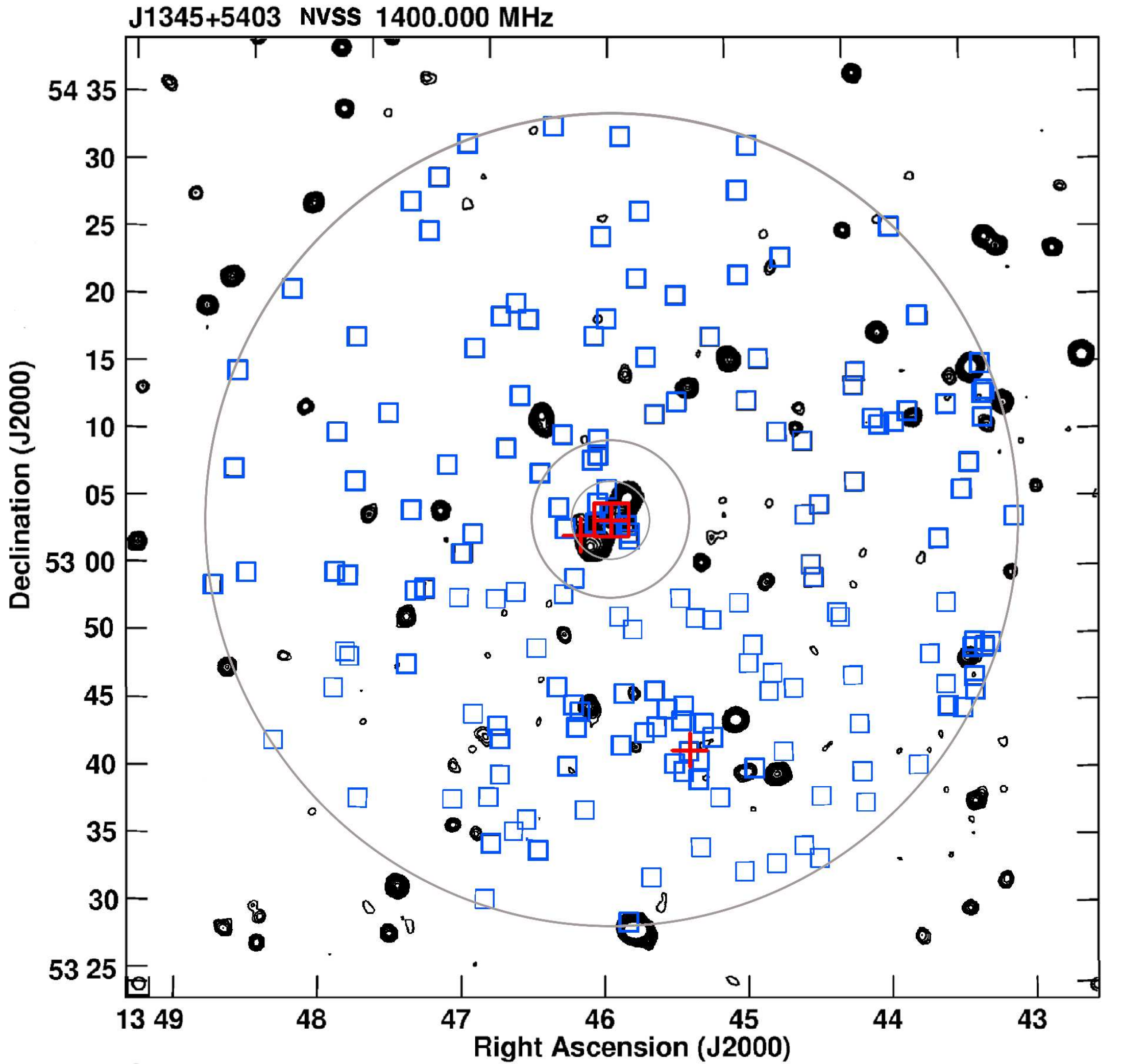} 
    \includegraphics[width=0.45\columnwidth]{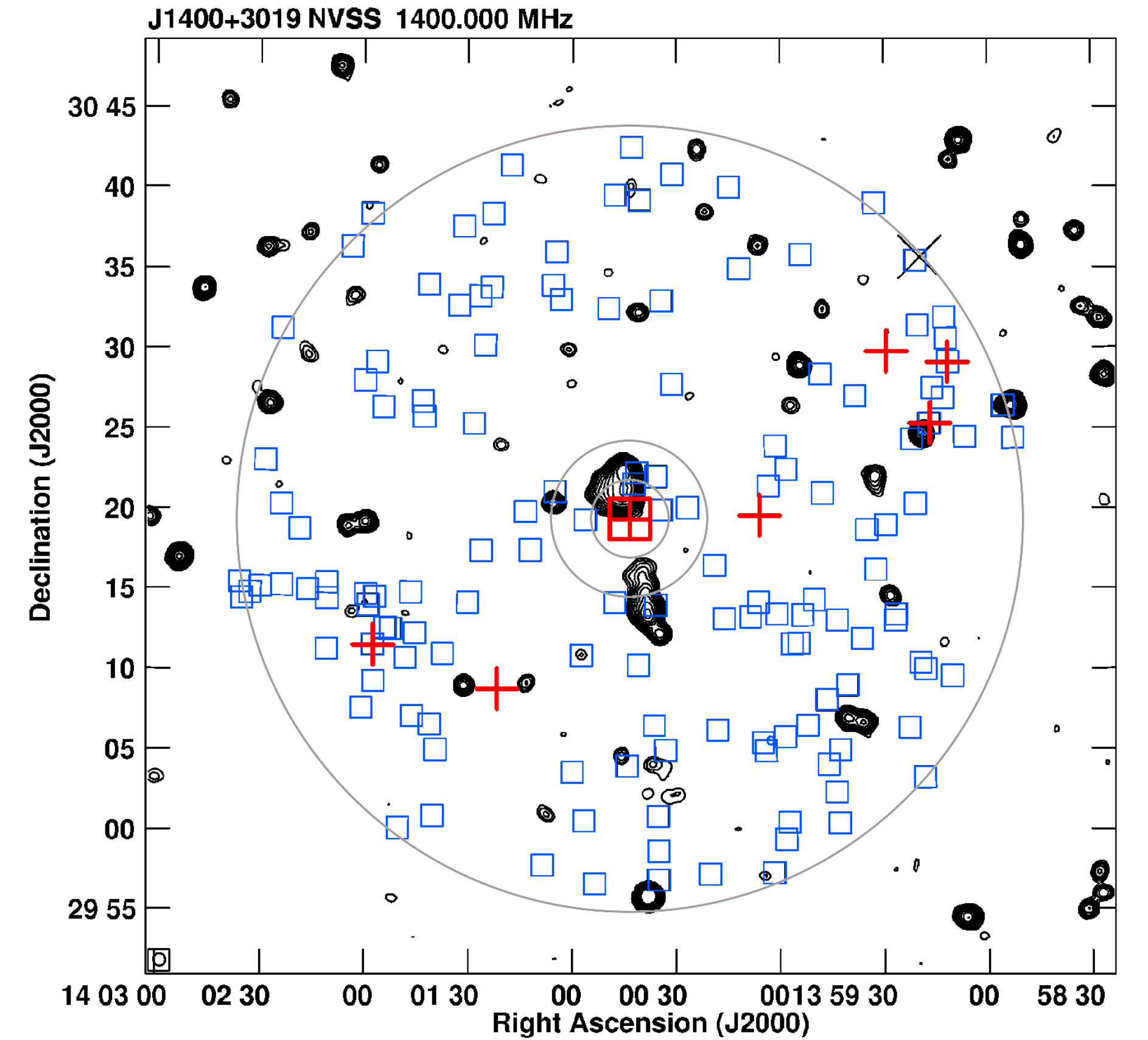}\\\\
    \includegraphics[width=0.45\columnwidth]{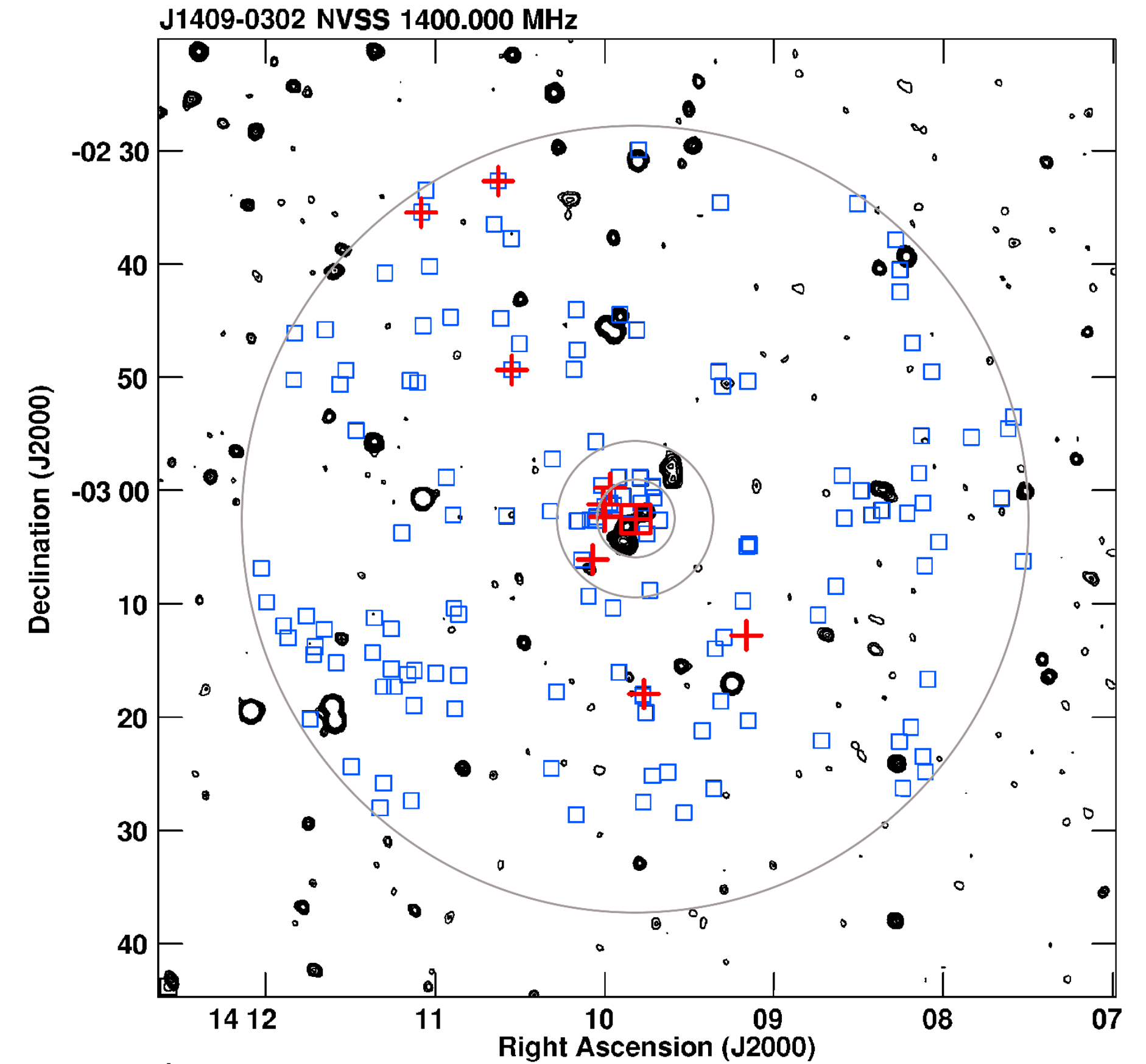}
    \includegraphics[width=0.45\columnwidth]{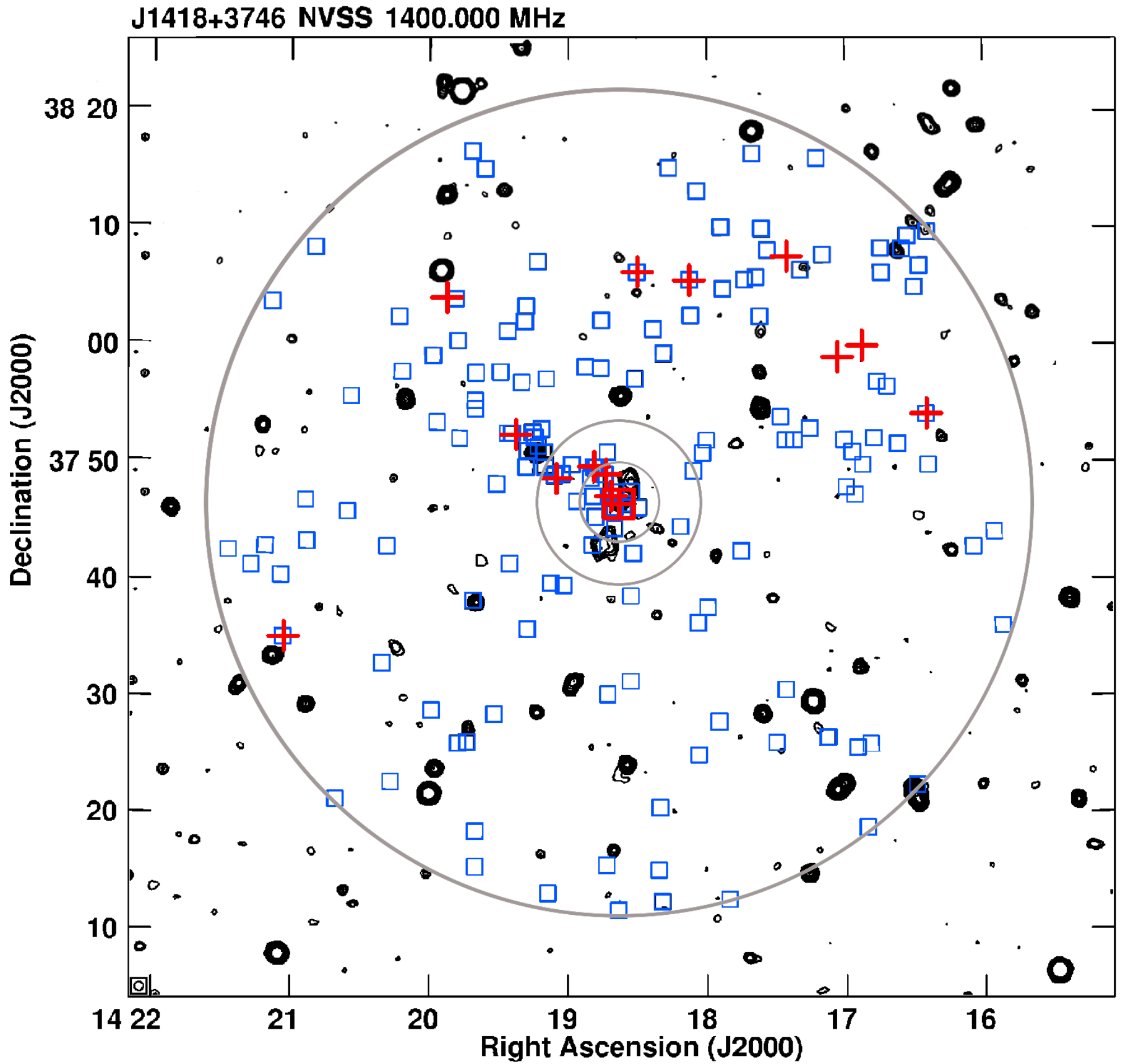} \\\\
\end{figure}

\begin{figure}
    \includegraphics[width=0.45\columnwidth]{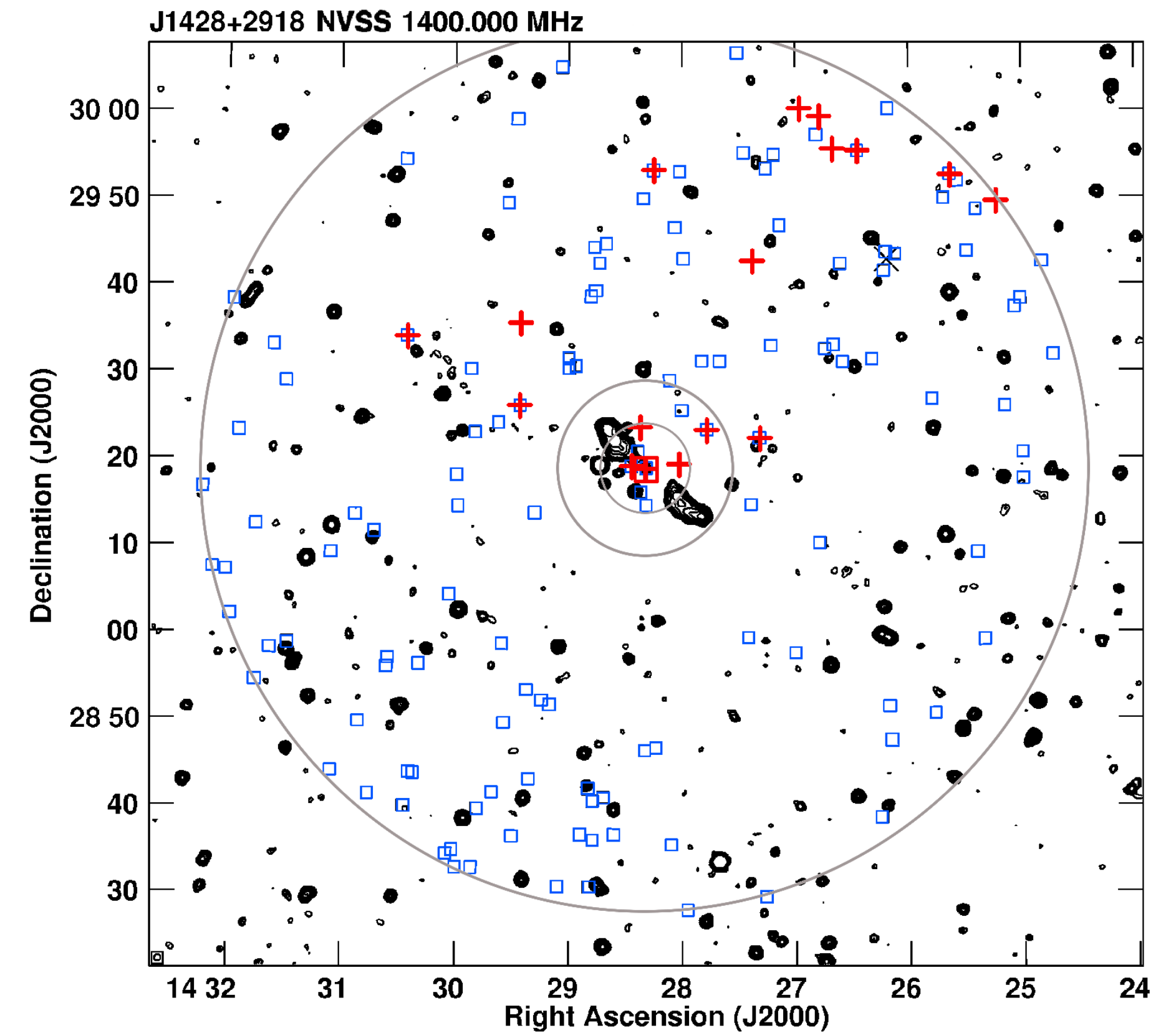}
    \includegraphics[width=0.45\columnwidth]{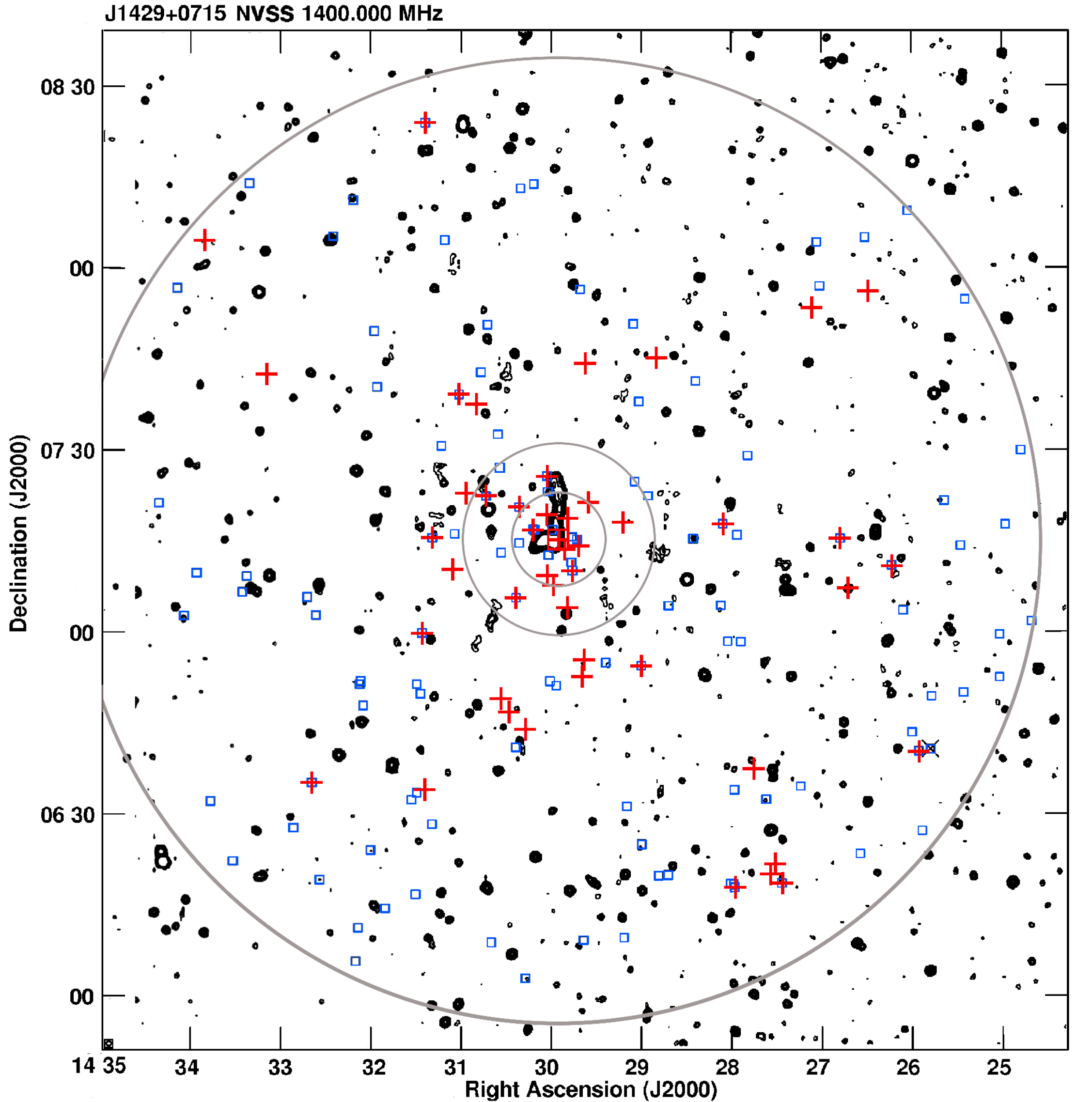}\\\\
    \includegraphics[width=0.45\columnwidth]{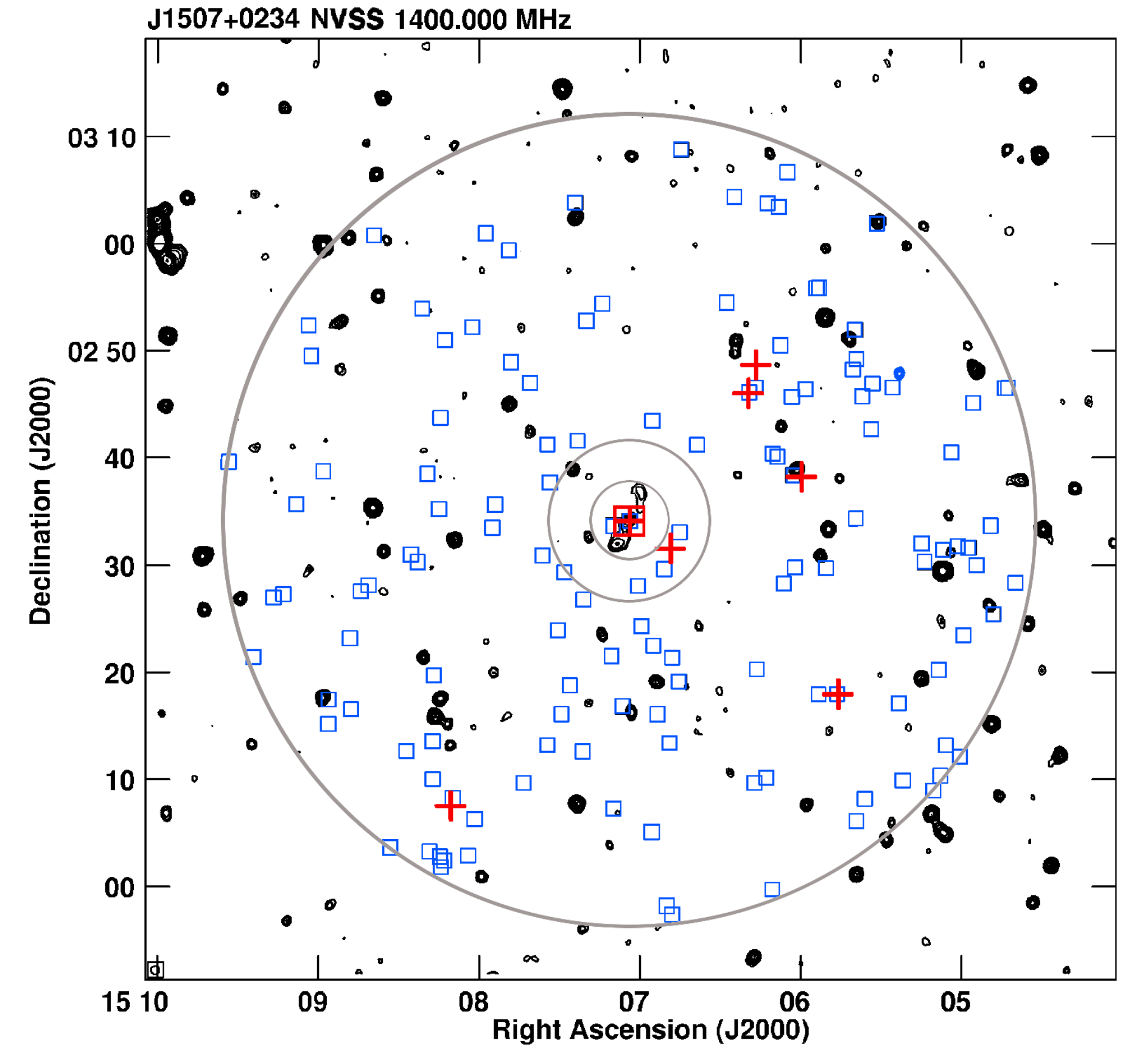} 
    \includegraphics[width=0.45\columnwidth]{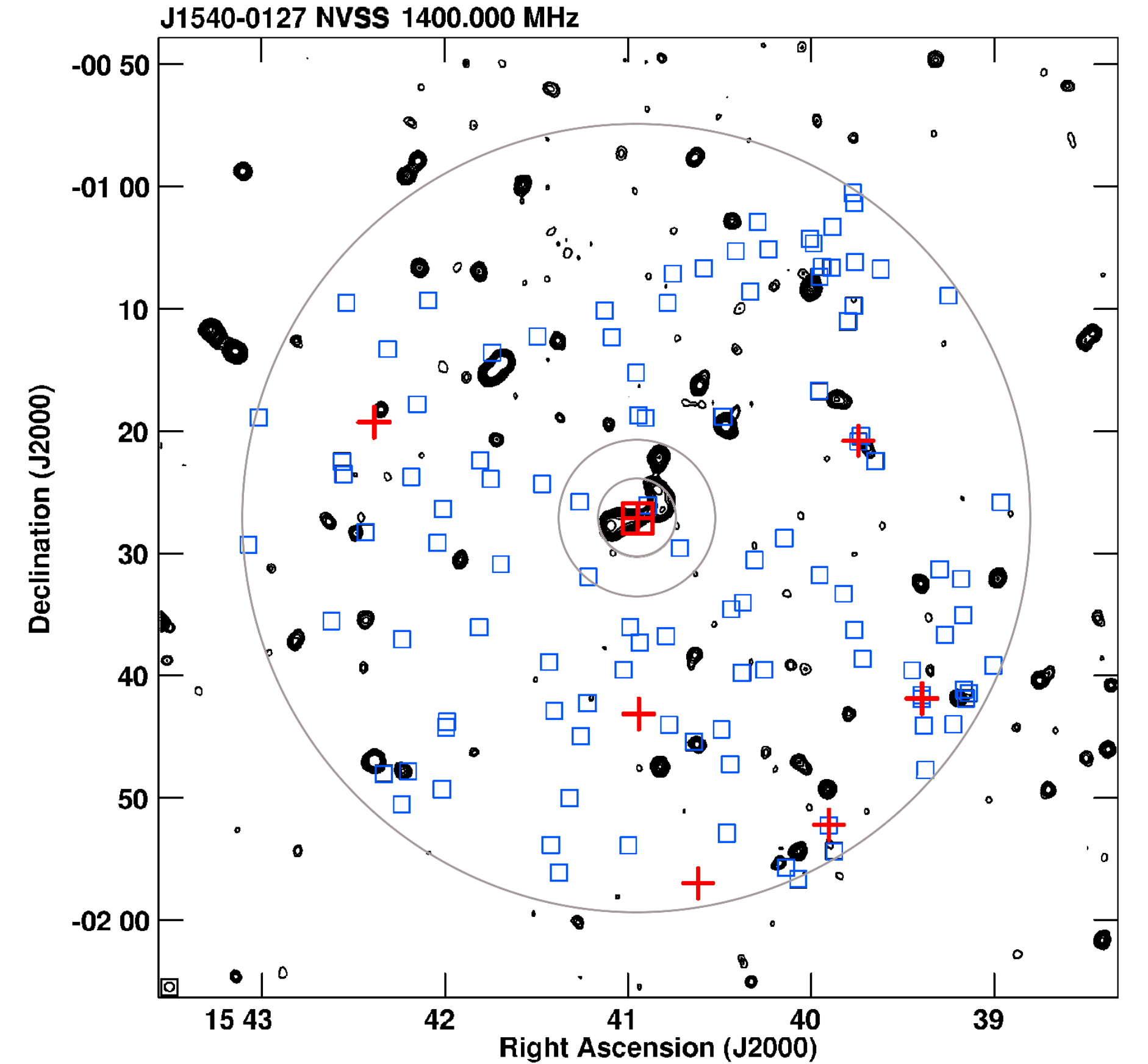}\\\\
    \includegraphics[width=0.45\columnwidth]{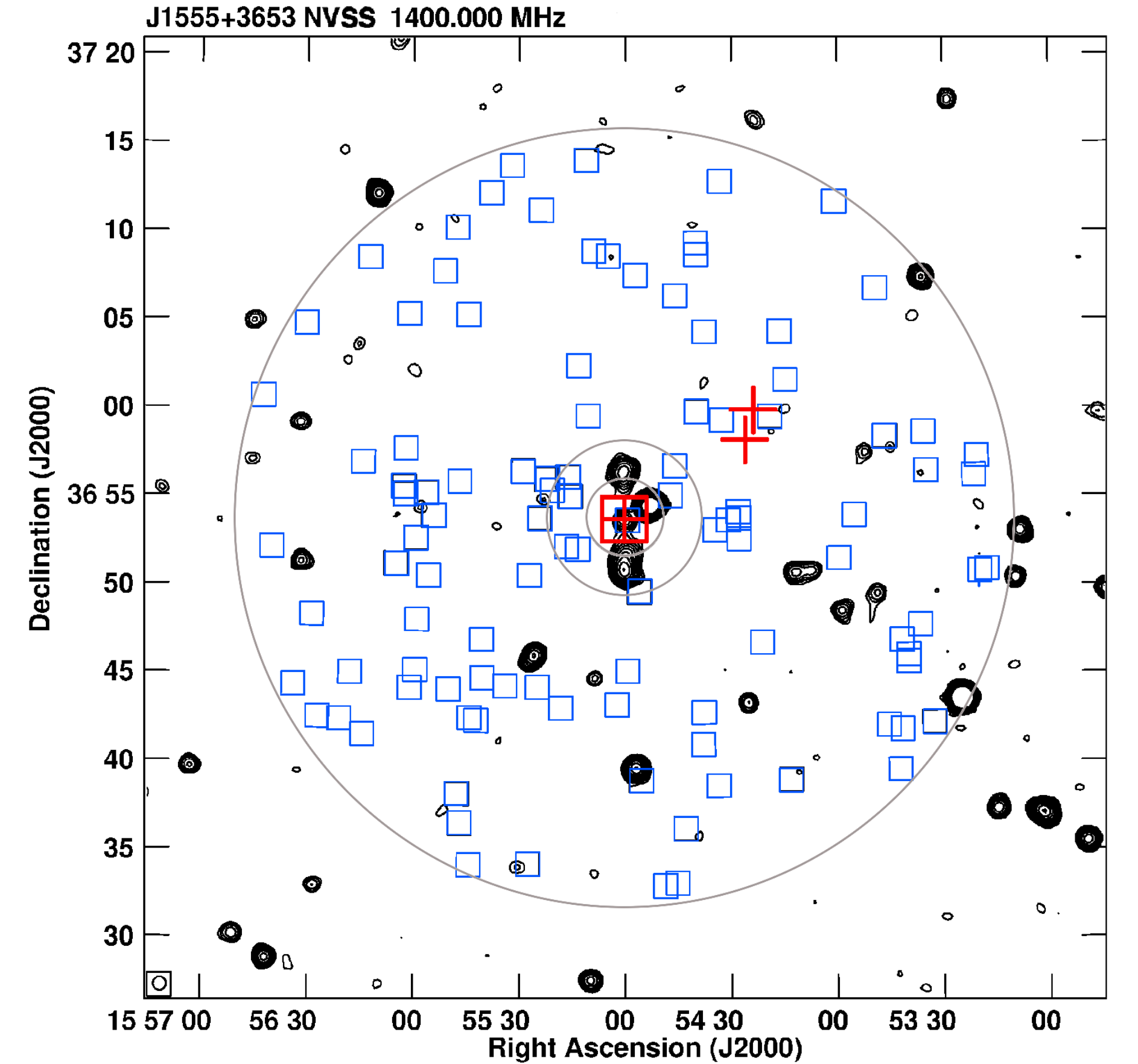}
    \includegraphics[width=0.45\columnwidth]{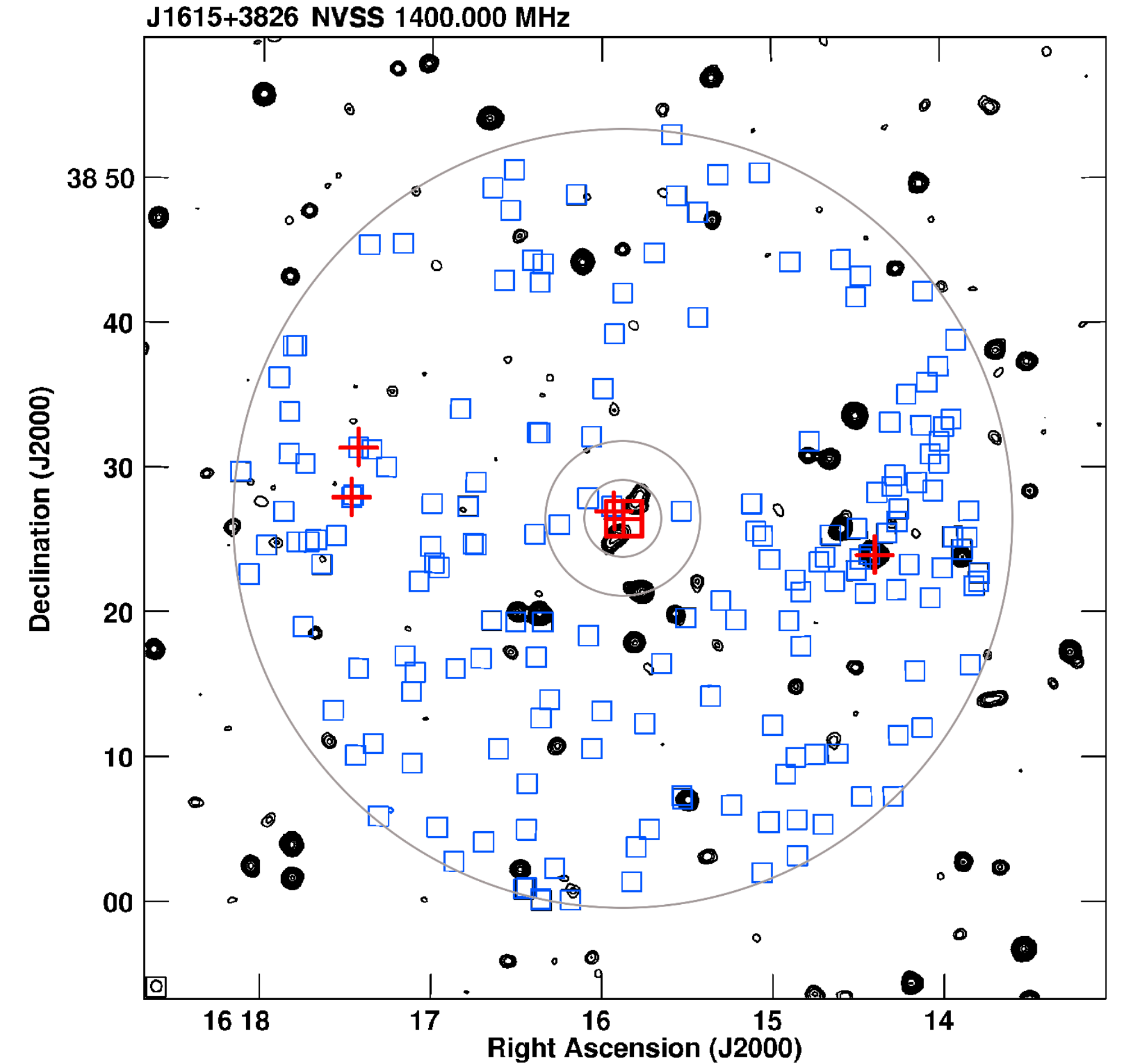} \\\\
\end{figure}
\begin{figure}
    \includegraphics[width=0.45\columnwidth]{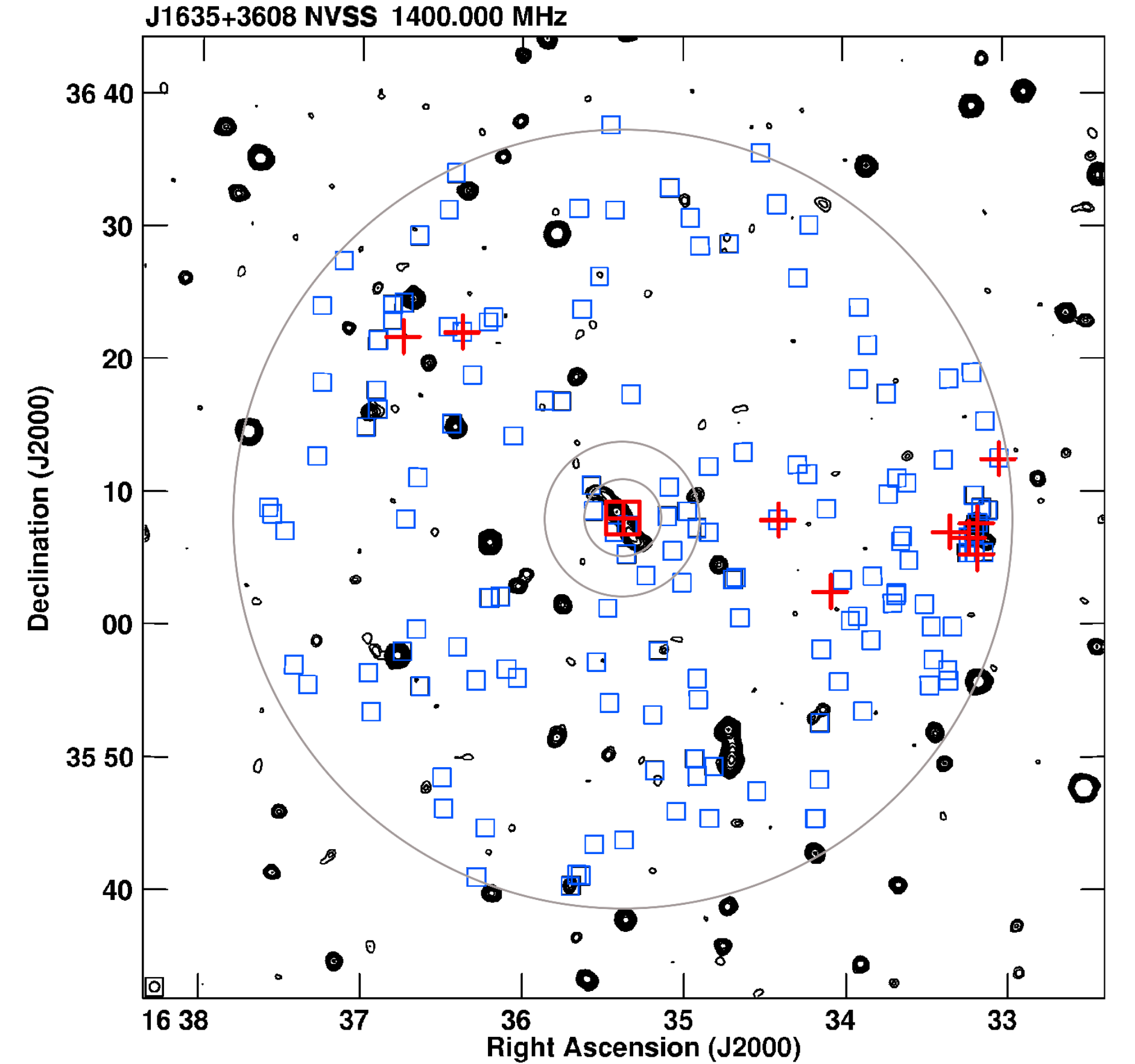}
\end{figure}
\end{onecolumn}
\end{appendix}

\end{document}